\documentclass[proc]{hch}
\usepackage{graphicx}
\begin{document}

\Course{Lectures given by both authors at the 72nd Les Houches Summer School 
on "Coherent Matter Waves", July--August 1999}
\title{Environment--Induced Decoherence and the  
Transition From Quantum to Classical}
\author{Juan Pablo Paz$^1$}
\author{Wojciech Hubert Zurek$^2$}
\address{$^1$Departamento de F\'{\i}sica J.J. Giambiagi, 
FCEN, UBA,\\ Pabell\'on 1, Ciudad Universitaria \\1428 Buenos Aires, 
Argentina\\
$^2$Theoretical Astrophysics, MS B288
Los Alamos National Laboratory, Los Alamos, NM87545, USA}
\authorsup{Juan Pablo Paz
\inst[Departamento de F\'{\i}sica J.J. Giambiagi, 
FCEN, UBA, Pabell\'on 1, Ciudad Universitaria, 1428 Buenos Aires, 
Argentina]{1}, Wojciech Hubert Zurek \inst[Theoretical Astrophysics, MS B288 
Los Alamos National Laboratory, Los Alamos, NM87545, USA]{2}}

\runningtitle{Decoherence}
\maketitle

\begin{abstract}{
We study dynamics of quantum open systems, paying special attention to those
aspects of their evolution which are relevant to the transition from 
quantum to
classical. We begin with a discussion of the conditional dynamics of simple
systems. The resulting models are straightforward but suffice to illustrate
basic physical ideas behind quantum measurements and decoherence. To discuss 
decoherence and {\it environment-induced superselection} ({\it einselection})
in a more general setting, we sketch perturbative as well as exact derivations
of several master equations valid for various systems. Using these equations 
we study einselection employing the general strategy of the predictability 
sieve. Assumptions that are usually made in the discussion of decoherence are
critically reexamined along with the ``standard lore'' to which they lead.
Restoration of quantum-classical correspondence in systems that are
classically chaotic is discussed. The dynamical second law ---it is shown---
can be traced to the same phenomena that allow for the restoration of the
correspondence principle 
in decohering chaotic systems (where it is otherwise lost
on a very short time-scale). Quantum error correction is discussed
as an example of an anti-decoherence strategy. Implications of decoherence
and einselection for the interpretation of quantum theory are briefly
pointed out. }\end{abstract}

\section{Introduction and overview}

The quantum origin of the classical world was so difficult to imagine 
for the forefathers of quantum theory that they were often willing 
to either postulate its independent existence (Bohr),
or even to give up quantum theory and look for something with more fundamental 
classical underpinnings (de Broglie, and, to a lesser extent, also Einstein).
The source of the problem is the quantum principle of superposition, which, 
in effect, exponentially expands the set of available states to all of the
conceivable superpositions. Thus, coherent superpositions of dead and alive 
cats have ---in the light of the quantum theory--- 
the same right to exist as either of the two classical alternatives.
Within the Hilbert space describing a given system ``classically legal''
states are exceptional. The set of all states in the Hilbert 
space is enormous as compared with the size of the set of 
states where one finds 
classical systems. Yet, it is a fact of life that classical objects are only 
found in a very small subset of all possible (and in principle, allowed) 
states.
So, one has to explain the origin of this apparent ``super-selection'' rule 
that prevents the existence of most states in the Hilbert space of 
some physical systems. Decoherence and its principal consequence 
---environment-induced super-selection or {\sl einselection}--- account for 
this experimental fact of life.

Decoherence is caused by the interaction between the system and its 
environment.
Under a variety of conditions, which are particularly easy to satisfy for 
macroscopic objects, it leads to the einselection of a small subset of 
quasi-classical states from within the enormous Hilbert space. The 
classicality 
is then an emergent property, induced in the system by its interaction with 
the environment.  Arbitrary superpositions are dismissed, and a preferred 
set of
``pointer states'' emerges. These preferred states are the candidate classical 
states. They correspond to the definite readings of the apparatus pointer
in quantum measurements, as well as to the points in the phase space of 
a classical dynamical system.

The role of the process of decoherence in inducing classicality has become
clear only relatively recently ---within the past two decades. The key idea 
is relatively simple: An environment of a quantum system can, in effect, 
monitor its states through continuous interaction. The imprint of the 
system left on the environment will contain information about selected 
states of the system. The states that leave the imprint without getting 
perturbed in the process are the preferred states. Thus, the key property
of quasi-classical pointer states is their insensitivity to monitoring by
---and consequently their resistance to the entanglement caused by---
interaction with the environment: states that entangle least are most stable. 
They are also, almost by definition, the only states that remain
an accurate description of the the system alone: All other states evolve into
joint system-environment states, preserving their purity (and, consequently,
the information the 
observer has about them) only when both the system and the 
environment are included in a larger ``super system". 

The fact that the interaction between quantum systems produces entanglement
was well known almost since the beginning of quantum theory. Indeed, because 
the ideas of decoherence and einselection rely on quantum theory, and on
quantum theory alone, it may be useful to ask why  it took so long to 
arrive at a natural explanation of the quantum origins of classicality. There 
are several possible
explanations for this delay. We shall return to them later in the paper. 
But, for the moment, it is useful to note that the ability of  
environment-induced decoherence to result in the same set of preferred 
states, essentially independently of the initial state of the system and the 
environment, is crucial. This was not appreciated until relatively recently 
\cite{Zurek81, Zurek82}. It is precisely this stability of the set of 
preferred states that allows
them to be regarded as good candidates for the quantum counterparts of
classical reality. Indeed, only still more recent research on the 
predictability sieve has allowed for more fundamental and general
understanding of the emergent classicality (see \cite{Zurek93,ZHP93} 
and also \cite{Gallis}).

The prejudice that seems to have delayed serious study of the role of the
``openness" of a quantum system in the emergence of classicality is itself 
rooted in the classical way of thinking about the Universe. 
Within the context 
of classical physics, all fundamental questions were always settled in the 
context of closed
systems. The standard strategy to ensure isolation involved enlarging a system 
---i. e., by including the immediate environment. The expectation was that
in this manner one can always reduce any open system to a larger
closed system. This strategy does indeed work in classical physics, where 
the enlargement can help in satisfying conservation laws for quantities such
as energy or momentum. It fails in the quantum case under discussion, 
because now
it is the information (about the state of the system) that  
must be prevented from spreading. Information is much harder to contain when 
the system in question becomes larger. Thus, in the end, the only truly 
isolated macroscopic system is the Universe as a whole. And we, the 
observers, are certainly not in a position to study it from the outside.

In what sense is the preferred set of states preferred? It is clear that 
generic superposition of the members of this preferred set will decay into 
mixtures. On the other hand, if the initial state is just one of the 
members of 
the preferred set, the temporal evolution will minimally affect the state, 
which will resist becoming entangled with the environment. 
Einselection can thus be thought of as a process by which a ``record'' 
of the state of the system is created dynamically (through  
interaction) in the state of the environment. It is this ongoing process
by which the system is being continuously monitored by the environment 
that leads to the emergence of a natural set of preferred states that 
are the least affected by the interaction. 

As sketched above, the physical principles of decoherence and einselection
appear, in retrospect, rather straightforward. How much can be accomplished
by exploring their consequences?
There are several interesting and important questions
that naturally arise in this context and that have been asked (and 
answered, in most cases) over the last two decades. First, one naturally 
asks how much can we explain with these ideas (i.e., is it 
consistent to think that all objects that are known to behave classically
are doing so because of decoherence?). A closely related question is
the one concerning natural time-scales associated with decoherence.  
How fast does decoherence take place? This is a very important
question because a first look at the decoherence process may leave us
wondering if decoherence may be consistent with the existence of a 
``reversible'' classical world. Thus, if one believes that classicality 
is really an emergent property of quantum open systems one may be tempted
to conclude that the existence of emergent classicality will always be
accompanied by other manifestations of openness such as dissipation of 
energy into the environment (this would be a problem because, as we know, 
there are many systems that behave classically while conserving 
energy). Second, one also wonders how, in detail, is the preferred set 
of states
dynamically selected through the interaction with the environment. In 
particular, it is interesting to know how this   
pointer basis is determined by the structure of the interaction 
Hamiltonian between system and environment and/or to the other details of the  
physics involved. Third, a related question arises in 
this context: are there observable manifestations of decoherence other than 
einselection?

A remarkable characteristic of the current debates on the nature of the 
quantum to classical transition and on the problem
of quantum measurement is that for the first time in history there have
been actual experiments probing the boundary between the quantum and
the classical domains in a controlled way 
\cite{Bruneetal,Myattetal,Raymer,XXX,Raizen}. Controlled decoherence 
experiments (which are very difficult because nature provides us with classical
or quantum systems but not with objects whose interaction with the 
environment can be controlled at will) were recently carried on for the 
first time and help us in understanding the nature of this process. Some 
of the most notable experiments in this area were performed at the Ecole 
Normale Superieure in Paris and are be part of Dr. Brune's lectures. 

Our lectures start with an introduction to quantum conditional dynamics
using two-state systems. Conditional dynamics is responsible both for 
setting up the problem of measurement, and for the decoherence and einselection
that solve it. The resulting models are straightforward and can serve 
in the idealized studies of the measurement process. However, they are clearly 
too simple to be realistic ---classicality is, after all, a property of 
essentially every sufficiently macroscopic object. To discuss decoherence 
and einselection in this more general setting, we shall therefore study 
dynamics 
of quantum open systems. Section 3 is devoted to the derivation of 
the key tool ---a master
equation for the reduced density matrix. This basic tool is
immediately used in section 4, where environment-induced superselection
is studied, including, in particular, the predictability sieve.
Section 5 analyzes some ``loose ends" ---that is, essentially technical issues
that are usually omitted in the derivations of the master equations. 
We show there that although the qualitative conclusions arrived at on the basis
of the ``naive" master equation approach are essentially unaffected by 
the detailed examination of some of the idealized assumptions that go 
into its derivation, quantitative estimates can change quite significantly when
a more realistic approach is adopted. Section 6 is devoted to the 
study of the effect of decoherence on the quantum-classical correspondence in
systems that are classically chaotic. We show there that decoherence not only 
explains the origin of classical dynamics, but that it may be 
responsible for the loss of information that accounts for the 
second law of thermodynamics as well. Section 7 is devoted to quantum error
correction ---to the strategies which can be used to suppress decoherence.
The summary and conclusions are briefly stated in Section 8.

\section{Quantum Measurements}

In this section we shall introduce the measurement problem ---the issue that 
has dominated the discussion of the relation between quantum and classical 
for a very long time. This will afford us the opportunity to study conditional 
dynamics that will be employed in one form or another throughout this 
review. 
Such interactions are necessary to achieve entanglement between quantum 
systems that set up the measurement problem. They are necessary for  
accomplishing  
decoherence, which leads to environment-induced superselection (or
einselection), and thus resolves many of the problems arising on the border 
between quantum and classical. Last, but 
not least, quantum conditional dynamics
and entanglement underlie quantum logic and will be of importance in 
the latter part of the paper devoted to quantum error correction.

Predictability is rightly regarded as one of the key attributes of classical 
dynamics. On the other hand, the defining feature of quantum mechanics 
is thought to be its probabilistic nature, which manifests itself in 
measurements. This discord between classical determinism and quantum 
randomness is often blamed for the difficulties with interpretation of 
quantum theory. Yet, the fundamental equations of either classical or 
quantum theory allow them 
---indeed, demand of them--- to be perfectly predictable: It 
is just that what 
can be predicted with certainty, especially in the quantum case, cannot be 
often accessed by measurements. And, conversely, what can be measured 
in an evolving quantum system cannot usually be 
predicted, except in the probabilistic sense. 

The 
Schr\"odinger equation allows one to predict the state of an isolated system 
at any subsequent moment of time. In an isolated quantum system, dynamical 
evolution is strictly deterministic. This perfect quantum predictability could 
be of use only if one were to measure observables that have the resulting
evolving state as one of its eigenstates. These observables are generally 
inaccessible to reasonable measuring devices, and therefore are 
of no interest. 

Quantum determinism is of little use for an observer who is only a part 
of the whole system. The overall quantum determinism could have predictive
power only for someone who is (i) monitoring quantum systems 
from the outside. 
Moreover, it would help if the observer was endowed with (ii) enough memory to 
measure and store data, and (iii) sufficient ability to compute and to model 
deterministic evolution of the system of interest. For an observer trapped 
inside of the quantum universe, this is obviously not the case.

The universe is all there is. Therefore, by definition, it is a closed quantum 
system. Given the deterministic nature of the Schr\"odinger evolution, one may 
be surprised that there is a problem with the interpretation of quantum theory.
After all, the interpretational ideal often mentioned in such discussions is 
deterministic Newtonian dynamics. However, the interpretation problem stems
from the fact that deterministic unitary evolution of quantum theory 
is incompatible with classical determinism. Indeed, as the studies 
of chaotic systems demonstrate, classical dynamics has more 
room for randomness than quantum physics. 

States of the quantum systems are perturbed by the very act
of monitoring them. The elemental unpredictability associated with the act
of observation cannot be avoided unless the observer knows in advance which
observables can be measured with impunity. This feature of quantum information 
is essential to guarantee the security of quantum cryptography 
(see lectures of A. Ekert; also \cite{Bennett}) ---the state of 
a quantum system cannot be found out by the eavesdropper unless the observation
is carried out on the same basis as the one used by the intended recipient of 
the message. The ``no cloning theorem'' \cite{WoottersandZurek, Dieks} 
prevents duplication of quantum information ---amplification is associated 
with breaking the symmetry associated with the superposition principle.

Environment-induced superselection rules allow the observer to be a succesfull
eavesdropper, and to extract useful information from the quantum systems 
without the environment getting in the way because (in contrast to the 
strategies employed in quantum cryptography) the measurements carried out 
by the environment are restricted to few observables. The state of the system
is therefore of necessity ``precollapsed'' and commutes with these observables.
Further measurements carried out by the observer will only reveal (rather than
perturb) the pre-existing state of affairs. Thus, environment-induced  
decoherence supplies a justification for the persistent impression 
of ``reality''. In contrast to the observables encountered in the microscopic
realm, macroscopic quantum systems can appear only in one of the preselected
(pointer basis) set of quantum states. The ``collapse of the wavepacket'' 
viewed in this way is just a familiar classical process of finding out which 
of the possible outcomes has actually occurred. The danger of interference 
between the alternatives was eradicated by decoherence long before the observer
became involved.  
 
How can one ever hope for a resolution that would allow for the familiar 
combination of classical determinism and classical randomness to emerge? 
At the 
risk of anticipating results that will be justified in detail only later, we 
note that quantum determinism may be relevant only for an observer who knows
the initial state of an isolated quantum system. For a quantum observer 
immersed
in a quantum universe this is a very rare exception, attainable only in 
carefully controlled laboratory experiments, and only for rather small quantum 
systems. The information capacity, memory, and information processing 
abilities 
of an observer that is a (macroscopic, yet comparatively small) subsystem of 
the universe are miniscule compared to the task of simulating even a small 
quantum system, let alone the universe as a whole. And as soon as the idea of 
the observer knowing the entire state of the universe is recognized as 
not 
feasible, ``environmental monitoring'' of both the state of the observer and 
of the observables he recorded begins to matter. An observer with decohering 
memory can keep reliable records only in the einselected states of 
his/her memory bits \cite{Zurek91,Zurek93,Zurek98a,Zurek98b,Tegmark99}. 
Records will have predictive power only when they correlate with the 
einselected observables in the rest of the universe.

\subsection{Bit-by-bit measurement and quantum entanglement}

This problem of transition from quantum determinism to classical 
definiteness is illustrated most vividly by the analysis of quantum 
measurements. An answer to a ``generic'' question about the state of a quantum
system (and the outcome of a measurement of the corresponding observable) is 
{\it not} deterministic. In the usual textbook discussions, 
this random element 
is blamed on the ``collapse of the wavepacket'' that is invoked 
whenever a quantum system comes into contact with a classical apparatus. 
In a fully quantum discussion of the problem, this issue still 
arises, in spite (or rather because) of the overall deterministic 
quantum evolution of the state vector of the universe. Indeed, as carefully
pointed out by von Neumann \cite{vonNeumann32} in his quantum analysis of 
measurements, there seems to be no room for ``real collapse'' in purely 
unitary models of measurements. 

To illustrate the ensuing difficulties, we consider with von Neumann
a quantum system ${\cal S}$ initially in a state $|\psi\rangle$ 
interacting with 
a quantum apparatus ${\cal A}$ initially in a state $|A_0\rangle$. The interaction 
will generally result in an entangled final state,
\begin{equation} |\Psi_0\rangle = |\psi\rangle |A_0\rangle = (\sum_i a_i |s_i\rangle)|A_0\rangle  
\longrightarrow 
\sum_i a_i |s_i\rangle |A_i\rangle = |\Psi_t\rangle.\label{two.one}
\end{equation}
Here $\{|A_i\rangle\}$ and $\{|s_i\rangle\}$ are states in the Hilbert 
spaces of the 
apparatus and of the system, respectively, and $a_i$ are complex
coefficients. This 
transition can be accomplished by means of a unitary Schr\"odinger evolution.
It leads to an uncomfortable conclusion. All that an 
appropriate interaction between ${\cal A}$ and ${\cal S}$ can achieve 
is putting the measuring apparatus (or an observer) in an EPR-like 
{\it entangled state} of all the possible outcomes consistent with 
the initial state \cite{Zurek81}. Operationally, this EPR-like nature 
of the state emerging from the pre-measurement (as the step achieved 
by Eq. (\ref{two.one}) is often called) can be made more explicit by 
rewriting the sum in a different basis 
\begin{equation}
|\Psi_t\rangle =  \sum_i a_i |A_i\rangle |s_i\rangle = \sum_i b_i |B_i\rangle |r_i\rangle = |\Psi_t\rangle \ .
\label{two.two}
\end{equation}
All we have done is use an alternative basis for both the apparatus and
the 
system, exploiting the freedom of choice guaranteed by the quantum principle 
of superposition.
Therefore, if one were to associate the state of the apparatus (observer) with 
a state in the decomposition of $|\Psi_t\rangle$, then even before one 
could start 
enquiring about the specific outcome of the measurement one would have 
to decide what decomposition of $|\Psi_t\rangle$ is to be used, because
the change 
of the basis corresponds to a redefinition of the measured quantity. 

One could make the clash between quantum and classical even more dramatic 
by making an additional measurement on the same quantum system after 
the premeasurement correlation is established. In accord with 
Eq. (\ref{two.two}),  
such additional measurement would have a power to select 
an arbitrary observable of the system ${\cal S}$ and would 
single out the corresponding states of the apparatus ${\cal A}$.
Yet, given the freedom to rewrite $|\Psi_t\rangle$ in an 
infinite number of ways, this state 
of ${\cal A}$ would be for almost any choice of the decomposition of the
sum of Eq. (\ref{two.two}) completely ``nonclassical'' in any reasonable 
sense, and it would depend on the initial state of the quantum system. 

In a quantum domain, such an entanglement must be commonplace, along with its 
disturbing consequences. Indeed, a ``Schr\"odinger kitten'' state recently 
implemented  by means of an atomic physics experiment (\cite{Monroeetal}
is an excellent illustration of the distinction between 
the quantum entanglement and the classical correlation in 
the context of quantum measurements). 
The NIST group in Boulder has managed ---manipulating 
a single ion inside a trap with lasers--- 
to establish a correlation between its internal 
state (designated here by $\{|\uparrow\rangle, |\downarrow\rangle\}$, 
respectively, for 
``excited'' and ``ground'') and its location ($|L\rangle$ or $|R\rangle$ for ``left'' 
or ``right''). The final correlated wavefunction has
a premeasurement, EPR-like form,
\begin{equation} |\Psi_A\rangle \ = \  (|+\rangle|L\rangle \  + \  |-\rangle|R\rangle)/\sqrt{2} \ , 
\label{two.three}\end{equation}
where
\begin{equation}|\pm\rangle \  = \  (|\uparrow\rangle \ \pm \  |
\downarrow\rangle)/\sqrt{2} \ \ , \label{two.four}\end{equation}
are superpositions of the ground and excited states. This very same 
$|\Psi_A\rangle$
can be written therefore as
\begin{equation} 
|\Psi_A\rangle = \ \{ |\uparrow\rangle(|L\rangle + |R\rangle) + 
|\downarrow\rangle(|L\rangle-|R\rangle)\}/\sqrt{2} \ . 
\label{two.five}\end{equation}
Thus, the same correlated state of the ``atom cat'' can be expressed in two 
very different--looking ways, implying the potential for still more kinds 
of ambiguous correlations. Expressed in the first way 
(see Eq. (\ref{two.three})), the atom can 
be in one of the two alternative locations, depending on its internal state
that is defined as a superposition of ground and excited states. In the second
way (given by 
Eq. (\ref{two.five})) the natural internal states of the atom are 
correlated with 
a very nonclassical state ---a superposition of an atom in two locations. 
Monroe et al. \cite{Monroeetal}
measure the internal state of the atom in the basis 
corresponding 
to the decomposition of Eq. (\ref{two.five}) and verify that it is indeed in a 
superposition of $|L\rangle$ and $|R\rangle$ with either a positive or a 
negative sign (an ``even'' or and ``odd'' Schr\"odinger cat).

Given the atomic size of this ``kitten'', its ability to appear in   
a superposition of two different widely separated locations may or may not be 
a surprise.  But the point this recent experiment
allows us to make is at the heart of the interpretation problem. If the 
quantum laws are universally valid, very nonclassical Schr\"odinger cat--like
states should be commonplace for an apparatus that measures a quantum system
and, indeed, for run-of-the-mill macroscopic systems in general.
One should be able to prepare such nonclassical states at will, by entangling 
arbitrarily large objects with quantum states of microscopic systems 
and then measuring these quantum objects in some arbitrary basis. If such
sequences of events were common, classical objects would  almost always 
be in very nonlocal superposition states. 

Quantum theory mandates this pandemonium. Yet, we never seem to encounter it, 
least of all in the course of measurements. The task of the interpretation of
quantum theory is to understand why. In the Copenhagen interpretation,  
this problem never arises, because the apparatus is by definition classical. 
However, if one insists on the universality of quantum theory, the difficulty 
described above is inevitable. It arises, for instance, in Everett's 
Many Worlds Interpretation, which was in fact originally called ``the Relative 
State Interpretation'' \cite{Everett57}. Everett and other followers of the 
MWI philosophy tried to occasionally bypass this question by insisting
that one should only discuss correlations. Correlations are indeed at the heart
of the problem, but it is not enough to explain how to compute them; for
that, quantum formalism is straightforward enough. What is needed instead is 
an explanation of why some states retain correlations, but 
most of them do not, 
in spite of the arbitrariness in basis selection that is implied by 
Eq. (\ref{two.two}).
Or, equivalently, what is needed is an explanation of the loss of general
quantum entanglement, but a selective retention of classical correlations
---correlations that are also quantum in their origin, but which consistently
single out the same basis of the quantum states violating the spirit of
the superposition principle.

\subsection{Interactions and the information transfer in quantum measurements}

The interaction required to accomplish the correlation between the measured 
system and the apparatus, Eq. (\ref{two.one}), can be regarded as 
a generalization of 
the basic logical operation known as a ``controlled not'' or a {\tt c-not}.
Classically, {\tt c-not} changes the state of the target bit when the control 
bit is in a state 1, and does nothing otherwise:
\begin{eqnarray} 
0_c \ {0_t\atop{1_t}}  &\longrightarrow& \ 0_c \ {0_t\atop{1_t}} \nonumber\\
%0_c \ 1_t  & \longrightarrow& \ 0_c \ 1_t \nonumber\\
1_c \ {0_t\atop{1_t}}  & \longrightarrow& \ 1_c \ {1_t\atop{0_t.}} 
\label{two.six}
%1_c \ 1_t  & \longrightarrow& \ 1_c \ 0_t \nonumber
\end{eqnarray}

Quantum {\tt c-not} is a straightforward quantum version of 
Eq. (\ref{two.six}). It 
differs from the classical case only because arbitrary superpositions of 
the control bit and of the target bit are allowed 
\begin{equation} (\alpha | 0_c \rangle ~ + ~ \beta | 1_c \rangle) 
| a_t \rangle 
\longrightarrow  \ 
\alpha | 0_c \rangle | a_t \rangle ~ + ~ \beta | 1_c \rangle | 
\neg a_t \rangle. \nonumber
%\label{two.seven}
\end{equation}
Above a ``negation'' of a state $|\neg a_t\rangle$ is a basis--dependent 
operation defined by 
\begin{equation}
\neg (\gamma|0_t\rangle+\delta |1_t\rangle) =\gamma| 1_t \rangle 
+ \delta | 0_t \rangle. 
\nonumber
%\label{two.eight}
\end{equation}
It suffices to identify $ |A_0\rangle = |0_t\rangle$, and
$|A_1\rangle = |1_t\rangle$ to have 
an obvious correspondence between the {\tt c-not} and a premeasurement. 

In the classical {\tt c-not} the direction of the information transfer is  
always 
consistent with the designations of the two participating bits. The state of
the control bit remains unchanged while it controls the state of the target 
bit, Eq. (\ref{two.six}). Written in terms of the logical 
$\{|0\rangle,|1\rangle\}$ basis, the truth table 
of the quantum {\tt c-not} is essentially ---that is, save for 
the possibility of superpositions--- the same as Eq. (\ref{two.six}). 
One might therefore anticipate that 
the direction of information transfer and the designations  
(``control/system'' and ``target/apparatus'') of the two qubits will also 
be unambiguous, as they are in the classical case.
This expectation however is incorrect, as can be readily demonstrated by 
expressing the process in 
the conjugate basis $\{|+\rangle, |-\rangle\}$ that, for
either control or target bit, is obtained through the Hadamard transform:
\begin{equation} 
|\pm\rangle = (|0\rangle \pm |1\rangle)/\sqrt{2}. \ \label{two.nine}
\end{equation}
The truth table of Eq. (\ref{two.six}) in conjunction with the principle 
of superposition (which allows one to write down Eq. (\ref{two.nine})) 
leads to a new complementary truth table
\begin{eqnarray}
|\pm\rangle|+\rangle& \longrightarrow& |\pm\rangle|+\rangle \nonumber\\
|\pm\rangle|-\rangle &\longrightarrow& |\mp\rangle|-\rangle . \label{(2.10)}
\end{eqnarray}
That is, in the complementary basis $\{ |+\rangle,|-\rangle \}$ the 
roles of the control and the target bit are reversed. 
The state of the former 
% (basis $\{|0\rangle,|1\rangle\}$)
target ---represented by the second ket in Eq. (\ref{(2.10)})--- 
remains unaffected in the new basis, and the state of the former 
control is conditionally ``flipped''.

In the above {\tt c-not} (or bit-by-bit measurement), the appropriate 
interaction Hamiltonian is
\begin{eqnarray} 
H_{int} &=& g|1\rangle\langle1|_{\cal S} |-\rangle\langle-|_{\cal A}\ 
=\ {g \over 2}|1\rangle\langle1|_{\cal S}
({\bf 1} - (|0\rangle\langle1| + |1\rangle\langle0|))_{\cal A}\nonumber \\
&=&g({1\over 2} -\sigma_z)_{\cal A}({1\over 2}-\sigma_x)_{\cal S}.
\label{(2.11a)} 
\end{eqnarray}
Above, $g$ is a coupling constant, $\sigma_i$ are Pauli matrices,  
and the two operators refer to the system
(i.e., to the former control), and to the apparatus pointer (the former 
target), respectively. It is easy to see that the states 
$\{|0\rangle,|1\rangle\}_{\cal S}$ of the system are unaffected by 
$H_{int}$, because 
\begin{equation} 
[H_{int},\  e_0|0\rangle\langle0| + e_1|1\rangle\langle1|] = 0.
\label{(2.12a)}
\end{equation}
Thus, the measured (control) observable 
$\hat \epsilon = e_0|0\rangle\langle0| + e_1|1\rangle\langle1|$
is a constant of motion under the evolution generated by $H_{int}$. 

The states $\{|+\rangle,|-\rangle\}_{\cal A}$ of the apparatus 
(which encode the 
information about the phase between the logical states) have exactly the 
same ``immunity''
\begin{equation} 
[H_{int},\  f_+|+\rangle\langle+| + f_-|-\rangle\langle-|] = 0. 
\label{(2.12b)}
\end{equation}
Hence, when the apparatus is prepared in a definite phase state (rather than 
in a definite pointer/logical state), it will pass its phase onto the 
system, as the truth table, Eq. (\ref{(2.10)}), shows. Indeed, $H_{int}$ 
can be rewritten in the Hadamard transformed basis  
\begin{eqnarray} 
H_{int} &=&  g|1\rangle\langle1|_{\cal S} |-\rangle\langle-|_{\cal A} 
\nonumber\\
&=& \ {g \over 2}({\bf 1} - (|-\rangle\langle+| + 
|+\rangle\langle-|))_{\cal S} 
                      |-\rangle\langle-|_{\cal A}, \label{(2.11b)} 
\end{eqnarray}
which, in comparison with Eq. (\ref{(2.11a)}), makes this ``immunity'' obvious.

This basis-dependent direction of the information flow in a quantum {\tt c-not}
(or in a premeasurement) is a direct consequence of complementarity. It can be
summed up by stating that although the information about the observable with 
the eigenstates $\{|0\rangle,|1\rangle\}$ travels from the measured system 
to the apparatus, in the complementary $\{|+\rangle,|-\rangle\}$ basis it 
seems to be the apparatus that is 
being measured by the system. This observation also clarifies the sense 
in which 
phases are inevitably ``disturbed'' in measurements. They are not really 
destroyed, but, rather, as the apparatus measures a certain observable of 
the system, the system ``measures'' the phases between the possible outcome 
states of the apparatus. 
These phases in a macroscopic apparatus coupled to the environment are 
fluctuating rapidly and uncontrollably, thus leading to the destruction of 
phase coherence. However, even if this consequence of 
decoherence were somehow prevented (i.e., by perfectly isolating the apparatus 
pointer from the environment), preexisting phases between the outcome  states 
of the apparatus would have to be known while, simultaneously, ${\cal A}$ is
in $|A_0\rangle$, the ``ready--to--measure--state''. This would require a 
simultaneous
knowledge of the two non-commuting observables, and is therefore impossible 
because of  Heisenberg indeterminacy.

It appears that even the question ``which of the two interacting systems is 
a measuring device?" (which should be decided by the direction of the flow of 
information) depends on the initial states.
In ``classical practice'' this ambiguity 
does not arise because the initial state of the apparatus can never be  
selected at the whim of the observer. Einselection limits the set of possible
states of the apparatus to a small subset of all the states available in 
Hilbert space.

\subsection{Monitoring by the environment and decoherence}

In this section, we shall see how the quantum-classical correspondence can be 
reestablished by decoherence and einselection, caused by the monitoring of 
the to-be-classical observables by the environment. The environment is defined 
as any set of degrees of freedom that are coupled to the system of interest,
and which can therefore `monitor' ---become entangled with--- its states. 
Environments can be external (such as particles of air or photons that  
scatter 
off, say, the apparatus pointer) or internal (e.g., collections of phonons or 
other excitations in the materials from which an apparatus is constructed). 
Often, environmental degrees of freedom emerge from the split of the original
set of degrees of freedom into the ``system of interest'' that is some 
collective observable (order parameter in a phase transition), and the 
``microscopic remainder''.

The superposition principle applies only when the quantum system is closed.
When the system is open, interaction with the environment will inevitably 
result
in an incessant ``monitoring'' of some of the observables by the environmental
degrees of freedom. This will result in the degradation of the pure 
states into 
mixtures. These mixtures will often ---remarkably often--- turn out to be 
diagonal in the same set of ``preferred states'' that are nearly 
independent of
the  initial state of the system and of the environment, but which are selected
with the crucial help of the interaction Hamiltonian. This decoherence process
determines the relative ``fitness'' of all the possible superpositions that 
exist in the Hilbert space. The resulting ``natural selection'' is responsible
for the emergence of classical reality. Its consequence is known as 
environment-induced superselection \cite{Zurek82}, or einselection. 

The set of habitually decohering states is often called ``the pointer basis'', 
in recognition of its role in the measurement problem. The criterion for 
the selection of pointer states goes well beyond the often--repeated 
characterizations based solely on the instantaneous eigenstates of the density 
matrix. What is of the essence is the ability of the einselected states
to survive monitoring by the external degrees of freedom. This heuristic
criterion can be made rigorous by quantifying the 
predictability of the evolution 
of the candidate classical states or of the associated pointer 
observables. To put it succinctly, measurement of the pointer observables 
yields an optimal initial 
condition. In spite of the openness of the system, its results can be employed 
for the purpose of prediction better than the other Hilbert space 
alternatives. 

The contrast between the resilience of the states associated with the 
preferred 
(pointer) observables and the fragility of their superpositions can be 
analyzed 
in terms of Heisenberg's principle of indeterminacy. The environment monitors
observables with the accuracy dictated by the interaction Hamiltonian. 
Thus, only a measurement that happens to commute with the observables 
monitored
by the environment will result in a useful record that can be successfully
employed for the purpose of prediction. In contrast, a system prepared by the 
measurement in an arbitrary superposition will  also be monitored by the 
environment, which will tend to correlate with the pointer observable. When 
the initial superposition prepared by the observer does not commute with 
the observables monitored by the environment, Heisenberg's indeterminacy 
implies that the records of the observer are of no use for the purpose of 
prediction. The monitoring continuously carried out by the environment on 
the pointer observables makes anything except
for the pointer states a poor choice.

Three quantum systems ---the measured system ${\cal S}$, the apparatus pointer
or the memory of the observer ${\cal A}$, and the environment ${\cal E}$---
and the correlations between them will be the subject of the discussion below.
In quantum measurements, ${\cal S}$ and ${\cal A}$ will be coupled. Their 
quantum
entanglement will be converted into an effectively classical correlation as a 
result of the interaction between ${\cal A}$ and ${\cal E}$. In measurements 
of 
classical systems, 
both ${\cal S}$ and ${\cal A}$ will interact with ${\cal E}$ 
and decohere. In either case, states einselected by the environment will be the
focus of attention. In ${\cal A}$, they will be the repository of information,
serving as pointer states of the apparatus or memory states of the observer.
The system ${\cal S}$ can also look effectively classical when it is subject
to einselection, and when ${\cal A}$ keeps records of its einselected states.

This ${\cal SAE}$ triangle (or a triangle much like it) is necessary for 
careful
study of decoherence and its consequences. By keeping all three corners of 
this triangle in mind, one can avoid the confusion about the relation of 
the instantaneous eigenstates of the density matrix (see, for example, 
the discussion following \cite{Zurek94}). This three-system context 
is necessary to keep track of the correlations between the memory of 
the observer and the state of the measured system. The evolution from a 
quantum 
entanglement to the classical correlation may be the easiest relevant theme 
to define operationally. 
In spite of this focus on the correlation, I shall often suppress one of
the corners of the above triangle to simplify the equations. All three 
parts of the triangle will however play a role in formulation of the questions 
we shall pose and in motivating of the criteria for classicality that 
we shall devise.

\subsection{One--bit environment for a bit-by-bit measurement}

The simplest discussion of a single act of decoherence involves just three 
one-bit systems \cite{Zurek81,Zurek83}. They are denoted by ${\cal S, \ A}$, 
and ${\cal E}$ in an obvious reference to their designated roles. 
The measurement starts with the interaction of a measured system with 
the apparatus, 
\begin{eqnarray}
|\uparrow \rangle | A_0 \rangle  &\longrightarrow& \ | \uparrow \rangle| 
A_1 \rangle \nonumber\\ 
|\downarrow \rangle | A_0 \rangle & \longrightarrow& \ | \downarrow 
\rangle| A_0 \rangle, \label{(4.1)}
\end{eqnarray} 
where $\langle A_0 | A_1 \rangle = 0 $. For a general state, 
\begin{equation} 
(\alpha | \uparrow \rangle ~ + ~ \beta | \downarrow \rangle ) 
|A_0 \rangle \ \longrightarrow
\alpha  | \uparrow \rangle | A_1 \rangle \ + \ \beta | \downarrow \rangle 
| A_0 \rangle = 
| \Phi \rangle. \label{(4.2)} 
\end{equation}
These formulae are an example of a {\tt c-not}--like premeasurement that has
already been discussed. As was noted previously, a correlated state of this 
form is not enough to claim that a measurement has taken place. 
The biggest problem 
with $| \Phi\rangle$ is the basis ambiguity. Equation (\ref{(4.2)}) 
represents only an entanglement, the same as in Bohm's version of the EPR 
state \cite{Bohm51}. 
The ambiguity of the basis selection in this simple example can be settled 
with the help of one additional system, ${\cal E}$, which performs 
a premeasurement on the apparatus. As a result,
\begin{eqnarray}
|\Psi(0)\rangle_{SAE}&=& ( \alpha | \uparrow \rangle | A_1 \rangle 
+ \beta | \downarrow \rangle 
| A_0\rangle )|\varepsilon_0\rangle 
\longrightarrow\nonumber\\
&\longrightarrow& \alpha | \uparrow \rangle | A_1 \rangle  
|\varepsilon_1\rangle 
 + \beta | \downarrow \rangle | A_0 \rangle |\varepsilon_0\rangle 
= | \Psi \rangle . \label{(4.3)} 
\end{eqnarray}
It may seem that very little can be accomplished by repeating the step that  
has led to the ${\cal S - A}$ correlation and the associated problems. 
But this is not the case. A collection of three correlated quantum systems 
is no longer subject to the basis ambiguity we have pointed out in 
connection with the
EPR-like state $|\Phi\rangle$, Eq. (\ref{(4.2)}). This is especially true 
when the states
of the environment are correlated with the simple products of the states
of the apparatus--system combination \cite{Zurek81}. 
In Eq. (\ref{(4.3)}) above, this can be 
guaranteed (irrespective of the value of $\alpha$ and $\beta$) 
providing that:
\begin{equation} 
\langle \varepsilon_0 | \varepsilon_1 \rangle = 0 \ . \label{(4.4)} 
\end{equation}
When this condition is satisfied, the description of the ${\cal A-S }$ pair 
can be readily obtained in terms of a reduced--density--matrix:
\begin{eqnarray}
 \rho_{\cal AS} &=&  Tr_{\cal E} | \Psi \rangle\langle \Psi | \nonumber\\
                          &=& |\alpha|^2 | \uparrow \rangle\langle \uparrow | 
|A_1 \rangle\langle A_1| \ + \ 
|\beta|^2 |\downarrow \rangle\langle \downarrow | |A_0 \rangle\langle A_0 
| \ . 
\label{(4.5a)}\end{eqnarray}
This reduced--density--matrix contains only terms corresponding to classical 
correlations. 

If the condition of Eq. (\ref{(4.4)}) did not hold ---that is, 
if the orthogonal states of the environment were not correlated with the 
apparatus in the basis in which the original premeasurement was carried 
out--- then the terms on the diagonal of the reduced density matrix 
$\rho_{\cal AS}$ would be the sum of products rather than simply products 
of states of ${\cal S}$ and ${\cal A}$. An extreme example of that situation 
is the pre-decoherence density matrix of the pure state:
\begin{eqnarray} | \Phi \rangle\langle \Phi | &=&   
|\alpha|^2 | \uparrow \rangle\langle \uparrow | |A_1 \rangle\langle A_1| +  
\alpha \beta^* | \uparrow \rangle\langle \downarrow | | A_1 \rangle 
\langle A_0 | +  \nonumber\\  
&+& \alpha^* \beta | \downarrow \rangle\langle \uparrow | | A_0 \rangle 
\langle A_1 |  +  
|\beta|^2 |\downarrow \rangle\langle \downarrow | |A_0 \rangle\langle A_0 | 
\label{(4.5)}
\end{eqnarray}
Its eigenstate is simply $|\Phi\rangle$. When expanded, 
$|\Phi \rangle\langle \Phi |$ contains terms that are off--diagonal when 
expressed in the natural basis consisting of  
tensor products of states in the two subspaces. Their disappearance as 
a result of tracing over the environment signals the disappearance of the 
basis ambiguity. There is of 
course a conceptual difference with the classical case. 
In classical mechanics,  
it was in principle possible to imagine that the outcome was predetermined. 
In quantum mechanics this is usually impossible even in principle. However, 
that distinction can be made only with a more complete knowledge than the one
typically available to the observer. 

The pointer observable that emerges from this simple case is easy to 
characterize. The interaction Hamiltonian between the apparatus and the
environment, $H_{\cal AE}$, should have the same structure as for 
the {\tt c-not}. It should be a function of the pointer observable 
\begin{equation} 
\hat A \ = \ a_1 |A_1\rangle\langle A_1| + a_0 |A_0\rangle\langle 
A_0|, \label{(4.6a)}
\end{equation}
of the apparatus. Consequently, the states of the environment will bear an
imprint of the pointer states $\{ |A_1\rangle,|A_0\rangle\}$. As was 
also noted in the discussion of {\tt c-not}s, 
$[ H_{\cal AE}, \hat A] = 0$ 
immediately implies that $\hat A$ is a control, and its eigenstates will 
be preserved.

Disappearance of quantum coherence because of a ``one--bit'' measurement 
has been verified experimentally in neutron and, more recently, 
in atomic interferometry 
\cite{RauchetalPhysScripta98,Pfauetal96,Chapmanetal96}. 
A single act of quantum measurement we have discussed here should be regarded 
as an elementary discrete instance of continuous monitoring, which is 
required to bring about the appearance of classicality.

\subsection{Decoherence of a single (qu)bit}

Another example of decoherence is afforded by a two-state apparatus 
${\cal A}$ interacting with an environment of $N$ other spins 
\cite{Zurek82}. 
We can think of it as just another two--state system, and, in that spirit, we 
shall identify in this section the two apparatus states as 
$\{|\Uparrow\rangle,|\Downarrow\rangle\}$. The process of decoherence is 
definitely not 
limited to states of the apparatus pointers, so these two generic candidate 
pointer states can belong to any system. 

The simplest, yet already quite illustrative example of this situation occurs 
when the self-Hamiltonian of the apparatus disappears, $ H_{\cal A} =0$, and 
the interaction Hamiltonian has the form:
\begin{equation} 
H_{\cal AE} = (|\Uparrow\rangle\langle\Uparrow| - |\Downarrow\rangle
\langle\Downarrow|) \otimes 
\sum_k g_k (|\uparrow\rangle\langle\uparrow| - |\downarrow\rangle\langle
\downarrow|)_k. \label{(4.7)}
\end{equation}
Under the influence of this Hamiltonian, the initial state
\begin{equation} |\Phi(0)\rangle = (a|\Uparrow\rangle + b 
|\Downarrow\rangle) ~ \prod_{k=1}^N 
(\alpha_k |\uparrow\rangle_k + \beta_k |\downarrow\rangle_k) \ \label{(4.8)}
\end{equation}
evolves into 
\begin{equation} 
|\Phi(t)\rangle = a |\Uparrow\rangle \otimes |{\cal E}_{\Uparrow}(t)\rangle +
b |\Downarrow\rangle \otimes |{\cal E}_{\Downarrow}(t)\rangle \ . 
\label{(4.9)} 
\end{equation}
Here:
\begin{equation} 
|{\cal E}_{\Uparrow}(t)\rangle = \prod_{k=1}^N 
(\alpha_k \exp (i g_k t) |\uparrow\rangle_k + \beta_k \exp (- i g_k t) 
|\downarrow\rangle_k) = |{\cal E}_{\Downarrow}(-t)\rangle \ .\label{(4.10)} 
\end{equation}
The reduced density matrix is then 
\begin{equation}
\rho_{\cal A} = |a|^2 |\Uparrow\rangle\langle\Uparrow| + 
ab^*r(t)|\Uparrow\rangle\langle\Downarrow|
+a^*br^*(t) |\Downarrow\rangle\langle\Uparrow| + |b|^2 
|\Downarrow\rangle\langle\Downarrow|. \label{(4.11)}
\end{equation} 
The coefficient $r(t)$ determines the relative size of the off-diagonal terms.
It is given by 
\begin{equation} 
r(t) = \langle{\cal E}_{\Uparrow}|{\cal E}_{\Downarrow}\rangle= \prod_{k=1}^N 
[\cos2g_kt + i (|\alpha_k|^2-|\beta_k|^2)\sin 2 g_k t ] \ . \label{(4.12)}
\end{equation}
Unless $k$'th spin of the environment is initially in an eigenstate of the 
interaction Hamiltonian, its contribution to the product will be less than
unity. Consequently, for large environments consisting of many ($N$) spins 
and at large times the off-diagonal terms are typically small, 
\begin{equation} 
|r(t)|^2 \simeq 2^{-N} \prod_{k=1}^N[1 + (|\alpha_k|^2 - 
|\beta_k|^2)^2].
\label{(4.13)}
\end{equation}
This effect can be illustrated with the help of the Bloch sphere.
The density matrix of any two-state system can be represented by a point 
in the 3-D space. In terms of the coefficients $a,\ b,$ and $r(t)$ that 
we have previously used, the coordinates of the point representing 
$\rho(t)$ are: $z=(|a|^2 - |b|^2)$, $x=\Re (ab^*r)$, and $y=\Im (a b^* r)$. 
When the state is pure, $x^2+y^2+z^2=1$ -- pure states lie on the surface 
of the Bloch sphere (Figure 1). When the state is mixed, the point 
representing it lies inside that sphere. 
Any conceivable (unitary or non--unitary) quantum evolution 
of the two--state system  
can be thought of as a transformation of the surface of the pure states into 
the ellipsoid contained inside the Bloch sphere. 
Deformation of the Bloch sphere caused by decoherence is a special case 
of such general evolutions. The decoherence process does 
not affect $a$ or $b$. Hence, evolution caused by decoherence alone 
occurs in a constant--$z$ plane. Such a ``slice'' through the Bloch sphere 
would show the point 
representing the state at a fraction $|r(t)|$ of its maximum distance. 
The complex number $r(t)$ can be expressed as the sum of the complex phase 
factors rotating with the frequencies given by differences $\Delta \omega_j$ 
between the energy eigenvalues of the interaction Hamiltonian, weighted with 
the probabilities  of finding these energy eigenstates in the initial state,
\begin{equation} 
r(t) = \sum_{j=1}^{2^N} p_j \exp(-i \Delta \omega_j t) \ . 
\label{(4.14)}
\end{equation}
The index $j$ now denotes partial energy eigenstates of the environment of 
the interaction Hamiltonian (tensor products of $\uparrow$ and $\downarrow$ 
states of the environmental spins). 
%, Eq. (\ref{(4.7)}), of the form:
%\begin{equation}|j\rangle =|\uparrow\rangle_1 \otimes |\downarrow\rangle_2 
%\otimes \dots \otimes |\uparrow\rangle_N \ \label{(4.15a)} \end{equation}
The corresponding eigenvalue differences between the two complete energy 
eigenstates $|\Uparrow\rangle |j\rangle$ and $|\Downarrow\rangle|j\rangle$ 
are 
\begin{equation} 
\Delta \omega_j = \langle\Uparrow|\langle j|
H_{\cal AE}|j\rangle|\Downarrow\rangle\ . 
\label{(4.15b)} \end{equation} 
There are $2^N$ distinct states $|j\rangle$, and, barring degeneracies, 
the same number of different $\Delta \omega_j$'s. The 
probabilities $p_j$ are given by 
\begin{equation} p_j = | \langle j | {\cal E}(t=0)\rangle |^2 \ , \label{(4.15c)} 
\end{equation}
which, in turn, is easily expressed in terms of the appropriate squares of 
the products of $\alpha_k$ and  $\beta_k$. 
\begin{figure}
\qquad\qquad\qquad\qquad
	\includegraphics[height=0.7\hsize]{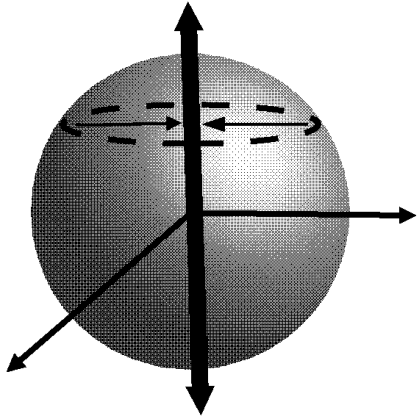}
\caption{Decoherence can be seen in the Bloch sphere as the process 
that induce the states to ``move towards the vertical axis'', which is 
defined by the two pointer states on the poles. Classical domain consists 
of just two pointer states. The classical core is the set of all mixtures 
of pointer states.}
\end{figure}

The evolution of $r(t)$ given by Eq. (\ref{(4.14)}) is a consequence of the
rotations of the complex vectors $p_k\exp (-i \Delta \omega_j t)$
with different frequencies. The resultant $r(t)$ will then start with
the amplitude $1$ and quickly ``crumble'' to a value approximately equal to 
\begin{equation}\langle|r(t)|^2\rangle = \sum_{j=1}^{2^N} p_j^2 \sim 2^{-N}.
\nonumber 
\end{equation}
In this sense, decoherence is exponentially effective ---the expected magnitude
of the off-diagonal terms decreases exponentially fast with the physical size 
$N$ of the environment--- with the number of systems (spins in our example).
In effect, any initial state asymptotically approaches the $z$-axis
as a result of decoherence. 

We note that the effectiveness of einselection depends on the initial state of 
the environment: When ${\cal E}$ is in the $k$'th eigenstate of $H_{\cal AE}$,
and $p_j = \delta_{jk}$, the coherence in the system will be retained because 
the environment is now in an eigenstate of the ``control''. This situation is, 
however, unlikely in realistic circumstances because 
the self-Hamiltonian of the
environment $H_{\cal E}$ will not commute, in general, with 
$H_{\cal AE}$. Moreover, even when $H_{\cal E}= 0$, finding an environment 
in an energy eigenstate of the Hamiltonian seems extremely unlikely ---the
eigenvalues of such eigenstates are bound to be dense in large systems, and 
therefore they will be easily perturbed by the interaction with their 
environments. Furthermore, the $2^N$ partial eigenstates of the interaction
Hamiltonian are exponentially rare among arbitrary superpositions.

The geometry of the flows induced by decoherence inside 
the Bloch sphere exhibits
characteristics that are encountered in more general physical situations,
involving decoherence in bigger Hilbert spaces as follows:

(i) Domain of pure quasi--classical states consisting of all the einselected 
pointer states ($\{|\Uparrow\rangle$, $|\Downarrow\rangle\}$ in our case). 
Pointer states are the pure states least affected (here, unaffected) 
by decoherence.

(ii) Classical core of probability distributions, i.e., all the mixtures 
of pointer states. In Figure 1 it
corresponds to the section [-1,+1] of the $z$-axis.

(iii) The rest of the space ---the rest of the volume of the Bloch sphere---
consists of more general density matrices. As a result of decoherence,  
that part of the Hilbert space is ``ruled out'' by einselection.  
  
Visualization of this decoherence-induced decomposition of the Hilbert space 
is still possible in the simple two-dimensional case studied here, but the 
existence of the elements (i)--(iii) is a general feature. It characterizes 
the emergence of classicality under all circumstances. We shall therefore 
appeal to the intuitions developed in the 
course of this discussion later. However, it may 
be useful to anticipate a few of the phenomena that can take place 
when decoherence combines with the evolution induced by the self-Hamiltonian 
of the system or when it is caused by more complicated couplings to the 
environment.

(a) Approach to equilibrium would affect elements diagonal in 
the pointer basis, so that the density matrix would asymptotically approach 
a time-independent distribution 
(such as $\rho \sim 1$ for a thermal equilibrium at infinite temperature or 
$\rho \sim |\Downarrow\rangle\langle\Downarrow|$ for decay).
This corresponds to a flow towards some specific point (i.e., the center 
or the ``south pole'' in the above two examples) within the Bloch sphere. 
However, when decoherence dominates, the flow would start somewhere within 
the Bloch sphere, and quickly (on the decoherence timescale) converge 
towards a point on the $z$--axis (the classical core). This would be 
followed by a much slower relaxation, a flow more or less along 
the $z$--axis (and therefore 
essentially within the classical core) on a {\it relaxation timescale}.

(b) Approximately reversible classical dynamics can coexist with decoherence 
when the self-Hamiltonian of the system can generate motions within the 
surfaces
of constant entropy inside the classical core. In the case considered here, 
the core is one-dimensional and the subspaces of constant entropy within 
it are zero-dimensional. Therefore, it is impossible to generate continuous 
isentropic 
motion within them. In multidimensional Hilbert spaces with richer dynamics 
that are
nearly isentropic, approximately reversible evolution is often possible and 
allows for the idealization of trajectories in the classical limit. 

(c) A sharp distinction between the classical core and the rest of the Hilbert 
space is possible only in idealized situations (or in an even more idealized 
``mathematical classical limit'', in which $\hbar \rightarrow 0$, 
mass$\rightarrow \infty$, etc.). In realistic situations,  
all that will be required is a clear contrast between the rates of the entropy 
production between the inside and the outside of the classical core. We shall 
refine such criteria in the discussion of the 
{\it predictability sieve} ---a criterion for the selection of the preferred 
pointer states, which in effect demands that the entropy production rate 
should be minimized for the einselected states. In the case discussed here, 
pointer states obviously satisfy this criterion, and the entropy production
vanishes in the classical domain.  In more general situations, 
we shall not be equally lucky. For instance, in the case of chaotic systems,  
entropy will also 
be produced in a classical core, but at a rate set by the
classical dynamics (i.e., by the self-hamiltonian rather than by the coupling 
with the environment) and much more slowly than outside of the classical core.

\subsection{Decoherence, einselection, and controlled shifts}

The above discussion of decoherence can be straightforwardly generalized 
to the situation where the system, the apparatus, and the environ\-ment have
many states, and where the interactions between them are much more complicated.
Here we assume that the system is isolated and that it interacts with the
apparatus only briefly. As a result of that
interaction, the state of the apparatus becomes entangled with the state
of the system, 
$(\sum_i\alpha_i |s_i\rangle)|A_0\rangle\ \rightarrow  
\sum_i \alpha_i |s_i\rangle |A_i\rangle $.
By analogy with a {\tt c-not}, we shall refer to this conditional operation
as a {\tt c-shift}.
This quantum correlation suffers from the basis ambiguity we have discussed 
previously:
The ${\cal S-A}$ entanglement implies that for any state of either of the two
systems there exists a corresponding pure state of the other. Indeed, when the
initial state of ${\cal S}$ is chosen to be one of the eigenstates of the
conjugate basis, $|r_l\rangle = N^{-{1 \over 2}} \sum_{k=0}^{N-1} 
\exp(2 \pi i k l 
/ N) |s_k\rangle$, this {\tt c-shift} would equally well represent 
a measurement of 
the apparatus state (in the basis conjugate to $\{ |A_k \rangle \}$) 
by the system
\cite{Zurek00}. Thus, it is not just the basis that is ambiguous, but 
also the roles of the control (system) and of the target (apparatus) can be 
reversed when the conjugate basis is selected. These ambiguities that exist 
for the ${\cal SA}$ pair can be removed by recognizing the role of 
the environment.

Decoherence is represented schematically in Figure  2 by a sequence of 
{\tt c-not}s (or {\tt c-shift}s) which, in some fixed basis, 
`measure' the state
of the apparatus and record the outcome of the measurement in the environment. 
The requirement for a good apparatus is to retain correlations between 
the measured observable of the system and some ``pointer observable''. 
This will happen 
when the {\tt c-shift} between ${\cal S}$ and ${\cal A}$ correlates the state 
of the system with the observable of the apparatus that is itself monitored 
(but not perturbed) by the environment. That is, in an idealized measurement, 
the measured observable of the system is playing the role of the control with 
respect to the ${\cal S-A}$ {\tt c-shift}. In a well--designed apparatus,
the pointer observable is a target of the ${\cal S-A}$ {\tt c-shift}, but 
a control of the ${\cal A-E}$ {\tt c-shift}s. Eigenstates of the pointer 
observable of the apparatus play the role of an alphabet of a communication
channel. They encode a state of the system and retain the correlation in spite
of the interaction with the environment.

\begin{figure}
\qquad\qquad (a)\includegraphics[height=0.3\hsize]{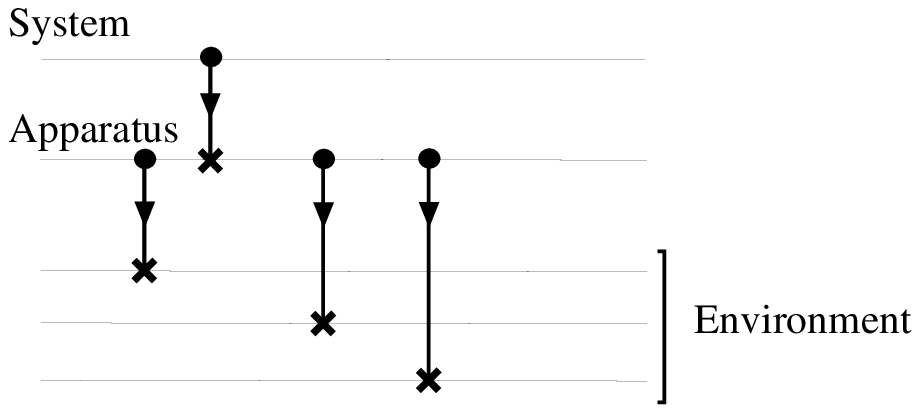}
\break 
\qquad\qquad (b)\includegraphics[height=0.4\hsize]{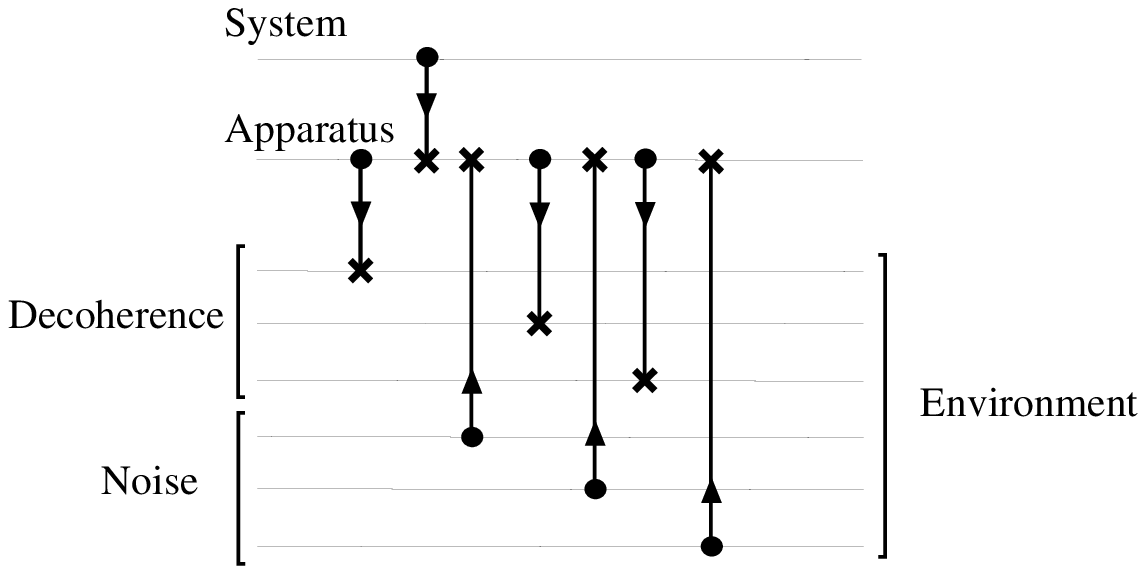}
\caption{(a) Decoherence can be viewed as the consequence of the  
monitoring of the state of the system by the environment. This is 
symbolically represented here by a sequence of c-not gates where the 
pointer states of the apparatus act as the control
and the environment is the target. (b) The distinction between decoherence
and noise depends on the direction of the information flow in the preferred
basis. Preferred states minimize the number of c--nots directed from the 
environment.}
\end{figure}

The graph in Figure 2 captures the essence of the idealized decoherence 
process, which yields ---in spite of the interaction with the environment--- 
a noiseless classical communication channel \cite{Lloyd96,Schumacher96}. 
This is 
possible because in the pointer basis, the ${\cal A-E}$ {\tt c-shift}s operate 
without disturbing the pointer observable, which is the constant of motion of 
the ${\cal A-E}$ interaction Hamiltonian.

The advantage of the graphical representation of the decoherence process as a
sequence of {\tt c-shift}s lies in its simplicity and suggestiveness. However, 
the actual process of decoherence is usually caused by a continuous 
interaction 
(so that it can be only approximately broken up into discrete {\tt c-shift}s). 
Moreover, in contrast to the {\tt c-not}s used in quantum logic circuits,
the record inscribed in the environment is more often than not distributed over
many degrees of freedom. Last but not least, the observable of the apparatus
(or any other open system) may be a subject to noise (and not just decoherence)
or it may evolve in a manner that will rotate pointer states into their
superpositions. 

The basic physics of decoherence is a simple premeasurement--like process 
carried out by the environment ${\cal E}$ as a result of the interaction with 
the apparatus, 
\begin{eqnarray} 
|\Psi_{\cal SA}\rangle |\varepsilon_0\rangle &=& (\sum_j \alpha_j 
|s_j\rangle |A_j\rangle) |\varepsilon_0\rangle  
\longrightarrow \nonumber\\
&\longrightarrow& \sum_j \alpha_j |s_j\rangle |A_j\rangle 
|\varepsilon_j\rangle \ = \ |\Phi_{\cal SAE}\rangle. \label{(4.18)}
\end{eqnarray}
Decoherence leads to the einselection when the states of the environment 
$|\varepsilon_j\rangle$ corresponding to different pointer states become 
orthogonal,
\begin{equation} 
\langle\varepsilon_i | \varepsilon_j \rangle = \delta_{ij}.  
\label{(4.19)}
\end{equation}
When this orthogonality condition is satisfied, Schmidt decomposition
of the state vector $|\Phi_{\cal SAE}\rangle $ into a composite subsystem 
${\cal SA}$ and ${\cal E}$ yields product states $|s_j\rangle |A_j\rangle$ as partners 
of the orthogonal
environment states. The density matrix describing the correlated but decohered 
${\cal SA}$ pair is then:
\begin{equation} 
\rho_{\cal SA}^D = \sum_j |\alpha_j|^2 |s_j \rangle\langle s_j||A_j
\rangle\langle A_j| 
= Tr_{\cal E} |\Phi_{\cal SAE}\rangle\langle\Phi_{\cal SAE}| \ . \label{(4.20)}
\end{equation}
The reduced density matrix of the ${\cal SA}$ pair is diagonal in the product
states.  
 
For notational simplicity,  
we shall often discard reference to the object 
that does not interact with the environment (here, the system ${\cal S}$). 
Never\-theless, it is
useful to keep in mind that the preservation of the ${\cal SA}$ correlations
is the criterion used to define the pointer basis. The density matrix of a
single object evolving in contact with the environment will be always diagonal 
in the same (instantaneous) Schmidt basis. This instantaneous 
diagonality should not be used as a sole criterion for classicality 
(although see \cite{Zeh73,Zeh90}; as well as \cite{Albrecht92,Albrecht93}). 
Rather, the ability of certain sets of states to retain correlations in 
spite of the coupling to the environment is decisive in the emergence 
of ``classical reality''. This is especially obvious in quantum measurements.

When the interaction with the apparatus has the form  
\begin{equation} 
H_{\cal AE} = \sum_{k,l,m} g^{\cal AE}_{klm} |A_k\rangle\langle A_k||
\varepsilon_l\rangle\langle\varepsilon_m| + h.c. \ , \label{(4.21)}
\end{equation}
the basis $\{|A_k\rangle\}$ is left unperturbed. Then, 
any correlation with the states $\{|A_k\rangle\}$ will be preserved. 
And, by definition, the states that preserve  
correlations will be the pointer states. Any observable $A$ co-diagonal with
the interaction Hamiltonian will be an effective pointer observable. 
For, when the Hamiltonian depends on $A$, it will commute with $A$,
\begin{equation} [ H_{\cal AE}(A), A] = 0. \label{(4.22)}\end{equation}
Moreover, the dependence of the interaction Hamiltonian on the observable 
is an obvious precondition for the monitoring of that observable by the 
environment.

\section{Dynamics of quantum open systems: master equations}

One of the most practical tools for analyzing the dynamics of a 
quantum open system is the evolution equation for the reduced 
density matrix, known as the "master equation". In this section we will
review some of the most common techniques to obtain such an equation.
As usual, we divide our uni\-verse into a system of interest ${\cal S}$ 
that interacts with an environment ${\cal E}$. 
The reduced density matrix of the system
is the operator that allows us to answer all physical questions 
that concern the system ${\cal S}$ only. We will denote the reduced density 
matrix as $\rho$, which is obtained from the total density matrix of 
the universe by tracing over the environment Hilbert space. Thus,
$$
\rho=Tr_{\cal E}\rho_{\cal T}, 
$$
where the total density matrix is denoted as $\rho_{\cal T}$. 

In principle, the evolution equation for $\rho$ could be obtained 
by solving Schr\"odinger (or von Neumann) equation for the total 
density matrix and then taking the trace. However, this task can 
be analytically completed in very few cases, and the study of the 
evolution of the reduced density matrix should be done by using 
some approximations. 
 
This section is divided in two parts. First we review some of the 
standard techniques used to obtain approximate master equations. 
Our plan is not to give a complete review of 
master equation techniques but to present some
useful tools to be applied later in studying decoherence. We do this not 
only to ensure that the paper is self--contained but also because we think 
it might be useful to present some simple and helpful results that are not
so well known. We focus on the simplest approximation scheme,  
obtaining master equations valid to a second order in a perturbative 
expansion in the system--environment coupling strength. We first review the 
general perturbative scheme and apply it to two 
physically interesting examples: (1) The  
Brownian motion of a particle coupled to an environment of independent
oscillators and (2) A quantum particle locally coupled to an environ\-ment 
formed by a quantum scalar field. As a further illustration of the 
way in which perturbative master equations can be obtained, we find the
corresponding equations for a two--level system coupled to a bosonic bath in 
two physically relevant cases (the decay of a two--level atom and the spin
boson model).

In the second part of this section we review the 
properties of an important model that is amenable to an exact solution. 
Thus, we concentrate on the linear quantum Brownian motion model analyzing the 
properties of its exact master equation. In particular, we stress the 
fact that in this simple but physically relevant model, the exact master 
equation has the same functional form as the one obtained using perturbation 
theory and can always be cast in terms of a local differential 
equation with time--dependent coefficients. 

\subsection{Master equation: Perturbative evaluation}

Here we present the  general procedure that can be used to derive 
the master equation, assuming that the system--environment 
coupling is small. Thus, we sketch a textbook derivation of the master
equation using perturbation theory. We think it is convenient 
to present this derivation just to stress the fact that perturbative
master equations can always be shown to be local in time.  
The calculation we follow is closely related to the one presented, 
for example, in \cite{WallsMilburn} and can be seen to be a 
variant of the time--convolutionless method discussed 
in \cite{Convolutionless}.

Let us consider the total Hamiltonian to be 
$$H=H_{\cal S}+H_{\cal E}+V,$$ 
where $H_{\cal S}$ and $H_{\cal E}$ are respectively the self--Hamiltonian of the system 
and the environment and  $V$ is the interaction term. The equation for the 
complete density matrix $\rho_{\cal T}$, in the interaction picture, reads (we use 
a tilde to denote operators in the interaction picture),
\begin{equation}
i\hbar\dot{\tilde{ \rho_{\cal T}}}=[\tilde V(t),\tilde\rho_{\cal T}],\label{VonNeumann}
\end{equation}
where the interaction potential and density matrix are
$\tilde V(t)=U^\dagger_0 V U_0$ and $\tilde\rho_{\cal T}=
U^\dagger_0\rho_{\cal T}U_0$, where $U_0=\exp(-i(H_{\cal S}+H_{\cal E})t/\hbar)$.
Solving equation (\ref{VonNeumann}) 
perturbatively is rather straightforward and leads to the Dyson series, 
\begin{equation}
\tilde\rho_{\cal T}(t)=\sum_{n\ge 1}\int_0^tdt_1\ldots\int_0^{t_{n-1}}dt_n
({1\over{i\hbar}})^n[\tilde V(t_1),\ldots,[\tilde V(t_n),\tilde\rho_{\cal T}(0)]].
\nonumber
\end{equation}
We can use this to compute the reduced density matrix to second order. To 
obtain the master equation we compute the time derivative of the resulting
expression and perform the trace over the environment. We get 
\begin{equation}
\dot{\tilde \rho}={1\over{i\hbar}} Tr_{\cal E}[\tilde V(t),\rho_{\cal T}(0)] -
{1\over{\hbar^2}} \int_0^t
dt_1 Tr_{\cal E}[\tilde V(t),[\tilde V(t_1),\rho_{\cal T}(0)]]. \label{rhodot}
\end{equation}
So far, the only assumption we made was the validity of a perturbative 
expansion up to second order. 
Now we will assume that the initial state is not entangled, i.e., that 
the total density matrix is a tensor product of the form
$\rho_{\cal T}(0)=\rho(0)\otimes\rho_{\cal E}(0)$. 
Substituting this into (\ref{rhodot}) we find, 
\begin{eqnarray}
\dot{\tilde\rho}&=&{1\over{i\hbar}} Tr_{\cal E}[\tilde V(t),\rho(0)\otimes\rho_{\cal E}(0)]
\nonumber\\ 
&-&{1\over{\hbar^2}} \int_0^t
dt_1 Tr_{\cal E}[\tilde V(t),[\tilde V(t_1),\rho(0)\otimes\rho_{\cal E}(0)]]. \label{rhodot3}
\end{eqnarray}
To finish the derivation, we make a rather trivial observation 
that enables us to rewrite the master equation in a very simple way: The 
initial state $\rho(0)$
that appears in the right-hand-side of equation (\ref{rhodot3}) could again 
be expressed in terms of $\tilde\rho(t)$ using the same perturbative expansion
that enabled us to obtain (\ref{rhodot3}). By doing this we can rewrite 
the right-hand-side of the master equation entirely in terms
of the reduced density matrix evaluated at time $t$. The resulting equation
is 
\begin{eqnarray}
\dot{\tilde\rho}&=&{1\over{i\hbar}} Tr_{\cal E}[\tilde V(t),\tilde\rho\otimes
\rho_{\cal E}(0)] 
-{1\over{\hbar^2}} \int_0^t
dt_1 Tr_{\cal E}[\tilde V(t),[\tilde V(t_1),\tilde\rho\otimes\rho_{\cal E}]]\nonumber\\
&+&{1\over{\hbar^2}} \int_0^t
dt_1 Tr_{\cal E}\Bigl([\tilde V(t),Tr_{\cal E}\bigl([\tilde V(t_1),\tilde 
\rho\otimes\rho_{\cal E}]\bigr)
\otimes\rho_{\cal E}]\Bigr).
\label{moth0}
\end{eqnarray}

This, when rewritten in the Schr\"odinger picture, is the basic 
master equa\-tion we will use in this section. It is important to keep  
in mind that to derive it, we only made two important assumptions: (a) 
we used a perturbative expansion up to second order in the system--environment
coupling constant and (b) we assumed uncorrelated initial conditions. 

Below, we will apply this equation to study three interesting
examples. Before doing that, let us stress that the 
master equation is local in time even though to obtain it, no 
Markovian assumption was made (see below). Moreover, this  
rather simple form can be simplified further by assuming that
the system--environment coupling is of the form 
\begin{equation}
V=\sum_n\bigl(S_nE_n+S^\dagger_nE^\dagger_n\bigl), 
\end{equation}
where $S_n$ ($E_n$) are operators acting on the Hilbert space of the
system (environment) only. In such case, the master equation in the 
Schr\"odinger picture can be written as 

\begin{eqnarray}
\dot\rho &=&{1\over{i\hbar}} [H_{\cal S} ,\rho]+ 
{1\over{i\hbar}} \sum_n[\langle E_n\rangle S_n+\langle E_n^\dagger
\rangle S_n^\dagger ,\rho]\nonumber\\
&-&{1\over{\hbar^2}} \sum_{nm}\int_0^tdt_1 \Bigl( K^{(1)}_{nm}(t,t_1)
[S_n,[S_m^\dagger(t_1-t),\rho]]\nonumber\\
&+&K^{(2)}_{nm}(t,t_1) [S_n,\{S_m^\dagger(t_1-t),\rho\}]
+K^{(3)}_{nm}(t,t_1)[S_n,[S_m(t_1-t),\rho]]\nonumber\\
&+&K^{(4)}_{nm}(t,t_1)[S_n,\{S_m(t_1-t),\rho\}]+h.c.\Bigr),\label{moth1}
\end{eqnarray}
where the bracket notation indicates the expectation value over the initial 
state of the environment and the kernels $K_{nm}^{(i)}$ are simply determined 
by the two time correlation functions of the environment as follows:
\begin{eqnarray}
K_{nm}^{(1)}(t,t_1)&=&{1\over 2}\langle\{E_n(t),E_m^\dagger(t_1)\}\rangle
-\langle E_n\rangle\langle E_m^\dagger\rangle\nonumber\\
K_{nm}^{(2)}(t,t_1)&=&{1\over 2}\langle [E_n(t),E_m^\dagger(t_1)]\rangle
\nonumber\\
K_{nm}^{(3)}(t,t_1)&=&{1\over 2}\langle\{E_n(t),E_m(t_1)\}\rangle
-\langle E_n\rangle\langle E_m\rangle\nonumber\\
K_{nm}^{(4)}(t,t_1)&=&{1\over 2}\langle [E_n(t),E_m(t_1)]\rangle. 
\end{eqnarray}

At this point, it is interesting to consider another important approximation 
that is usually employed in this context, i.e., the Markovian approximation 
that we have refrained from using so far. The Markovian approximation 
corresponds to considering cases for which the kernels $K^{(i)}$ are 
strongly peaked about $t=t_1$. When this is the case, i.e. when the 
environment has a very short correlation time, one can transform the 
temporal integrals into integrals over the variable $\tau=t-t_1$, which 
can then be extended over the entire interval $[0,\infty)$. As we mentioned
above, so far, we have not used the Markovian assumption and therefore the 
above equations are valid even if the environment has a long correlation 
time and the kernels $K^{(i)}$ are not strongly peaked. In the examples
below, we will mention some cases where this happens and use the above
equation to study decoherence produced by a non Markovian environment. 

It is also worth mentioning that to go one step beyond 
equation (\ref{moth1}), one needs to know the temporal dependence of 
the free Heisenberg operators of the system (i.e., $S_n(t)$) which obviously
depend on the Hamiltonian $H_S$ that we have not specified so far. We 
will do so in some concrete examples below. 

\subsection{Example 1: Perturbative master equation in Quantum 
Brownian Motion}

The system of interest is a quantum particle, which moves in a one 
dimensional space (generalization to higher dimensions is immediate). 
The environment is an ensemble of harmonic oscillators interacting bilinearly 
through position with the system. Thus, the complete Hamiltonian 
is $H=H_{\cal S}+H_{\cal E}+V$ where  
\begin{equation}
H_{\cal E}=\sum_n({1\over{2m_n}}p_n^2 + {1\over 2} m_n\omega_n^2 q_n^2)\nonumber
\end{equation}
and $V=\sum_n \lambda_n q_n x$. The Hamiltonian of the system will be
left unspecified for the moment (we will concentrate later on the 
case of a harmonic oscillator). The initial state of the environment 
will be assumed to be a thermal equilibrium state at temperature 
$T=1/k_B\beta$. Under these assumptions the first-order
term in the master equation disappears because $Tr_{\cal E}(\tilde V(t)\rho_{\cal E})=0$. 
Therefore, the master equation in the Schr\"oedinger picture is
\begin{equation}
\dot\rho ={1\over{i\hbar}} [H_{\cal S},\rho] 
-{1\over\hbar}\int_0^t dt_1 \Bigl(\nu(t_1) [x,[x(-t_1),\rho]]
-i \eta(t_1) [x,\{x(-t_1),\rho\}]\Bigr).\label{qbm0}
\end{equation}
The two kernels appearing here are respectively called the 
noise and the dissipation kernel and are defined as
\begin{eqnarray}
\nu(t)&=&{1\over{2\hbar}}\sum_n\lambda_n^2\langle\{q_n(t),q_n(0)\}\rangle=
\int_0^\infty d\omega J(\omega) \cos\omega t (1+2N(\omega))\nonumber\\
\eta(t)&=&{i\over{2\hbar}}\sum_n\lambda_n^2\langle[q_n(t),q_n(0)]\rangle
=\int_0^\infty d\omega J(\omega) \sin\omega t, \label{noisedis}
\end{eqnarray}
where $J(\omega)=\sum_n
\lambda_n^2\delta(\omega-\omega_n)/2m_n\omega_n$ is the spectral
density of the environment and 
$N(\omega)$ is the mean occupation number of the environmental 
oscillators (i.e., $1+2N(\omega)=\coth(\beta\hbar\omega/2)$). 

Equation (\ref{qbm0}) is already very simple but it can be further simplified 
if one assumes that the system is a harmonic oscillator. Thus, if
we consider the Hamiltonian of the system 
to be $H_{\cal S}=p^2/2M + M\Omega^2 x^2/2$,  
we can explicitly solve the Heisenberg equations for the system and 
determine the operator $x(t)$ to be
$x(t)=x \cos(\Omega t) + {1\over{M\Omega}} p\sin(\Omega t)$. 
Inserting this into (\ref{qbm0}), we get the final expression for the  
master equation,
\begin{eqnarray}
\dot\rho=&-&{i\over\hbar}\bigl[H_{\cal S}+{1\over2}M\tilde\Omega^2(t)x^2,
\rho\bigr]
-{i\over\hbar}\gamma(t)\bigl[x,\bigl\{ p,\rho\bigr\}\bigr]\nonumber\\
&-&D(t)\bigl[x,\bigl[ x,\rho\bigr]\bigr] 
 - {1\over\hbar}f(t)\bigl[x,\bigl[ p,\rho\bigr]\bigr].\label{hpz}
\end{eqnarray}
Here the time--dependent coefficients (the frequency renormalization
$\tilde\Omega(t)$, the damping coefficient $\gamma(t)$, and the two
diffusion coefficients $D(t)$ and $f(t)$) are
\begin{eqnarray}
\tilde\Omega^2(t)&=& -{2\over M}\int_0^t dt'\cos(\Omega t') 
\eta(t'), \ 
\gamma(t)={1\over M\Omega}\int_0^t dt'\sin(\Omega t') \eta(t')
\nonumber\\
D(t)&=&{1\over\hbar}\int_0^t dt'\cos(\Omega t') \nu(t'), \ 
f(t)=-{1\over M\Omega}\int_0^t dt'\sin(\Omega t') \nu(t').\label{coefhpz}
\end{eqnarray}

   From this equation it is possible to have a qualitative idea of the 
effects the environment produces on the system. First we observe that there
is a frequency renormalization. Thus, the ``bare'' frequency of the 
oscillator is renormalized by $\tilde\Omega^2$. This term does not 
affect the unitarity of the evolution. The terms proportional to $\gamma(t)$,
$D(t)$ and $f(t)$ bring about non--unitary effects. Thus, one can easily
see that the second term is responsible for producing friction ($\gamma(t)$
plays the role of a time--dependent relaxation rate). The last two  
are diffusion terms. The one proportional to $D(t)$ is the main 
cause for decoherence. 

Of course, the explicit time dependence of the coefficients can only 
be computed once we specify the spectral density of the environment. 
To illustrate their qualitative behavior, we will consider a typical 
ohmic environment characterized by a spectral density of the form
\begin{equation}
J(\omega)=2M\gamma_0{\omega\over\pi} 
{\Lambda^2\over{\Lambda^2+\omega^2}},\label{Drude}
\end{equation}
where $\Lambda$ plays the role of a high--frequency cutoff and $\gamma_0$
is a constant characterizing the strength of the interaction. For this 
environment, it is rather straightforward to find the following 
exact expressions for the coefficients $\tilde\Omega(t)$ and $\gamma(t)$:
\begin{eqnarray}
\gamma(t)&=&{\gamma_0}{\Lambda^2\over{\Lambda^2+\Omega^2}}
\left(1-\left(\cos\Omega t +{\Lambda\over\Omega}\sin\Omega t\right)
\exp(-\Lambda t)\right)\label{gammat}\\
\tilde\Omega^2(t)&=&-{2\gamma_0\Lambda}{\Lambda^2\over{\Lambda^2+\Omega^2}}
\left(1-\left(\cos\Omega t -{\Omega\over\Lambda}\sin\Omega t\right)
\exp(-\Lambda t)\right).\label{omegat}
\end{eqnarray}
      From these equations we see that these coefficients are initially zero 
and grow to asymptotic values on a timescale that is fixed by 
the high--frequency cutoff $\Lambda$. Thus, we see the relation between
this result and the one we would obtain by using a Markovian approximation
simply corresponds to taking the limit $\Lambda\rightarrow\infty$. In 
such a case both coefficients are not continuous at $t=0$ and jump into
constant values (the frequency renormalization 
diverges as it is proportional to the product $\gamma_0\Lambda$). 

The time dependence of the diffusion coefficients can also be studied 
for the above environment. However, the form of the coefficients for 
arbitrary temperature is quite complicated. To analyze the qualitative 
behavior, it is convenient to evaluate them numerically. In Figure 3 one  
can see the dependence of the coefficients (for both the long and 
short timescales) for several temperatures (high and low). 
We observe that both coefficients have an initial
transient where they exhibit a behavior that is essentially temperature 
independent (over periods of time comparable with the one fixed by 
the cutoff). 
The direct diffusion coefficient $D(t)$ after the initial transient 
rapidly settles into the asymptotic value 
given by $D_\infty=M\gamma_0\Omega
\coth(\beta\hbar\Omega/2)\Lambda^2/\hbar(\Lambda^2+\Omega^2)$. The anomalous 
diffusion coefficient $f(t)$ also approaches an asymptotic value (which 
for high temperatures 
is suppressed with respect to $D_\infty$ by a factor of $\Lambda$), but 
the approach is algebraic rather than exponential. 
More general environments can be studied using our equation. In fact, 
the behavior of the coefficients is rather different for environments 
with different spectral content. This has been analyzed in the 
literature, in particular in relation to decoherence \cite{HPZ}. 

It is interesting to mention that the master equation 
(\ref{hpz}) (although it has been derived perturbatively) can be 
shown to be very similar to its exact counterpart whose derivation 
we will discuss later in this section.

\begin{figure}
\qquad
	\includegraphics[height=0.9\hsize]{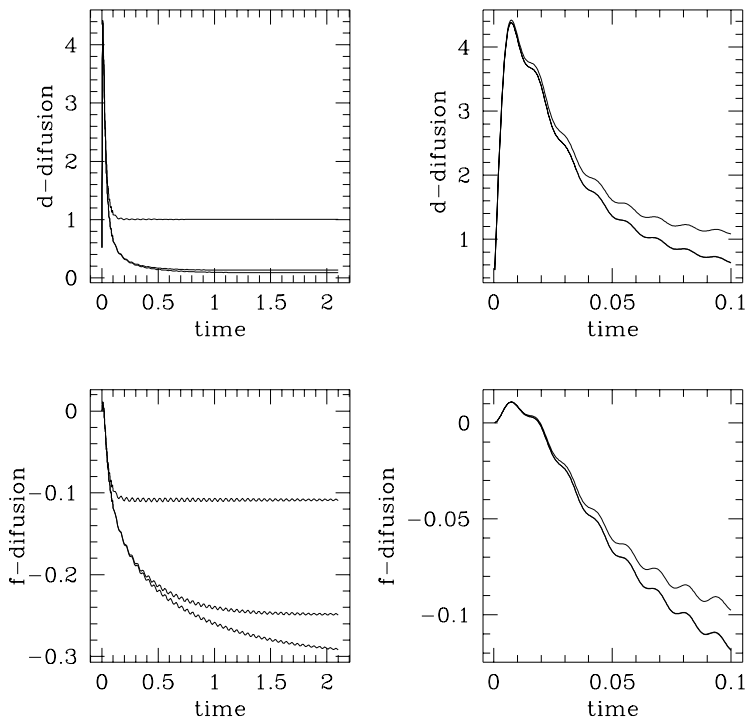}
\caption{Time dependence of the diffusion coefficients of the 
perturbative master equation for quantum Brownian motion. Plots on the right
show that the initial transient is temperature independent (different 
curves correspond to different temperatures, higher temperatures 
produce higher final values of the coefficients). Plots on the left show 
that the final values of the coefficients are strongly dependent on the 
temperature of the environment. The parameters used in the plot (where time
is measured in units of $1/\Omega$) are $\gamma/\Omega=0.05$, 
$\Lambda/\Omega=100$, $k_BT/\hbar\Omega=10, 1, 0.1$.}
\end{figure}

\subsection{Example 2: Perturbative master equation for a 
two--level system coupled to a bosonic heat bath}

As a second example we obtain the 
perturbative master equation for a two--level system coupled to an 
oscillator environment. We consider two different models characterized
by different interaction Hamiltonians. 
First, we discuss the model describing the physics
of the decay of a two--level atom (in the rotating wave approximation),
\begin{equation}
H={1\over 2}\hbar\Delta\sigma_z+
\sum_n\lambda_n\left(a_n\sigma_++a_n^\dagger\sigma_-\right)
+\sum_n\hbar\omega_na_n^\dagger a_n,\label{Hspin1}
\end{equation}
where $a_n$ and $a_n^\dagger$ are annihillation and creation operators
of the environment oscillators, and $\sigma_\pm$ are the raising and
lowering operators of the two--level system. The perturbative master equation 
obtained following the procedure described above is
\begin{eqnarray} 
\dot\rho &=&{1\over{i\hbar}} [H_{\cal S},\rho] \nonumber\\
&-&{1\over2 \hbar^2}\int_0^t dt_1 
k(t_1)\left( [\sigma_+,[\sigma_-(-t_1),\rho]]
+ [\sigma_+,\{\sigma_-(-t_1),\rho \}]\right) + h.c.),\nonumber
\end{eqnarray}
where the kernel $k(t)$ is defined as
\begin{equation}
k(t)=\sum_n\lambda_n^2\langle[a_n(t),a_n^\dagger]\rangle=
\sum_n\lambda_n^2\exp(-i\omega_nt).\nonumber
\end{equation}
Using the solution of the free Heisenberg equations for the spin 
operator (i.e., $\sigma_\pm(t)=\sigma_\pm\exp(\pm i\Delta t)$), 
we can deduce that the master equation is 
\begin{equation}
\dot\rho ={1\over{i\hbar}} [\hbar\left({\Delta\over 2}-c(t)\right)
,\rho] +a(t)\left(\sigma_+\sigma_-\rho
+\rho\sigma_+\sigma_--2\sigma_-\rho\sigma_+\right),\nonumber
\end{equation}
where the time--dependent coefficients are 
\begin{equation}
a(t)=2Re f(t),\qquad c(t)=Im(f((t)),\nonumber
\end{equation}
with 
\begin{equation}
f(t)={1\over{2\hbar^2}}\int_0^tds k(s)\exp(i\Delta s).\label{coefspin1}
\end{equation}
We recognize in this equation similar features to those present
in the one for quantum Brownian motion (QBM). 
The interaction with the environment on the one
hand renormalizes the Hamiltonian of the particle through the term
$c(t)$ (including thermal fluctuations, we could verify that $c(t)$ is
generally temperature dependent, as opposed to the QBM case). 
The nonhermitian part has a zero temperature contribution 
that is responsible for the spontaneous decay of the two--level 
system. The decay rate is determined by $b(t)$ and has a 
time dependence that is essentially the same as the one found for 
the diffusion coefficient in the zero temperature QBM case analyzed 
above. The finite temperature contributions can be shown 
to be responsible not only for the 
changes in the value of the decay rate $b(t)$ (which in that case would 
account also for the induced decay) but also for adding new terms to 
the master equation that take into account the induced absorption. 

Finally we obtain the perturbative master equation for the spin 
boson Hamiltonian, which is also widely used in various condensed--matter 
physics problems (and was thoroughly studied in the nonperturbative 
regime in \cite{Leggetal})
\begin{equation}
H={1\over 2}\hbar\Delta\sigma_x+ \sigma_z 
\sum_n\lambda_n q_n+\sum_n\hbar\omega_na_n^\dagger a_n,\label{Hspin2}
\end{equation}
where $q_n$ are the coordinates of the environmental oscillators.
The master equation can be shown to be 
\begin{equation}
\dot\rho ={1\over{i\hbar}} [H_{\cal S},\rho] 
-{1\over\hbar}
\int_0^t dt_1 \Bigl(\nu(t_1) [\sigma_z,[\sigma_z(-t_1),\rho]]
-i \eta(t_1) [\sigma_z,\{\sigma_z(-t_1),\rho\}]\Bigr),\nonumber
\end{equation}
where the two kernels are the same 
as defined above in the QBM case (\ref{noisedis}). Using the 
free Heisenberg operator $\sigma_z(t)=\sigma_z \cos \Delta t
+\sigma_y\sin \Delta t$ we obtain the master equation,  
\begin{equation}
%\dot\rho ={1\over{i\hbar}} [H_{\cal S},\rho] - 
%\tilde D(t) [\sigma_z,[\sigma_z,\rho]] +  \tilde f(t)[\sigma_z,
%[\sigma_y,\rho]] + i \tilde \gamma(t)) [\sigma_z,\{\sigma_y,\rho\}]
%\nonumber\\
\dot\rho ={1\over{i\hbar}} [H_{eff},\rho] - 
\tilde D(t) [\sigma_z,[\sigma_z,\rho]] 
+ z(t) \sigma_z\rho\sigma_y+z^*(t)\sigma_y\rho\sigma_z\ ,\nonumber
\end{equation}
where the effective Hamiltonian and the time--dependent coefficients
are now given by 
\begin{eqnarray}
H_{eff}&=&\hbar\left({\Delta\over 2}-z^*(t)\right)\sigma_x,\nonumber\\
\quad\tilde D(t)&=&\int_0^tds\nu(s)\cos \Delta s, \qquad
z(t)=\int_0^t ds \left(\nu(s)-i\eta(s)\right) \sin \Delta s.\nonumber
\end{eqnarray}
As before, the interpretation is quite straightforward. The effect of the 
environment is to renormalize the frequency as well as to introduce 
the decay of the system. This effect takes place only if the bare 
frequency $\Delta$ is nonzero (otherwise $z(t)$ vanishes). 
The other effect of the environment is to destroy the nondiagonal 
terms in the density matrix, a task that is carried out by the 
term proportional to $\tilde D$, which is present even when the bare
driving vanishes. As before, the expression for the time--dependent 
coefficients is qualitatively similar to the one observed in the QBM
model. 

\subsection{Example 3: Perturbative 
master equation for a particle interacting with a quantum field}

We consider the following simple model: The system is a
particle with position $\vec x$ (moving in a $3$-dimensional space) and
the environment is a quantum scalar field $\phi$. The interaction 
between them is local as described by the Hamiltonian
$V=e\phi(\vec x)$, 
where $e$ is the coupling constant (the "charge" of the particle). 
Expanding the scalar field in normal modes, the Hamiltonian can be
written as $V=\int d{\vec k}(h_{\vec k}\exp(i\vec k\vec x)+ h.c.)$ 
where the Fourier components $h_{\vec k}$ are proportional to 
annihillation operators of the quantum field (i.e., 
$h_{\vec k}=e\ a_{\vec k}/(2\pi)^{/2}(2\omega_k)^{1/2}$).
More generally, we could consider models in which the
particle--field interaction is slightly nonlocal taking into account
the finite extent of the particle (thus, a nonrelativistic treatment
of the quantum particle would only give consistent results if we do 
not attempt to localize it beyond its Compton wavelength). 
In this case, the interaction Hamiltonian
$\tilde H_{int}=e\int d \vec y W(\vec x -\vec y)\phi(\vec y)$
depends upon the window function $W(\vec r)$ whose 
support lies inside a sphere of radius $R$ 
(the Compton radius of the particle) centered around the origin. 
This nonlocal interaction corresponds to a Hamiltonian whose Fourier components
$h_{\vec k}$ are multiplied by $\hat W(\vec k)$ (the Fourier transform 
of $W(\vec r)$). As we can see, the net effect of taking into account the 
finite size of the particle is to introduce an ultraviolet cutoff in 
the scalar field (the particle does not interact with the field modes with 
frequencies higher that its rest mass). 

It is interesting to note that 
for this class of models we can also derive a master equation for the 
reduced density matrix of the particle. Thus, using the perturbative 
approach described above, we simply obtain (assuming the initial 
state of the quantum field is thermal equilibrium) the master equation  
as follows:
\begin{eqnarray}
\dot\rho &=&-{i\over\hbar}[H,\rho]
-{e^2\over\hbar^2}\int d{\vec k}\int_0^tdt_1
\Bigl( G_H(\vec k,t_1)\bigl[e^{i\vec k\vec x},
\bigl[e^{-i\vec k\vec x(-t_1)},\rho\bigr]\bigr]-\nonumber\\
&-&i G_R(\vec k,t_1)\bigl[e^{i\vec k\vec x},\bigl\{e^{-i\vec k\vec
x(-t_1)},\rho\bigr\}\bigr]\Bigr).\label{masterfield}
\end{eqnarray}
Here, $\vec x(t)$ is the Heisenberg position operator for the 
particle (evolved with the free Hamiltonian $H$) and
$G_{R,H}(\vec k,t)$ are the Fourier transforms of
the retarded and symmetric two--point functions 
of the scalar field (multiplied by the appropriate window function if 
the interaction is nonlocal). When the environment is a free field, 
we have
\begin{eqnarray}
G_R(\vec k, t)&=&W(\vec k)\sin(\omega_{\vec k}t)/2\omega_{\vec k}\ ,
\nonumber\\
G_H(\vec k, t)&=&W(\vec k)\cos(\omega_{\vec k}t)(1+2N_k)/2
\omega_{\vec k}\ ,
\end{eqnarray}
where $N_k$ is the number density of particles in the 
initial state of the quantum field (the above result is valid
if the field is not free, in which case the propagators are appropriately
dressed). This master equation is extremely rich.
Here, we will use it for two main purposes. On the one hand, 
we can see that the Quantum Brownian Motion case is a special limit 
of this particle--field model that arises in the so--called dipole 
approximation. This is the most widely used approximation 
in this context and is valid whenever the dominant wavelengths in the
environment are much larger than the lengthscale over which the position
of the particle varies. If this is the case, we can expand the exponentials
up to second order ($\vec k \vec x \ll 1$) and obtain: 
\begin{eqnarray}
\dot\rho=-{i\over\hbar}[H,\rho]-{e^2\over\hbar^2}\int_0^t&dt_1& \Bigl(
F_H(t_1)\bigl[\vec x,\bigl[\vec x(-t_1),\rho\bigr]\bigr]\nonumber\\
&-&i F_R(t_1)\bigl[\vec x,\bigl\{\vec x(-t_1),\rho \bigr\}\bigr]\Bigr),
\nonumber\end{eqnarray}
where $F_{R,H}(t_1)=\int d{\vec k} \vec k^2 G_{R,H}(\vec k,t_1)/N(2\pi)^{3/2}$.
Thus, our first example of a linear Brownian particle coupled to an 
oscillator environment arises as the dipole approximation of the particle
field model. With this in mind, we will use the particle field model as an
example to show that some of the results obtained in the QBM case are
just artifacts of the dipole approximation. In particular, this will 
be the case with the dependence of the decoherence rate on distance. Using
the master equation of our particle--field model we will easily show that
the decoherence rate does not indefinitely grow with distance but exhibits 
saturation. 

\subsection{Exact master equation for Quantum Brownian Motion}

After presenting some simple perturbative master equations one may 
wonder under what circumstances 
are they a reasonable approximation. To partially address this issue,  
it is interesting to compare these equations with the ones
that can be obtained for exactly solvable problems. In particular, 
we describe the master equation for a model that has been 
thoroughly studied in connection with decoherence, i.e., the linear
quantum Brownian motion. Thus, because the Hamiltonian is 
quadratic both in the coordinates of the system and the 
environment, it is not surprising that it can be exactly solved. 
In this subsection, we will describe a simple derivation of the exact master 
equation, discuss its main features, and show that its functional form
is the same as the one obtained by using perturbation 
theory. Indeed, the exact master equation has the same functional form 
as (\ref{hpz}), the only difference being that the time dependence of the  
coefficients is different in general, as expected. 

It is interesting to 
note that the exact master equation for QBM has only been found 
recently in spite of the simplicity of the model (in particular, the 
fact that it can always be written as an equation that is local in 
time was not appreciated until very recently \cite{HPZ}). Unfortunately, 
the derivation of the exact master equation is not so simple and, to say
the least, the original one presented in \cite{HPZ} is indeed rather 
complicated. Here we will present the simplest derivation of the exact
master equation that we know of, 
which is done following the method proposed
first in \cite{PazMazag}. Previous studies of the master equation for 
QBM, obtained under various approximations, include the celebrated paper 
by Caldeira and Legget \cite{Caldeira} among others 
(see also \cite{Haake,UZ}).

The derivation will focus on properties of the evolution operator for
the reduced density matrix. This operator will be denoted as $J$ and 
is defined as the one that enables us to find the reduced density matrix
at some arbitrary time from the initial one. Thus, by definition, this
operator satisfies:
\begin{equation}
\rho(x,x',t)=\int dx_0\int dx'_0 J(x,x',t;x_0,x'_0,t_0)
\rho(x_0,x'_0,t_0)\ . \label{Jdef}
\end{equation}

The derivation of the exact master equation has two essential steps. 
The first step is to find an explicit form for the evolution operator 
of the reduced density matrix. The second step is to use this explicit 
form to obtain the master equation satisfied by the reduced density matrix.  
To make our presentation simpler, we postpone the proof of the first 
step, which will be done below using path integral techniques. Here, we
first want to demonstrate how to obtain the master equation once we know the
explicit form of the evolution operator. So, let us show what the 
evolution operator for the reduced density matrix looks like. For 
linear QBM we will show later that it can always be written as 
\begin{eqnarray} 
J(X,Y,t; X_0,Y_0,t_0)&=&{b_3\over 2\pi}\exp i\left(b_1XY+b_2X_0Y-b_3XY_0-b_4
X_0Y_0\right)\nonumber\\
&\times&\exp\left(-a_{11}Y^2-a_{12}YY_0-a_{22}Y_0^2\right),\label{jQBM}
\end{eqnarray}
where for notational convenience we are using sum and difference 
coordinates (i.e., $X=x+x'$, $Y=x-x'$, etc) and the coefficients $b_i$ 
and $a_{jl}$ are time--dependent functions whose explicit form will 
be given below (and depend on the properties of the environment). Thus, 
the evolution operator (\ref{jQBM}) is simply a Gaussian function of 
its arguments with time--dependent coefficients. This comes as no 
surprise because the problem is linear. 

Knowing the propagator for the reduced density matrix, it is easy to obtain 
the master equation following the simple method described
in \cite{PazMazag}. This is the second step of the derivation of the 
master equation and is done as follows. We compute the temporal derivative 
of the propagator $J$ noting that the only time dependence is through 
the coefficients $b_i$ and $a_{jl}$. Thus, we obtain
\begin{eqnarray}
\dot J&=&\Bigl({\dot b_3\over b_3}+i(\dot b_1XY +\dot b_2 X_0Y+\dot b_3 XY_0
+\dot b_4 X_0Y_0)\nonumber\\
&-&\dot a_{11}Y^2-\dot a_{12} YY_0-\dot a_{22}Y_0^2\Bigr) J\ . 
\label{tderivJ}\end{eqnarray}
Using this equation, we can try to find the master equation through 
multiplying
by the initial density matrix and integrating this 
over the initial coordinates. 
The master equation would be trivially obtained in this way if, 
after multiplying by the initial density matrix, we could 
integrate over all the initial coordinates. 
This is straightforward, with some of the terms appearing in (\ref{tderivJ})
but it is not so obvious how to handle terms that explicitly 
depend upon the initial coordinates $X_0$ and $Y_0$. 
Fortunately, there is a simple trick that we can use: because 
we know that 
the propagator (\ref{jQBM}) is Gaussian, we can make use of this fact to 
obtain the following simple relations:
\begin{eqnarray}
Y_0J&=&\left({b_1\over b_3} Y + {i\over b_3} \partial_X\right) J,
\ {\rm and}\nonumber\\
X_0J&=&\left(-{b_1\over b_2} X-{i\over b_2} \partial_Y-i({2a_{11}\over b_2}
+{a_{12}b_1\over{b_2b_3}}Y)+{a_{12}\over{b_2b_3}}\partial_X\right)
J\ .\nonumber
\end{eqnarray}
These two equations can be used in (\ref{tderivJ}) and in this way
we can express the right hand side of this equation entirely in terms
of the reduced density matrix. The resulting master equation is 
\begin{eqnarray}
\dot\rho(x,x')={1\over i\hbar}\langle x|[H_R(t),\rho]|x\rangle -
\gamma(t)(x-x')(\partial_x-\partial_x')\rho(x,x')\nonumber\\
-D(t)(x-x')^2\rho(x,x') +if(t)(x-x')(\partial_x+\partial_x')\rho(x,x')\ .
\label{hpzexact}
\end{eqnarray}
The coefficients appearing in this equation are determined by $b_i$ and
$a_{jl}$ as follows:
\begin{eqnarray}
\Omega^2(t)&=&2(\dot b_2b_1/b_2-\dot b_1)\qquad \gamma(t)=-\dot b_2/2b_2-b_1
\nonumber\\
D(t)&=&\dot a_{11}-4a_{11}b_1+\dot a_{12}b_1/b_2-\dot b_2(2a_{11}+
a_{12}b_1/b_3)/b_2\nonumber\\
2f(t)&=&\dot a_{12}/b_3-\dot b_2a_{12}/b_2b_3-4a_{11}\ .\label{coefexact}
\end{eqnarray}

Thus, we showed that the exact master equation is a simple consequence
of the Gaussian form of the evolution operator (\ref{jQBM}). 
To complete our derivation of this equation we need to 
explicitly show how to obtain equation (\ref{jQBM}) and 
also find the explicit form of the time--dependent coefficients (which 
is also required to simplify the expressions leading to the master 
equation (\ref{hpzexact}). 

To obtain the explicit form of the evolution operator we will follow 
a derivation based on the use of path integral techniques 
(see \cite{Grabert, HPZ, HPZ2, PazMazag, Romero}). To understand it, 
very little previous knowledge of path integrals is required. The main 
ingredient is the path integral expression for the evolution operator
of the complete wave function. Thus, if the action of the combined 
system is $S_T[x,q]$,  the matrix elements of the evolution operator
$U$ can be written as
\begin{equation}
U(x, q, t; x_0, q_0, t_0) = \int Dx  Dq \ 
{\rm e}^{ i S_T[x, q] }\ ,\label{U}.
\end{equation}
where the integration is over all paths that satisfy the boundary conditions,
\begin{equation}
x(0) = x_0,\ x(t) = x,\ q(0) = q_0,\ q(t) = q\ .  \label{bc}
\end{equation}
In the above and following equations, to avoid the proliferation of 
sub-indices we use $q$ to collectively denote all the coordinates
of the oscillators $q_n$ (we will not write the subscript $n$ that 
should be implicitly assumed). 
Using this equation, one can obtain a path integral representation of 
the evolution operator of the complete density matrix and, after taking
the final trace over the environment, we find a path integral representation
of the propagator for the reduced density matrix. 
It is clear that the resulting expression 
will involve a double path integral (one to 
evolve kets and another one to evolve bras). For a generic initial state
$\rho_{\cal T}$,  
the propagator is a somewhat complicated--looking expression. To simplify
our presentation, we will only consider here factorizable initial states
(and refer the reader to \cite{Romero} for the most general situation
where initial correlations are included). Thus, if the initial state
can be factored we can express the reduced density matrix at arbitrary
times as a function of the reduced density matrix at initial time 
using a (state--independent) propagator that 
has the following path integral representation:
\begin{equation}
 J(x,x',t;x_0,x'_0,t_0)=\int Dx\int Dx' \exp(iS[x]-iS[x']) F[x,x']\ .
\label{pathQBM}
\end{equation}
where the integral is over paths satisfying the above boundary conditions,
$S[x]$ is the action for the system only, and $F[x,x']$ is the so--called 
``Influence Functional''
first introduced by Feynman and Vernon \cite{FeynmanVernon}. This functional
is responsible for carrying all the physical effects produced by the 
environment on the evolution of the system. In fact, if there is no 
coupling between the system and the environment, the Influence Functional 
is equal to the identity, and the above expression reduces to the one
corresponding to the free Schr\"odinger evolution for the isolated system. 
The Influence Functional is defined as 
\begin{equation}
F[x,x']=\int dq dq_0 dq'_0 \rho_{\cal E}(q_0,q'_0)\int Dq Dq' 
\exp(i(S_{\cal SE}[x,q]-S_{\cal SE}[q',x'])),\label{influence}
\end{equation}
where $\rho_{\cal E}$ is the initial state of the environment and $S_{\cal SE}[q,x]$
is the action of the environment (including the interaction term with 
the system). 
It is easy to see that if there is no interaction (or if the
two systems trajectories are the same, i.e., $x=x'$), then the influence
functional is equal to one. 

Calculating the Influence Functional for an environment formed 
by a set of independent oscillators coupled linearly to the system 
is a rather straightforward task (and, to our knowledge, was first done
by Feynman and Vernon in \cite{FeynmanVernon}). Assuming the initial 
state of the environment is thermal equilibrium at temperature 
$T=1/k_B\beta$, the result is
\begin{eqnarray}
F[x,x']&=&\exp(-i\int_0^tdt_1\int_0^{t_1}dt_2 Y(t_1)\eta(t_1-t_2)
X(t_2)\nonumber\\
&-&\int_0^tdt_1\int_0^t dt_2 Y(t_1)\nu(t_1-t_2)Y(t_2))\ ,
\label{influenceQBM}
\end{eqnarray} 
where $X=x+x'$, $Y=x-x'$, and the two kernels $\nu(s)$ and $\eta(s)$ are
the so--called noise and dissipation kernels that were defined above 
in (\ref{noisedis}). Thus, all the influence of the environment on 
the evolution of the system is encoded in the noise and dissipation 
kernels (two different environments that produce the same kernels would
be equivalent as to the impact they have on the system). To obtain 
the above expression is a simple exercise in path integrals. However, 
the calculation can also be done by a more straightforward procedure 
that makes no reference to path integrals. 
Indeed, one can notice that the influence functional can always be 
expressed in operator language as 
\begin{eqnarray}
F[x,x']=Tr_{\cal E}&(&T(e^{-i\int_0^tdt_1V_{int}[x'(t_1),q(t_1)]}
) \rho_{\cal E} \times\nonumber\\
&&\tilde T(e^{i\int_0^tdt_1V_{int}[x(t_1),q(t_1)]}
))\ ,\nonumber
\end{eqnarray} 
where $T$ ($\tilde T$) denotes the time ordered (antitime ordered) 
product of the corresponding Heisenberg operators, and $V_{int}$ is the 
interaction term between the system and the environment. 
If the interaction is bilinear and the initial state of the environment
is thermal, one can easily realize that 
the result should be a Gaussian functional of both $x$ and $x'$. Therefore, 
one can just write down such most general Gaussian functionals in 
terms of unknown kernels. These kernels could be identified by 
using the above expression, taking functional derivatives with respect 
to $x$ and $x'$ and evaluating the result when $x=x'$. In this way,  
one realizes that the result is given by (\ref{influenceQBM}), where
the noise and dissipation kernels are given by expectation of 
symmetric and antisymmetric two--time correlation functions of the 
environment oscillators, exactly as in (\ref{noisedis}). 

Knowing the Influence Functional enables us to compute the exact 
expression for the evolution operator of the reduced density matrix. 
In fact, all we need is to perform the path integral in (\ref{pathQBM}). 
If the system is linear we see that the integrand is Gaussian and, 
therefore, the integral can also be explicitly computed. To perform 
this integral is not so trivial because the integrand is not separable
into a product of functions of $x$ and $x'$. However, the integral can 
be calculated simply by changing variables. First we should integrate
over sum and difference coordinates $X$ and $Y$. Then, we should change
variables writing $X=X_c+\tilde X$ and $Y=Y_c+\tilde Y$ where $X_c$ 
and $Y_c$ satisfy the equations obtained by varying the phase of the
integrand and imposing the corresponding boundary conditions. In this
way, we show that the result of the path integral is simply the 
integrand evaluated in the trajectories $X_c$, $Y_c$, multiplied by 
a time--dependent function that can be determined by normalization. The
only nontrivial part of this derivation is to realize that the 
trajectories $X_c$ and $Y_c$ can be chosen as the ones extremizing only 
the phase of the integrand, (and not the entire exponent that, as we
saw, has a real part coming from the noise). For more details on this
derivation the interested reader can look in \cite{HPZ,Grabert,Romero}. 
Therefore, the final result is given in equation (\ref{jQBM}) 
where the coefficients $b_i$ and $a_{jl}$ are time--dependent
functions that are determined in the following way. Let the functions 
$u_{1\atop 2}$ be two solutions of the equation,
\begin{equation}
\ddot u(s)+\Omega^2 u(s) +2\int_0^s ds'\eta(s-s') u(s') =0\ ,\nonumber
\end{equation}
satisfying the boundary conditions $u_1(0)=u_2(t)=1$ and $u_1(t)=u_2(0)=0$.
Then, the coefficients appearing in (\ref{jQBM}) are simply given by
\begin{eqnarray}
b_{1\atop 2}&=&{1\over 2} \dot u_{2\atop 1}(t),\qquad b_{3\atop 4}={1\over 2}
\dot u_{2\atop 1}(0)\nonumber\\
a_{jl}&=&(1+\delta_{jl})^{-1}\int_0^tds\int_0^tds' u_j(s) u_k(s')
\nu (s-s')\ .\label{biajl}
\end{eqnarray}

The time dependence of the coefficients of the master equation 
can be investigated after specifying the spectral density and the 
temperature of the environment. This has been done
in great detail in a series of papers \cite{HPZ, HPZ2, PazMazag, PHZ}. 
We will not review these results in detail but would just like to mention
that for the case that is most interesting for studying decoherence, which
is the underdamped (i.e., weakly coupled) harmonic oscillator, the time
dependence of the exact coefficients is very similar to the one
obtained by analyzing the coefficients appearing in the perturbative 
master equation. Indeed, the perturbative coefficients obtained above
can be recovered by solving the equation for the functions $u_1$ perturbatively
and replacing these equations inside (\ref{biajl}) and
(\ref{coefexact}). Thus, to 
get a qualitative idea about the behavior of the coefficients, we restrict
ourselves to the analysis already made for the perturbative ones (see 
Figure 3). 

It will be useful to analyze decoherence not only using the reduced
density matrix but also the Wigner function that is the phase space
distribution function that can be obtained from the density matrix as
\cite{Wigner}
\begin{equation}
W(x,p)=\int_{-\infty}^{+\infty}{{dz}\over{2\pi\hbar}}
\hbox{e}^{ipz/\hbar}\rho(x-z/2,x+z/2). \label{wigdef}
\end{equation}
It is simple to show that for the case of the harmonic oscillator, 
the evolution equation for the Wigner function can be obtained 
from the master equation and has the form of a Fokker Planck equation
\begin{equation}
\dot{W} = -\left\{ H_{ren}(t), W\right\}_{PB} + \gamma (t) \partial_p(p W) +
D(t) \partial^2_{pp} W - f(t) \partial^2_{px} W. \label{qbmwignereq} 
\end{equation}
The form of the evolution equation for the Wigner function 
for more general (nonlinear) systems will be discussed in Section 6. 

As a final remark, it is worth pointing out that the exact master equation
does not have the so--called ``Lindblad form''. A master equation is 
of the Lindblad form \cite{Lindblad} if it can be written as
\begin{equation}
\dot\rho={1\over i\hbar}[H,\rho]-\sum_n \gamma_n(L_n^\dagger L_n\rho +
\rho L_n^\dagger L_n -2 L_n\rho L_n^\dagger)\ ,\nonumber
\end{equation}
for some operators $L_n$ and some (positive) constants $\gamma_n$. As 
shown by Lindblad, this is the most general master equation with the 
property of being Markovian and preserving the positivity of the density
matrix. The fact that the exact master equation does not have 
the Lindblad
form may be puzzling but after some thinking becomes natural. Of course, 
the exact evolution also preserves positivity of the density matrix, but
it does so in a more subtle way than through a Lindblad master equation. 
The true evolution is not Markovian (but in a very weak sense). The only 
memory effect relies on the fact that the system remembers the initial 
time when the (factorizable) initial conditions were imposed. This 
effect appears in the time dependence of the coefficients that is 
responsible also for enforcing positivity in an interesting way (see 
\cite{Romero, PHZ} for some discussion on the way positivity follows
from the exact master equation). As a final comment, we would like to 
mention the fact that exact master equations are rather rare, but the 
above equation for QBM is not the only interesting exact master equation
known. For example, it is possible to derive an exact master equation
that has strong similarities with the one for QBM (i.e., an equation
that is local in time and has time--dependent coefficients) for the 
model of a two--level system coupled to a bosonic bath through the 
Hamiltonian (\ref{Hspin1}) (this equation was derived first in 
\cite{Garraway} and rediscovered by other means in \cite{AnasHu}).

\section{Einselection in quantum Brownian motion}

\subsection{Decoherence of a superposition of two coherent states}

We will analyze here the decoherence process in a simple example:
the linear quantum Brownian motion model whose exact master equation 
is given in (\ref{hpzexact}). For this we will first set up an initial state
that is delocalized in position (or momentum) space and examine
its temporal evolution, paying special attention to the fate of interference
effects. Thus, we will consider a state of the form \cite{Zurek86,PHZ}
\begin{equation}
\Psi(x,t=0)=\Psi_1(x) + \Psi_2(x)\ ,            \label{init}
\end{equation}
where
\begin{equation}
\Psi_{1,2}(x) = N\exp\left(-{(x\mp L_0)^2\over
2\delta^2}\right)~\exp\left(\pm iP_0 x\right), 
\end{equation}
\begin{equation}
N^2 \equiv {{\bar{N}}^2\over\pi\delta^2} = {1\over
2\pi^2\delta^2}\left[1+\exp\left(-{L_0^{~2}\over\delta^2}-\delta^2
P_0^2\right)\right]^{-1}.                  
\end{equation}
Note that we assumed (just for simplicity) that the two wave
packets are symmetrically located in phase space. 
%These manifests in Fig. 4 where we have plotted the initial
%Wigner function for two special initial conditions: Fig. 1.a 
%corresponds to the case $L_0=5\delta,~P_0=0$ which, in what follows, 
%will be called condition A while Fig. 1.a$^{\prime}$ corresponds 
%to $L_0=0,~P_0=5/\delta$ (condition A$^{\prime}$). 
The above expression allows us to study two extreme cases: the 
coherent states are
%One can consider 
%these as two extreme conditions that allow us to study the dependence 
%of decoherence on the initial condition of the system (the coherent 
separated in position or in momentum. In both cases, as a
consequence of quantum interference, the Wigner function oscillates and
becomes negative in some regions of phase space (and therefore cannot
be interpreted as a probability distribution). When the coherent states
are separated in position (momentum), the fringes are aligned along 
the $p$ ($x$) axis.  

\begin{figure}
\vspace{-2.0cm}
\qquad\qquad\quad
	\includegraphics[height=1.\hsize]{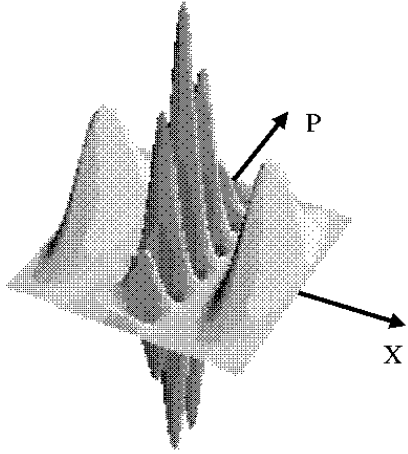}
\vspace{-4.0cm}
\caption{Wigner function for a quantum state which is a superposition 
of two Gaussian wave-packets separated in position. The interference 
fringes are alligned along the $p$ axes.}
\end{figure}

To evolve this initial state, we should solve the master equation 
(\ref{hpzexact}). Rather than doing this, one can use the explicit 
form of the evolution operator (\ref{jQBM}) and obtain the exact 
form of the reduced density matrix or the Wigner function at any time. 
We will adopt this strategy but will use the master equation 
(\ref{hpzexact}) and the equation for the Wigner function 
(\ref{qbmwignereq}) as a guide to interpret our results and to obtain 
simple estimates for the most important effects that take place as a
result of the interaction between the system and the environment. 
The exact evolution of the above initial state is such that the 
Wigner function can be written always as the sum of two Gaussian
peaks and an interference term,    
\begin{equation}
W(x,p,t) = W_1(x,p,t) + W_2(x,p,t) + W_{int}(x,p,t)\ , \label{wigtot}
\end{equation}
where
\begin{eqnarray}
W_{1,2}(x,p,t) &=& {{\bar{N}^2}\over \pi}{\delta_2\over\delta_1} 
\exp\left(-{(x\mp x_c)^2\over\delta_1^{~2}}\right)
\exp\left(-\delta_2^{~2}(p\mp p_c-\beta(x\mp x_c))^2\right), \nonumber\\
W_{int}(x,p,t) &=& 2 {{\bar{N}^2}\over \pi}{\delta_2\over\delta_1} 
\delta_2^{~2}(p-\beta x)^2)\nonumber\\
&\times& \cos{\left(2\kappa_p p + 2(\kappa_x-\beta \kappa_p) x\right)}\ . 
\label{wigint}
\end{eqnarray}
All the coefficients appearing in these expressions are somewhat 
complicated functions of time that are determined by the coefficients
that appear in the propagator (\ref{jQBM}) and the initial state 
(in the same way, they also depend on temperature and on the spectral
density of the environment). Their explicit form can be found in 
\cite{PHZ}. The initial state is such that 
$\delta_1^{~2}=\delta_2^{~2}=\delta^2$, $\kappa_x=P_0=p_c$, 
$\kappa_p=L_0=x_c$ and $A_{int}=0$.

       From the form of the exact solution, it is clear what the 
qualitative behavior of the quantum state is. The two Gaussian peaks 
follow the two 
classical trajectories (which get distorted by the interaction with the 
environment) and change their width along their evolution. On top of this,
the interference fringes change their wavelength and also rotate somewhat
following the rotation of the two wavepackets. The effect of decoherence
is clearly manifested in the damping of the interference fringes that, 
in the above formulae, is produced by the exponential term 
$\exp(-A_{int})$.
Thus, we will look carefully at this term, which can be seen to be the
``fringe visibility factor'' defined as 
\begin{equation}
\exp\left(-A_{int}\right)={1\over 2} 
{{W_{int}(x,p)|_{peak}}\over{\left(W_1(x,p)|_{peak}
W_2(x,p)|_{peak}\right)^{1/2}}}\ .\label{ppeak}
\end{equation}
A close analysis of the definition of $A_{int}$ shows that it 
vanishes initially and is always bounded from above, i.e.,  
\begin{equation}
A_{int}\leq {{L_0^{~2}}\over{\delta^2}}+\delta^2P_0^{~2}=
A_{int}|_{max}\ .\label{aintmax}
\end{equation}
The value of $A_{int}$ cannot grow to infinity as a consequence of
the fact that the two Gaussian initial states have a finite overlap
that is proportional to $\exp{(-A_{int}|_{max})}$. 

To understand qualitatively and quantitatively the time dependence
of the fringe visibility factor, it is interesting to 
obtain an evolution equation for $A_{int}$. Using its definition, 
we know that
\begin{equation}
\dot{A}_{int}={\dot{W}_{int}\over W_{int}}|_{peak} - {1\over
2}\left({\dot{W}_1\over W_1}+{\dot{W}_2\over W_2}\right)|_{peak}\ .
\label{deriva} 
\end{equation}
This, after using the form of the Wigner function together with the 
evolution equation, can be transformed into 
\begin{equation}
\dot A_{int} =4 D(t) 
\kappa_p^{~2} - 4 f(t) \kappa_p (\kappa_x -\beta\kappa_p)\ . 
\label{adot}
\end{equation}

This equation enables us to obtain a clear picture of the time
evolution of the fringe visibility function. Thus, we can see that the
first term on the right--hand side is always positive and corresponds
to the effect of normal diffusion. The normal diffusion will tend
to wash out interference. The initial rate at which $A_{int}$ grows 
is determined by the diffusion coefficient and by the initial wavelength
of the fringes in the momentum direction (remember that initially we 
have $\kappa_p=L_0/\hbar$. As time goes by, we see that the effect
of this term will be less important as the effective wavelength
of the fringes grows (making $\kappa_p$ decrease). 

Various simple estimates 
of the temporal behavior of the fringe visibility factor can be 
obtained from this equation. The most naive one is to neglect the 
time dependence of the diffusion coefficient and assume that the 
fringes always stay more or less frozen, as in the initial state. 
In such a case, we have
$A_{int}\approx 4L_0^2 D t/\hbar^2$. 
Thus, if we use the asymptotic 
expression of the diffusion coefficient, we obtain  
(at high temperatures) $A_{int}\approx \gamma t 4L_0^2/\lambda_{DB}^2$
where $\lambda_{DB}$ is the thermal de Broglie wavelength. Consequently,
we find that decoherence takes place at a rate 
\begin{equation}
t_{dec}=\gamma_0^{-1}(\lambda_{DB}/L_0)^2\ ,\label{tdec}
\end{equation}
which is the relaxation rate 
multiplied by a factor that could be very large
in the macroscopic domain (this is the result originally obtained by one of us,
see \cite{Zurek86} where it is shown that for 
typical macroscopic parameters, i.e., room temperature, $cm$--scale distances
and masses on the order of a gram, the factor $4L_0^2/\lambda_{DB}^2$ can be
as large as $10^{40}$). 

By analyzing the temporal behavior of $A_{int}$ obtained using the 
exact solution, we can check that this naive estimate is an excellent 
approximation in many important situations. However, it may fail in 
other important cases. Here, we want to stress a message that we 
believe is very important (see \cite{Decodeco}): It may be rather 
dangerous to draw conclusions that are too general 
from the theoretical analysis
of simple models of decoherence (like the one of linear QBM). The reason
is that simple estimates like the one corresponding to the decoherence
timescale (\ref{tdec}) are just that: simple estimates that apply to 
specific situations. They do not apply in other circumstances,  some 
of which we will describe here (and in the next section). For example,
the above simple estimate of the decoherence timescale fails in the simple 
case of ``ultrafast'' decoherence. For, in the high--temperature 
approximation of the master equation we neglected (among 
other things) initial transients occurring in the 
timescale fixed by the cutoff. Nothing (not even decoherence) can happen 
faster than the cutoff timescale since only after such timescale  
the diffusion coefficient reaches a sizable value. Thus, studying the 
initial time behavior of the normal diffusion coefficient one realizes
that for very short times, $A_{int}$ always grows quadratically (and not
linearly). In fact, we have
\begin{equation}\label{tdec'}
A_{int} \simeq {4M\gamma_0 k_BT L_0^2 \over\hbar^2}\Lambda t^2\;.
\nonumber\end{equation}
       From this expression one sees that in this case $A_{int}$ is 
smaller than the one obtained under the assumption of a constant diffusion 
coefficient (at least for times $t\le \Lambda^{-1}$). In this case, 
the decoherence timescale may be longer than the one corresponding 
to the high temperature approximation, 
\begin{equation}
t'_{dec} = {\hbar\over 2L_0\sqrt{M\gamma_0\Lambda k_BT}}\;.
\end{equation}

On the other hand, the above estimate for $A_{int}$ also 
fails to take into account the fact that $A_{int}$ does not grow 
forever because 
it finally saturates to the value fixed by equation (\ref{aintmax}). 
Saturation is achieved in a timescale that can be estimated to be
$t_{sat}\approx \gamma_0^{-1}(\hbar\Omega/k_BT)$. At approximately this
time the saturation of $A_{int}$ takes place (it is clear that this
is a very short time, much shorter than any dynamical timescale). 

The high--temperature approximation to the behavior of $A_{int}$ will 
clearly fail at very low temperatures (however, it is quite remarkable
how robust an approximation this is; see \cite{PHZ} for a detailed 
analysis). We will comment in the next section about the effects arising
at low temperatures giving more accurate estimates for $A_{int}$ in 
such a domain. 

\subsection{Predictability sieve and preferred states for QBM}

The most important consequence of the decoherence process is the 
dynamical selection of a set of stable, preferred states. These are, by 
definition, the least affected by the interaction with the environment
in the sense that they are the ones that become less entangled 
with it. To obtain these states, a systematic (``predictability
sieve'') criterion has been proposed \cite{Zurek93,ZHP}. 
The basic idea is the following: To find the pointer states, one
should consider all possible pure initial states for the system and 
compute the entropy associated with its reduced density matrix after some
time $t$. The pointer states are the ones that minimize the entropy 
production for a dynamic range of times. 

The predictability sieve can be applied to the simplest models  
of a quantum measurement, for which the Hamiltonian of the 
system can be completely neglected. In such a case, the pointer states
are directly associated with the eigenstates of the interaction 
Hamiltonian (actually, to its eigensubspaces that may be degenerate).
In other more realistic situations where the self--Hamiltonian of 
the system is not negligible the pointer states are not going to be
picked only by the interaction Hamiltonian  but by the interplay between
it and the evolution produced by the systems own Hamiltonian. The 
best example where we can explicitly compute these pointer states
is the QBM model we have been studying in this section. To do this, the
master equation is, as we will see, a very convenient tool. 

To find pointer states, we should minimize the entropy production 
at some time (varying over times to find a stable answer). However, 
to make our task simpler, instead of using von Neumann entropy, we will 
simply study the evolution of the purity of the system as 
measured by $\varsigma=Tr\rho^2$. This 
quantity is equal to one for a pure state and decreases when the state
of the system gets mixed because 
entanglement is generated by the evolution. 
The master equation directly enables us to write down an evolution equation 
for the purity $\varsigma$. Thus, using the definition of $\varsigma$ 
and the equation (\ref{hpz}) we obtain\cite{Zurek93}:
\begin{equation}
\dot\varsigma=2\gamma\varsigma
-4DTr(\rho^2x^2-\rho x\rho x) - 2f Tr(\rho^2(xp+px)-
2\rho x\rho p).\label{chidot}
\end{equation}
To simplify our treatment, 
we will once again use a perturbative approximation
and substitute in the right hand side of this equation the expression for 
the free Heisenberg operators: $x(t)=x\cos\Omega t+p/M\Omega\sin\Omega t$
and $p(t)=p\cos\Omega t-M\Omega x\sin\Omega t$. Moreover, we will average 
over one period of the harmonic motion, assuming that the coefficients
of the master equation do not vary during that time (clearly, this 
is a crude approximation, and we will comment later about 
what happens when we 
relax it). We also assume that the initial state is pure (and use the fact
that in that case $\rho^2=\rho$). Moreover, we neglect the effect of the 
friction term because, as we see, this term will always try to increase the 
purity in a way that is not sensitive to the state itself (thus, friction
always tries to localize the state competing against diffusion that has
the opposite effect). Doing this, we find out that the change in purity 
over one period is simply given by
\begin{equation}
\varsigma(T)-
\varsigma(0)=-2D (\Delta x^2+\Delta p^2/M^2\Omega^2)\ .\label{deltachi}
\end{equation}
where $\Delta x$ and $\Delta p$ are respectively the position and 
momentum dispersion of the initial state. 
The anomalous diffusion term does not produce
any net entropy increase (or purity decrease) because its effect averages
out over one oscillation. The term responsible for purity decrease is 
simply coming from diffusion, and to minimize it, we should vary over
all possible initial states. This can easily be implemented by varying
over all values of the initial dispersion in position and momentum 
in such a way that the right--hand side of (\ref{deltachi}) is minimized.
Because $\Delta x\Delta p\ge\hbar/2$ must always be satisfied, it is clear
that the minimum is obtained when the state saturates  uncertainty
relations. From the resulting equation we obtain  
the pointer states as having $\Delta x^2=\hbar/2M\Omega$ and 
$\Delta p^2=\hbar M\Omega/2$. Therefore, the pointer states are simply 
given by coherent states with minimum uncertainty. This result is simple
and satisfying. In fact, coherent states are the closest we can get 
to points in phase space. They are preferred states in QBM because they 
turn out to be the most robust and most effectively resist the combined 
effect of
the system and the environment. They are also 
well localized in position and, therefore, 
are not significantly perturbed by the environment monitoring their position.
Moreover, because of their symmetry, they are also not drastically altered by 
the evolution induced by the Hamiltonian of the system. 

\subsection{Energy eigenstates can also be selected by the environment!}

So far, we have discussed two regimes in which the predictability sieve 
can be successfully applied. We first mentioned the case of a 
measurement (where the Hamiltonian of the system is negligible), and we 
just studied the case where both the system and the environment induce 
nontrivial evolution. There is a third regime that is interesting 
to study and is one in which the evolution of the environment is 
very slow as compared with the dynamic timescales of the system. 
If this is the case, it is possible to show \cite{energy} 
that the preferred states are simply the eigenstates of the Hamiltonian
of the system. However, it is interesting to note that to find out this 
result, it is not possible to use a model like the linear QBM we described
before. In fact, in such a model we can see that if we consider a very 
slow environment (with frequencies much smaller than the one belonging to 
the system) the master equation (\ref{hpz}), which is still applicable,  
has time--dependent coefficients that are oscillatory
functions of time with no well--defined sign. Therefore, the predictability
sieve criterion does not give a robust set of states in this case. 

However, the third regime of einselection can be examined  
using a simple argument based on an adiabatic solution of the full 
Schr\"odinger equation. The main ingredient we need is, as will be shown
below, a slow environment that couples to the system through an interaction
Hamiltonian that has a nonzero expectation value in the energy eigenstates
of the system. To see this, we will solve the full Schr\"odinger equation
treating the environment adiabatically. 
Suppose that the initial state of the universe given
as $|\Psi(0)\rangle=\sum_nc_n|\phi_n\rangle|\epsilon_0\rangle$ where the 
states $|\phi_n\rangle$ are nondegenerate
eigenstates of the Hamiltonian of the system
(with distinct
energies $E_n$), and $|\epsilon_0\rangle$ is a state of the environment
that, for simplicity, we will consider as a coherent state (the vacuum,
for example). We 
can solve the full Schr\"odinger equation in the adiabatic approximation
and show that this state evolves into $|\Psi(t)\rangle=\sum_nc_n
\exp(-iE_nt/\hbar)|\phi_n\rangle |\epsilon_n(t)\rangle$ where the 
state $|\epsilon_n(t)\rangle$, that gets correlated with the $n$th energy 
eigenstate of the system, obeys the following Schr\"odinger equation
\begin{equation}
i\hbar{d\over dt}
|\epsilon_n\rangle=\langle\phi_n|H_{int}|\phi_n\rangle |\epsilon_n
\rangle.\nonumber
\end{equation}
Note that in this equation the operator $\langle\phi_n|H_{int}|\phi_n\rangle$
acts on the Hilbert space of the environment and depends parametrically
on the energy eigenstates of the system. We will assume that 
the interaction is such that the Hamiltonian is of the form 
$H_{int}=S\otimes \Pi_{\cal E}$, where the operator $S$ acts on the 
system Hilbert 
space, and the environment operator $\Pi_{\cal E}$ acts 
on the environment as a translation generator (it could be 
the momentum operator, for example, but from our discussion it will be clear
that the choice of momentum here is not crucial). 

The  decoherence in energy eigenbasis can easily be established as 
follows. Because $\Pi_{\cal E}$ is a momentum operator and the initial 
state of the 
environment is a coherent state, the evolution turns out to be simply such 
that $|\epsilon_n(t)\rangle=|\epsilon_0+S_{nn} t\rangle $, where 
$S_{nn}=\langle \phi_n|S|\phi_n\rangle $. 
Therefore, the overlap between the two states that  
correlate with different energy eigenstates can be estimated 
as $\langle\epsilon_n(t)|\epsilon_m(t)\rangle\approx 
\exp(-t^2(S_{nn}-S_{mm})^2\hbar^2)$. Consequently, in this case,  
we see einselection of energy eigenstates (superpositions of energy 
eigenstates are degraded while pure energy eigenstates are not affected). 
 For this reason, pointer states are energy eigenstates. This result has
a rather natural interpretation. It just tells us that the environment is 
not able to react before the system has time to evolve and therefore only 
probes time--averaged quantities of the system. Energy, being the only 
observable that does not average out to zero is therefore the preferred
observable. The conditions for energy eigenstates to become 
the pointer basis are  
the following: the environment must behave adiabatically (and be slow
as compared with the dynamics of the system), and the interaction with
the system must be through an observable with a nonvanishing expectation
value in energy eigenstates. 

\section{Deconstructing decoherence: landscape beyond the standard 
models}

Simple models of decoherence, like the one we discussed so far (linear
quantum Brownian motion) are important to illustrate the simplicity 
and high efficiency of the decoherence process (two characteristics
that may be interpreted as indicating its generality). However, it is 
important to keep 
in mind that no generic conclusions should be drawn from simple estimates. 
This is especially important in view of the possibility of 
carrying out 
experiments to test decoherence in a controlled manner. In such cases, it
is essential to study specific models of the decoherence process in 
the correct context. Estimates of the decoherence timescale, nature 
of pointer 
states, and other characteristics of decoherence obtained in models like 
QBM should be taken as indications rather than as strong predictions. 

In this section, we would
like to stress the fact that some of the simple features 
that have became identified as ``standard lore'' 
in the decoherence process for the simplest case of 
linear QBM are not generic by showing explicitly how they
fail in two specific examples. We will address basically two 
issues. First we will consider the status of one of the simplest predictions 
arising from studying decoherence in linear QBM: the ``decoherence rate 
grows quadratically with distance''. We will show that this is 
not the case for more realistic models where local interactions between 
particles and fields (rather than oscillators) are taken into account. 
Second, we will consider the status of predictions of the decoherence
timescale like the ones in Eq. (\ref{tdec}) at low temperature. In this case,
by analyzing the same linear QBM at low temperatures, we will show that
the decoherence process may be more complicated, allowing even for 
nonmonotonic behavior. 

\subsection{Saturation of the decoherence rate at large distances}

One of the results obtained studying the decoherence process in linear
QBM models is that the decoherence rate grows quadratically with the
separation between different pieces of the system wavefunction. 
This result is natural (delocalized wavepackets decohere faster) but
would certainly not be physical if it held for arbitrarily large separations.
Apart from any arguments involving cutoff (see discussion following 
Eq. (\ref{tdec}), it is clear that 
the environment should have a coherence length
so that separations that are bigger than this natural lengthscale should
be equivalent and therefore induce the saturation of the decoherence
rate. 

However, saturation is not present in the linear QBM model, as  is 
clear from the discussion above. One therefore
asks what kinds of models predict saturation. We will describe here the 
simplest of such models. The environment is formed
by a quantum scalar field; the system is a quantum particle, and the 
interaction between them is local. This is the model whose 
perturbative master equation we derived in section 3.4. It is 
important to stress once more that the linear QBM model is obtained  
from the particle--field model by means of the dipole approximation.  
Thus, saturation in this context arises only if we do not make  
the dipole approximation (which is certainly not well justified for
large separations). The issue of 
the saturation of the 
decoherence rate was analyzed first in \cite{FlemingGallis}
and also discussed in \cite{Decodeco}. In this review we present a 
simpler discussion than the one of \cite{Decodeco} that captures the 
main ingredients necessary for saturation and enables us to obtain the
principal results without complicated calculations (some experimental 
results related to these issues were reported in \cite{Raymer}). 

As we discussed in 3.4, the reduced density matrix of the particle 
obeys the perturbative master equation (\ref{masterfield}). In this
equation, the Heisenberg operator of the particle $x(t_1)$ appears. 
To simplify our argument, we will consider the system that is a free and 
very massive particle and therefore replace $x(t)=x(0)$ in 
(\ref{masterfield})
(corrections to this approximation can be computed also). In the 
simplest example, we will consider as environment a 
massless scalar field (and replace the corresponding expressions for the
Fourier transform of the two point functions; see Eq. (\ref{moth1})). 
In this case,   
we can express the master equation in the position representation as
\begin{equation}
\dot\rho(x,x\prime)=-\Gamma(x-x\prime)\rho(x,x\prime) +\ldots
\ ,\label{mastersat}
\end{equation}
where only the term producing decoherence has been written out, and 
the function $\Gamma(x,x\prime)$ is defined as
\begin{equation}
\Gamma(x,x\prime)=-8\pi{e^2\over\hbar^2}\int_0^\infty dk W(k)\sin kt 
\coth(\beta 
k/2)(1-{\sin kr\over kr})\ ,\label{gammaxx}
\end{equation}
where $r=|x-x\prime|$ (and, as before, $W(k)$ is the Fourier transform
of the window function that introduces a natural high--frequency cutoff). 
It is simplest to analyze the high--frequency limit of the above
expression. In that case, the integral can be exactly computed and
turns out to be
\begin{eqnarray}
\Gamma(r)&=&8\pi^2{e^2\over\beta\hbar^2}({\sinh \Lambda r\over\Lambda r}-1)
\exp(-\Lambda t)\qquad {\rm if\ }r\le t\nonumber\\
&=&8\pi^2{e^2\over\beta\hbar^2}(1-{t\over r}-\exp(-\Lambda t)+
{\sinh \Lambda t\over\Lambda r}\exp(-\Lambda r)), {\rm if\ } r\ge t
\label{gammar}
\end{eqnarray}
   From this expression we clearly see the saturation. Thus, the solution 
of the master equation in the ``decoherence--dominated'' approximation 
(neglecting all terms except  the one producing decoherence) is 
simply $\rho(x,x\prime)\approx\exp(-\int_0^t dt_1 
\Gamma(x-x\prime, t_1))\rho(x,x\prime)$. 
The dependence of $\Gamma(r)$ for long distances is given by the 
second instance in equation (\ref{gammar}) that approaches a constant 
as $r$ grows larger than $1/\Lambda$ and $t$. On the other hand, the 
quadratic dependence of the decoherence rate is recovered for small 
distances: by 
expanding the function $\Gamma(r)$ around $r\approx 0$,  we obtain 
a quadratic behavior.

\subsection{Decoherence at zero temperature}

A simple estimate
for the low temperature behavior of the fringe visibility function 
can be obtained as follows. Use the asymptotic form
of the diffusion coefficient for low temperatures given by perturbation
theory and integrate the equation for $A_{int}$, neglecting both its time 
dependence as well as the temporal evolution of the wavelength of the 
fringes. In this way, we obtain 
$A_{int}\approx \gamma_0t (4L_0^2/\Delta x^2)\coth\beta\Omega/2$. 
However, this is not always a good approximation. 
On the one hand, if this behavior were correct, we could  
estimate the saturation time (the time for which $A_{int}$ would approach
its maximum value) to be on the order of $t_{sat}\approx \gamma_0^{-1}
\tanh\beta\Omega/2$, which for very low temperatures can 
be very close to, or 
even larger than, a dynamical timescale. Note that this does not 
imply that decoherence occurs in a dynamical timescale: For that, the 
important fact is the actual value of $A_{int}$ and not how close to the
maximum value we are. The decoherence time-scale at low temperatures 
is on the order of $t_{dec}\approx \gamma_0^{-1}(\Delta x/2L_0)^2$, which
is still much shorter than $\gamma_0^{-1}$ for macroscopic parameters). 
The fact that the naive estimate for the saturation timescale becomes 
larger than typical dynamical times means that $A_{int}$ does not have
a monotonic behavior in time. In fact, it turns out that at very low 
temperatures, the role of the anomalous diffusion term in the master 
equation starts to be relevant (its value is of the same order of 
magnitude as the normal diffusion coefficient). The contribution of 
this term to the evolution of $A_{int}$ is clearly seen in equation 
(\ref{adot}) where we see that the second term (associated with 
anomalous diffusion) does not have 
a well--defined sign (its sign changes as the interference
fringes rotate in phase space). From this observation, one expects that if
at low temperatures the fringe visibility factor does not saturate, its
time dependence should exhibit some oscillatory behavior (modulating an 
overall increase dictated by normal diffusion). The periods of 
slower decoherence coincide with the moments when fringes get oriented 
along the position axis 
(this coincides with the instant when the two wavepackets are
most 
separated in momentum). This qualitative prediction concerning the behavior
of $A_{int}$ is confirmed by the 
exact numerical calculations shown in Figure
5. In this figure, the oscillations are clearly seen. 

A very simple and interesting expression for $A_{int}$ can be obtained
for the QBM model. Thus, in \cite{Romero} it has been shown that the 
fringe visibility factor can always be written as follows:
\begin{equation}
A_{int}={1\over 2}\left({2L_0\over \Delta x}\right)^2\coth^2(\beta\Omega/2)
\left(1-\left(\ddot S^2/\Omega^4+ \dot S^2/\Omega^2\right)\right)\ ,
\label{aintnice}
\end{equation}
where $\Delta x^2$ is the position dispersion in thermal equilibrium
(i.e., $\Delta x^2=\hbar\coth(\beta\Omega/2)/M\Omega$) 
and $S$ is the normalized position autocorrelation function defined as
\begin{equation}
\Delta x^2 S(t)={1\over 2}\langle\{x(t),x\}\rangle -
\langle x(t)\rangle\langle x\rangle\ .\nonumber
\end{equation}

This equation enables us to obtain very simple qualitative estimates 
of the efficiency of decoherence. More interestingly, it clearly shows
that decoherence has the same physical origin as other dissipative
effects (and is closely related to the decay of the autocorrelation 
function through Eq. (\ref{aintnice})). However, in spite of their common 
origin, the decay of correlations and the decoherence process  
have very different timescales. In fact, from the above equation we 
can estimate how much the correlation functions 
have to decay in order for the 
system to decohere. Thus, at the time for which $A_{int}$ approaches
unity, the spatial correlations in the system should have decayed 
by a factor $S(t_{dec})/S(0)=\sqrt{1-\Delta x^2/4L_0^2}$, which is 
indeed very small (note that $\Delta x$ approaches the thermal de Broglie 
wavelength at high temperatures and the spread of the ground state at 
zero temperature). 

On the other hand, the above formula (\ref{aintnice}) 
can also allow us to estimate
correctly $A_{int}$ both at high and low temperatures for the 
underdamped Brownian motion model. In fact, we just need to 
obtain a reasonable approximation for the position correlation function. 
For example, assuming a simple exponential decay would lead us to 
conclude that 
$$A_{int}={1\over 2} {4 L_0^2\over \Delta x^2} \coth^2(\beta\Omega/2)(1-
\exp(-\gamma_0 t)).$$
This is a crude but very reasonable approximation that is, for example, 
not only very good at high temperatures and very early times but also 
exhibits the correct saturation behavior for long times. It can be 
further improved by better
approximating the position correlation function. For example, 
computing $S(t)$ in the highly underdamped regime we obtain
\begin{eqnarray}
A_{int}&=&{1\over 2} {4 L_0^2\over \Delta x^2} \coth^2(\beta\Omega/2)
\nonumber\\
&\times&\left(1-\exp(-\gamma_0 t)
\left(1+\gamma_0^2\sin^2\Omega t/2\Omega_0^2-\gamma_0
\sin 2\Omega_0t/2\Omega_0\right)\right),
\end{eqnarray}
which is a very good approximation for the low--temperature (low--damping) 
behavior exhibited in Figure 5. 

\begin{figure}
\qquad\qquad\qquad\qquad
	\includegraphics[height=0.7\hsize]{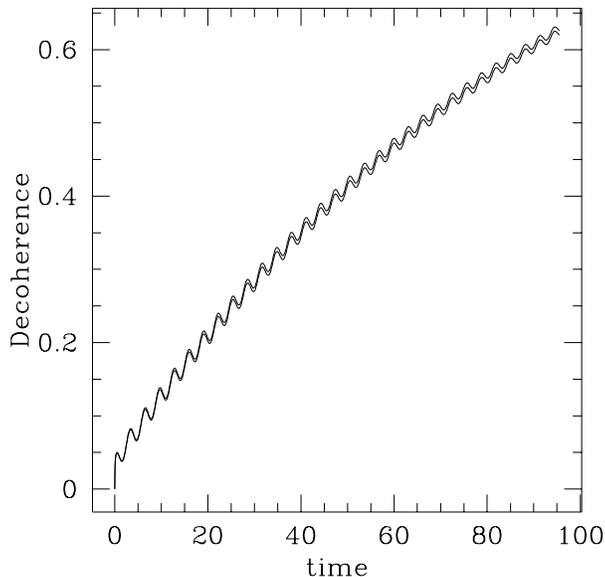}
\caption{Decoherence at zero temperature proceeds nonmonotonically. Here, 
the time dependence of $A_{int}$ for a harmonic oscillator interacting
with a zero temperature environment is displayed. Oscillations correspond to
the change in orientation of the interference fringes.}
\end{figure}

\subsection{Preexisting correlations between the system and 
the environment.}

Almost all papers concerning decoherence assume that the initial 
state has no correlations between the system and the environment (i.e., 
that the state can be factored). In this section we will analyze what 
happens if we consider more general initial conditions. In particular, 
we are interested in analyzing initial conditions that are closer 
to what we encounter experimentally. Thus, we consider a situation 
in which the system and the environment are initially in a thermal 
 equilibrium state at some temperature (which could be zero) and  
at the initial time we perform a measurement on the system to prepare 
an initial state. This measurement could be imperfect (i.e., may be 
characterized not by a projection operator, but by a POVM). 
After this measurement, we consider the evolution of the system 
coupled to the environment in the usual way. Under these
circumstances, the initial state of the universe is generally not a 
product. Moreover, in the case when the initial state is a product
(i.e., when the measurement performed on the system is perfect), 
the state of the environment depends functionally on the state of the 
system. This type of initial states can generally be written as
\begin{equation}
\rho_o = \sum_j A_j \rho_ \beta A'_j\ , \label{OO'}
\end{equation}
where $A_j$ and $A'_j$ are Krauss operators (not necessarily projectors) 
acting on the Hilbert space of the system (see \cite{Schumacher96} 
for a 
good review).

We are not going to present any details of the calculations leading 
to the (exact) solution of this model. Our presentation follows 
closely the one in \cite{Romero} where the influence of initial 
correlations on decoherence was examined. Here we present a summary
of the results obtained in that paper. 

First, it is worth stressing the fact that it is still possible
to find a relatively simple 
master equation for the reduced density matrix of the 
system. However, the existence of initial correlations prevents us
from expressing this equation entirely in terms of the reduced 
density matrix. Thus, the evolution of $\rho$ not only depends on 
$\rho$ itself but also on initial correlations between the system 
and the environment. Interestingly 
enough, for the case of the linear QBM model, an exact master 
equation that is very similar to (\ref{hpz}) can be obtained. It reads
as follows:
\begin{eqnarray}
\dot \rho(q, q', t) &=& i \left( 
{1\over 2} \left(\partial^2_q- \partial^2_{q'}\right) - 
{1\over 2} \Omega^2(t)(q^2 - {q'}^2) \right)\rho(q, q', t) \nonumber \\
&-& \gamma (t) \ (q - q') \ \left(\partial_q - \partial_{q'} \right)
\rho(q, q', t) 
\nonumber \\
&-& D_1 (t)\  (q - q')^2 \ \rho(q, q', t)\nonumber\\
&-& i D_2 (t) \ (q - q') \ 
\left(\partial_q + \partial_{q'}\right) \rho(q, q', t) \nonumber \\
&+& i {\tilde C}_1 (t) \ (q -q') \ \rho_{11} (q, q', t) \nonumber\\
&-& i {\tilde C}_2(t) \ (q - q') \ \rho_{12} (q, q', t)\ . \label{mastereq}
\end{eqnarray}
It is important to stress that this equation is exact and 
valid for all spectral densities and initial temperatures. The 
time--dependent coefficients appearing in (\ref{mastereq}) are functions 
of time and temperature (and of the spectral density of the 
environment, of course). 
Explicit formulae are given in \cite{Romero}. The interpretation of the
first three lines of this equation is identical to the ordinary case
where no correlations are present. The initial correlations appear in the
time dependence of the coefficients but, for realistic environments, this 
dependence is very weak (thus, these coefficients are qualitatively the 
same as before). 
The last two lines make this equation nonhomogeneous. In fact, these
terms are present because of the correlated nature of the initial state. 
Thus, in that case, the master equation cannot be entirely 
written in terms of the reduced density matrix. 
It can be shown that the two density matrices $\rho_{11}$ 
and $\rho_{12}$ are 
obtained by propagating two different initial states
given by the ``density matrices" 
$\rho_{11}=\{q,\rho\}$ and $\rho_{12}=i[q,\rho]$. The evolution of 
$\rho_{1i}$ can also be studied with this formalism 
because (apart from not being normalized) they belong to the class of 
initial conditions defined by (\ref{OO'}). 
Therefore, the evolution equation obeyed by these operators is 
also (\ref{mastereq}), with new inhomogeneous terms. Thus, a 
hierarchy of equations, which are coupled because of the initial 
correlations,
can be derived in this way (see \cite{Romero} for more details).

The time dependence of all the coefficients has been studied in detail 
in \cite{Romero} and the conclusion is that, for an ohmic environment at 
arbitrary temperatures, the coefficients $\tilde C_1$ and $\tilde C_2$, 
entering in the inhomogeneous terms of the master equation are exceedingly 
small and become negligible after a time that is on the order of the 
cutoff timescale. After this short initial transient, the impact of the
initial correlations on the future evolution of the system can be 
entirely neglected. Of course, in less--realistic situations, it is 
possible to show that these coefficients have an important effect. For
example, the formalism we described could be applied to the case of 
two coupled oscillators in which one considers one of them 
as the system and the other one as the environment. 
In this case, when the size of the system and 
environment are comparable, initial correlations play an important role. 
The time dependence of the other coefficients of the master equation
is also affected by the correlations but they all behave qualitatively
in a similar way as in the absence of such initial correlations 
(see \cite{Romero} for a detailed study of these coefficients). 
%The exact form of this coefficients is displayed in Figure 3 for the ohmic
%environment at zero temperature. 

It is interesting to analyze the evolution of a delocalized initial state
to see how decoherence takes place in this model, which includes the 
effect of initial correlations. For this, we consider the initial 
condition  (\ref{OO'} with the operators associated to a projection 
onto a Schr\"odinger cat state (say, a superposition of two coherent
states separated in position). Thus, we take
\begin{equation}
\rho =  {{\hat P \rho_{\beta} \hat P}\over{Tr(\rho_\beta \hat P)}}\ ,
\label{rhoin2}
\nonumber
\end{equation}
where $\hat P$ is a projector onto a pure state of the system 
$\hat P=|\Psi\rangle\langle\Psi|$ and the state $|\Psi\rangle$ is itself a 
Schr\"odinger's cat state (i.e., a superposition of two Gaussian packets), 
\begin{equation}
|\Psi\rangle = |\Psi_+\rangle + |\Psi_-\rangle  \label{psi}\ ,
\end{equation}
where $|\Psi_\pm\rangle$ are such that
\begin{equation}
\langle x|\Psi _\pm\rangle\, = N \ \exp \biggl[ - {(q \mp L_o)^2 \over {2 \delta^2}} 
\pm i P_o q \biggr]\ .  \label{psicoord}
\end{equation}

The decoherence process for this initial state has been analyzed 
in the previous section in the absence of initial correlations. The fate
of this state is not very different from the behavior we described 
before but there are some subtle differences. Thus, initial correlations 
distort the Gaussian peaks in the initial Wigner function as well as
the intermediate interference fringes. An exact solution of the 
problem is possible (see \cite{Romero}) and it turns out that it is no 
longer true that the Wigner function can be written as the sum of two
Gaussian peaks plus interference fringes. In fact, it turns out that
each Gaussian peak is distorted in such a way that it can be written
as the sum of two nearby Gaussians with a term between them. The same
is true for the interference fringes, which get distorted and split into
several (actually ten) terms. However, for realistic (ohmic) environments, 
this effect is very small (as discussed in \cite{Romero}), and  
the decoherence process goes qualitatively in the same way as described
in the previous section (in fact, in Figure 5, the two curves for the 
decoherence factor are almost indistinguishable from each other: one 
corresponds to an initially uncorrelated state while the other to the 
case described in this section).

\section{Decoherence and chaos}

Here we investigate environment induced superselection in the context of 
quantum chaos (i.e., quantum dynamics of systems that are classically 
chaotic). We first argue \cite{ZP94} 
that the evolution of a chaotic macroscopic (but, ultimately, quantum)
system is not just difficult to predict (requiring accuracy exponentially 
increasing with time) but quickly ceases to be deterministic in principle as  
a result of the Heisenberg indeterminacy (which limits the resolution 
available 
in the initial conditions). This happens after a time $t_{\hbar}$, which 
is only 
logarithmic in the Planck constant. A definitely macroscopic (if somewhat
outrageous) example \cite{ZP95} 
is afforded by various components of the solar system 
that are chaotic, with the Lyapunov timescales ranging from a bit 
more then a month 
(Hyperion, a prolate moon of Saturn\cite{Wisdom}) 
to millions of years (planetary system 
as a whole \cite{Laskar,Sussman}). On the timescale 
$t_{\hbar}$ the initial minimum uncertainty wavepackets corresponding 
to celestial bodies would be smeared over distances of the order of 
the radii of 
their orbits into ``Schr\"odinger cat--like'' states, and the concept of  
a trajectory would cease to apply. In reality, such paradoxical states are 
eliminated by decoherence that helps restore quantum-classical 
correspondence. We shall also see that 
the price for the recovery of classicality is the loss of predictability.  
In the classical limit (associated with effective decoherence, and not just
with the smallness of $\hbar$) the rate of increase of the von Neumann
entropy of the decohering system is independent of the strength of the 
coupling to the environment and equal to the sum of the positive 
Lyapunov exponents. 

\subsection{Quantum predictability horizon: How the correspondence is lost.}

As a result of chaotic evolution, a patch in the phase space that  
corresponds to some regular (and classically ``reasonable'') initial condition 
becomes drastically deformed. Classical chaotic dynamics is characterized by 
the exponential divergence of trajectories. Moreover, conservation of the 
volume
in the phase space in the course of Hamiltonian evolution (which is initially
a good approximation for sufficiently regular initial conditions, even in
cases that are ultimately quantum) implies that 
the exponential divergence in some of the directions must be balanced 
by the exponential squeezing ---convergence of trajectories--- in  other 
directions. It is that squeezing that forces a chaotic system to explore the
quantum regime. As the wavepacket becomes narrow in the direction 
corresponding to momentum,
\begin{equation} 
\Delta p (t) \ = \ \Delta p_0 ~ \exp (- \lambda t)\ ,  
\label{(2.1)}
\end{equation}
(where $\Delta p_0$ is its initial extent in momentum, and $\lambda$ is the 
relevant Lyapunov exponent) the position becomes delocalized: The wavepacket 
becomes coherent over the distance $\ell(t)$ that can be inferred from 
Heisenberg's principle,
\begin{equation} 
\ell (t) \geq (\hbar / \Delta p_0) \exp (\lambda t) \ . \label{(2.2)}
\end{equation}
Coherent spreading of the wavepacket over large domains of space is disturbing
in its own right. Moreover, it may lead to a breakdown of the correspondence 
principle at an even more serious level. Predictions of the classical 
and quantum dynamics concerning some of the expectation values no longer
coincide after a time $t_{\hbar}$ when the wave-packet coherence length 
$\ell (t)$ reaches the scale on which the potential is nonlinear.

Such a scale $\chi$ can usually be defined by comparing the classical force 
(given by the gradient of the potential $\partial_x V$) 
with the leading order nonlinear contribution $\sim \partial^3_x V$,
\begin{equation}\chi \ \simeq \ \sqrt {{\partial_x V} 
\over {\partial^3_x V}} \ . \label{(2.3)}
\end{equation}
For instance, for the gravitational potential $\chi \simeq R / \sqrt{2}$, 
where $R$ is 
a size of the system (i.e., a size of the orbit of the planet). The reason
for the breakdown of the correspondence is that when the coherence length
of the wavepacket reaches the scale of nonlinearity, 
\begin{equation} \ell (t) \ \simeq \chi \ , \label{(2.4)}\end{equation}
the effect of the potential energy on the motion can no longer be 
represented by the classical expression for the force \cite{ZP94}, 
$F(x) = \partial_x V(x)$, because 
it is not even clear where the gradient is to be evaluated for a delocalized 
wave-packet. As a consequence, after a time given by 
\begin{equation} 
t_{\hbar} \ = \ \lambda^{-1} \ln {{\Delta p_0 \chi} \over \hbar} \ ,
\label{(2.5)}
\end{equation}
the expectation value of some of the observables of the system may even begin 
to exhibit noticeable deviations from 
the classical evolution \cite{HSZ98}. 

This is also close to the time beyond which the combination of classical 
chaos and Heisenberg's indeterminacy makes it impossible {\it in principle} 
to employ the concept of a trajectory. Over the time $\sim t_{\hbar}$ a 
chaotic 
system will spread from a regular Planck-sized volume in the phase space into 
a (possibly quite complicated) wavepacket with the dimensions of its envelope
comparable to the range of the system. This timescale defines the quantum 
predictability horizon ---a time beyond which the combination of classical
chaos and quantum indeterminacy makes predictions not just exponentially 
difficult, but impossible in principle. The shift of the origin of the loss 
of predictability from classical deterministic chaos to quantum 
indeterminacy   
amplified by exponential instabilities is just one of the symptoms of the
inability of classical evolution to track the underlying quantum dynamics.

This breakdown of correspondence can be investigated more rigorously by 
following the evolution of the Wigner function (defined in (\ref{wigdef}))
for the possibly macroscopic, yet ultimately quantum system. 
Dynamics of the Wigner function is generated by the {\it Moyal bracket} 
(that is simply the Wigner transform of the right--hand side of von Neumann 
equation for the density matrix). This Moyal bracket can be expressed 
through the familiar classical Poisson bracket:
\begin{equation} 
\dot W\ 
= \{H,W\}_{MB} \ = \ -i \sin(i \hbar \{H,W\}_{PB})/\hbar \ . \label{(2.6)}
\end{equation}
Above, $H$ is the Hamiltonian of the system, and $W$ is the 
Wigner transform of the density matrix. 

When the potential $V$ in $H$ is analytic, the Moyal bracket can be expanded 
in powers of the Planck constant. Consequently, the evolution of the 
Wigner function is given by
\begin{equation} \dot W \ = \ \{H,W\}_{PB} ~ + \sum_{n \geq 1}
{{\hbar (-)^n \over{2^{2n} (2n+1)!}}} 
\partial_x^{2n+1} V (x) \partial_p^{2n+1} W (x,p). 
\label{(2.7)}
\end{equation}
Correction terms above will be negligible when $W(x,p)$ is a reasonably
smooth function of $p$, that is, when the higher derivatives of $W$ with
respect to momentum are small. However, the Poisson bracket alone predicts 
that, in the chaotic system, they will increase exponentially quickly 
as a result of 
the ``squeezing'' of $W$ in momentum, Eq. (\ref{(2.1)}). Hence, after 
$t_{\hbar}$, 
quantum ``corrections'' will become comparable to the first classical term on 
the right--hand side of Eq. (\ref{(2.7)}). At that point, 
the Poisson bracket will no 
longer suffice as an approximate generator of evolution. The phase space 
distribution will be coherently extended over macroscopic distances, and 
interference between the fragments of $W$ will play a crucial role.

The timescale on which the quantum--classical correspondence is lost in 
a chaotic system can also be estimated (or rather, bounded from above) by the
formula \cite{Chirikov,Berman} 
\begin{equation} t_r \ 
= \ \lambda^{-1} \ln (I / \hbar) \ , \label{(2.8)}\end{equation}
where $I$ is the action.

\subsection{Exponential instability vs. decoherence}

In a quantum chaotic system weakly coupled to the environment, the process of 
decoherence briefly sketched above will compete with the tendency for coherent
delocalization, which occurs on the characteristic timescale given by the 
Lyapunov exponent $\lambda$. Exponential instability would spread the 
wave-packet
to the ``paradoxical'' size, but monitoring by the environment will attempt 
to limit its coherent extent by smoothing out interference fringes.
The two processes shall reach {\it status quo} when their rates are 
comparable, 
\begin{equation} \tau_D (\delta x)~ \lambda \ \simeq \ 1 . \label{(5.1)}
\end{equation}
Because the decoherence rate 
depends on $\delta x$, this equation can be solved 
for the critical, steady state coherence length, which yields 
$\ell_c \sim \Lambda_{dB}(T) \times \sqrt{\lambda/\gamma}$. 

A more careful analysis can be based on the combination of the Moyal bracket 
and the master equation approach to decoherence we have just sketched. 
In many cases, (including the situation of large bodies immersed in the typical
environment of photons, rarefied gases, etc.) an effective approximate 
equation 
can be derived and translated into the phase space by performing a Wigner 
transform of the master equation. Then:
\begin{eqnarray} \dot W \ = \ \{H,W\}_{PB} &+&
\sum_{n\geq 1}{{\hbar^{2n} (-1)^n} \over {2^{2n} (2n+1)!}} 
\partial_x^{2n+1} V (x) \partial_p^{2n+1} W (x,p)\nonumber\\
&+&2\gamma\partial_p pW+D \partial_p^2 W\ . 
\label{(5.2)}
\end{eqnarray}
As before, we are interested in the regime where we can neglet the 
term that causes relaxation, which, in the macroscopic limit, 
can be made very small without decreasing the effect of decoherence 
caused by the last, diffusive term. As we saw in the previous Section, 
the role of this decoherence term is to destroy the quantum coherence 
of the fragments of the wavefunction between spatially separated regions. 
Thus, in effect, this {\it decoherence term} can esure that the Poisson 
bracket is always reasonably accurate. Diffusion prevents the wavepacket 
from becoming too finely structured in momentum, which would have caused 
the failure of the correspondence principle. In the case of the thermal 
environment, the diffusion coefficient $D= \eta k_B T$, where $\eta$ 
is viscosity. The competition between the squeezing resulting from 
the chaotic instability and spreading resulting from diffusion leads to 
a standoff when the Wigner function becomes coherently spread over
\begin{equation} 
\ell_c \ = \ \hbar \sqrt {\lambda \over {2D}} \ = \ \Lambda_{dB}(T) 
\ \sqrt{\lambda / 2 \gamma} \ . \label{(5.3)}\end{equation}
This translates into the critical (spatial) momentum scale of 
\begin{equation} 
\sigma_c \ = \ \sqrt{{2 D} \over \lambda}\ , \label{(5.4)}\end{equation}
which nearly coincides with the quick estimate given by equation 
(\ref{(5.1)}).

Returning to an outrageous example of the solar system, for a planet of the 
size of Jupiter a 
chaotic instability on the four--million--year timescale and the consequent 
delocalization would be easily halted even by a very rarefied medium 
($0.1$ atoms/cm$^3$, comparable to
the density of interplanetary gas in the vicinity of massive outer planets) 
at a temperature of $100$K (comparable to the surface temperature of 
major planets): The resulting
$\ell_c$ is on the order of $10^{-29}$~cm! Thus, decoherence is exceedingly 
effective in preventing the packet from spreading; $\ell_c << \chi$, by 
an enormous margin. Hence, the paradox we have described in the first part of
the paper has no chance of materializing.

The example of quantum chaos in the solar system is a dramatic illustration of
the effectiveness of decoherence, but its consequences are, obviously, not 
restricted to celestial bodies: Schr\"odinger cats, Wigners friends, and, 
generally, all of the systems that are in principle quantum but sufficiently 
macroscopic will be forced to behave in accordance 
with classical mechanics as a 
result of the environment--induced superselection \cite{Zurek81,Zurek82}. 
This will be the case whenever
\begin{equation} \ell_c \ll \chi \ , \label{(5.5)}\end{equation}
because $\ell_c$ is a measure of the resolution 
of ``measurements'' carried out by the environment. 

This incredible efficiency of the environment in monitoring (and, therefore,
localizing) states of quantum objects is actually not all that surprising.
We know (through direct experience) that photons are capable of maintaining 
an excellent record of the location of Jupiter (or any other 
macroscopic body). 
This must be the case, because we obtain our visual information about 
the universe by intercepting a minute fraction of the reflected (or emitted) 
radiation with our eyes. 

Our discussion extends and complements developments that go back more than 
a decade \cite{Ottetal}. We have established a simple criterion for the 
recovery of the correspondence, Eq. (\ref{(5.5)}), which is generously 
met in the 
macroscopic examples discussed above. And, above all, we have demonstrated that
the {\it very same} process of decoherence that delivers ``pointer basis'' 
in the measuring apparatus can guard against violation of the 
quantum-classical correspondence in dynamics.

\subsection{The arrow of time: a price of classicality?}

Decoherence is caused by the continuous measurement-like interactions between 
the system and the environment. Measurements involve the 
transfer  of information, 
and decoherence is no exception: The state of the environment acquires 
information about the system.  For an observer who has measured the state of 
the system at some initial instant the information he will still have 
at some later time will be  influenced (and, in general, diminished) by 
the subsequent interaction between  the system and the environment. 
When the observer and the environment monitor the same set of observables, 
information losses will be minimized. This is in fact the idea behind the 
{\it predictability sieve}\cite{Zurek93,ZHP93} ---an information-based 
tool which allows 
one to look for the einselected, effectively classical states under quite
general circumstances. When, however, the state implied by 
the information acquired by the observer either differs right away
from the preferred basis selected by the environment, or ---as will be 
the case
here--- evolves dynamically into such a ``discordant'' state, 
the environment
will proceed to measure it in the preferred basis, and, from the observer's
point of view, information loss will ensue.

This information loss can be analyzed in several ways. The simplest is to 
compute the (von Neumann) entropy increase in the system. This will be our 
objective in this section. However, it is enlightening to complement this 
``external'' view by looking at the consequences of decoherence from 
the point of view of the observer, who is repeatedly monitoring the system 
and updating his records. \cite{Caves}
The loss of information can be quantified by the increase of 
the von Neumann entropy, 
\begin{equation} {\cal H} \ = \  -Tr~ \rho \ln \rho \ ,\label{(6.1)}
\end{equation}
where $\rho$ is the reduced density matrix of the system. 
We shall now focus on the rate of increase of the von Neumann entropy in 
a dynamically evolving system subject to decoherence. As we have seen 
before, decoherence restricts the spatial extent of the quantum--coherent 
patches to the 
critical coherence length $\ell_c$, Eq. (\ref{(5.3)}). A coherent 
wavepacket 
that  overlaps a region larger than $\ell_c$ will decohere rapidly, 
on a time-scale $\tau_D$ shorter than the one associated with the classical 
predictability loss 
rate given by the Lyapunov exponent $\lambda$. Such a wavepacket 
will deteriorate into a mixture of states, each of which is coherent over
a scale of dimension $\ell_c$ by $\sigma_c=\hbar/\ell_c$. Consequently, the
density matrix can be approximated by an incoherent sum of reasonably 
localized
and approximately pure states. When $N$ such states contribute more or less 
equally to the density matrix, the resulting 
entropy is ${\cal H}  \ \simeq  \ \ln N.$ 

The coherence length $\ell_c$ determines the resolution with which the 
environment is monitoring the position of the state of a chaotic quantum 
system. That is, by making an appropriate
measurement on the environment, one could in principle localize the system to 
within $\ell_c$. As time goes on, the initial phase space patch 
characterizing the observer's information about the state of the system will 
be smeared over an exponentially increasing range of the 
coordinate, Eq. (\ref{(2.2)}).
When the evolution is reversible, such stretching does not matter, at least
in principle: It is matched by the squeezing of the probability density in 
the complementary directions (corresponding to negative Lyapunov exponents). 
Moreover, in the quantum case folding will result in the interference 
fringes ---telltale signature of the long range 
quantum coherence, best visible in the structure of the Wigner functions. 

Narrow wavepackets, and, especially, small-scale interference fringes are 
exceedingly susceptible to monitoring by the environment. Thus, the
situation changes dramatically as a result of decoherence.  
In a chaotic quantum system, the number of independent eigenstates of 
the density matrix will increase as
\begin{equation} 
N \ \simeq \ \ell(t)/\ell_c \ \simeq \ {\hbar \over { \Delta p_0 \ell_c}} ~
\exp(\lambda t)\ . \label{(6.2)}\end{equation}
Consequently, the von Neumann entropy will grow at the rate:
\begin{equation} \dot {\cal H} \simeq \ {d \over {dt}} 
\ln (\ell(t)/\ell_c)  \ \simeq \ \lambda\ .\label{(6.3)}
\end{equation}
This equation emerged as a ``corollary'' of our discussion, but perhaps 
it is even its key result: Decoherence will help restore the 
quantum-classical correspondence. But we have
now seen that this will happen at a price. Loss of information is an 
inevitable consequence of the eradication of the ``Schr\"odinger cat'' 
states that were otherwise induced by the chaotic dynamics. They disappear 
because the environment is ``keeping an eye'' on the phase space, 
monitoring the location of the system with an accuracy set by $\ell_c$.

Throughout this section we have ``saved'' on notation, using ``$\lambda$'' 
to denote (somewhat vaguely) the rate of divergence of the trajectories 
of the hypothetical chaotic system. It is now useful to become a bit more 
precise. A Hamiltonian system with ${\cal D}$ degrees of freedom will have in 
general many (${\cal D}$) pairs of Lyapunov exponents with the same 
absolute value but 
with opposite signs. These global Lyapunov exponents are obtained by 
averaging
local Lyapunov exponents, which are the eigenvalues of the Jacobian of the
local transformation,
and which describe the rates at which a small patch centered on a trajectory 
passing through a certain location in the phase space is being deformed. 

The evolution of the Wigner function in the phase space is governed by the 
local dynamics. However, over the long haul, and in the macroscopic case, the
patch that supports the probability density of the system
will be exponentially stretched. This stretching and folding will produce 
a phase--space structure that differs from the classical probability 
distribution because of the presence of the interference fringes, with the fine
structure whose typical scale is on the order of $\hbar/\ell^{(i)}(t)$. In an 
isolated system, 
this fine structure will saturate only when the envelope of the Wigner 
function fills in the available phase space volume. Monitoring by the 
environment destroys these small--scale interference fringes and keeps 
$W$ from becoming narrower than $\sigma_c$ in momentum. As a result ---and in 
accord with Eq. (\ref{(6.3)}) 
above--- the entropy production will asymptotically 
approach the rate given by the sum of the positive Lyapunov exponents, 
\begin{equation} 
\dot {\cal H} \ = \ \sum_{i=1}^{\cal D} \lambda^{(i)}_+  
\ . \label{(6.4)}
\end{equation}
This result \cite{ZP94} 
is at the same time familiar and quite surprising. It is 
familiar because it coincides with the Kolmogorov-Sinai formula for 
the entropy production rate for a {\it classical} chaotic system. Here we have 
seen underpinnings of its more fundamental quantum counterpart. All the same, 
it is surprising because it is independent of the strength of the coupling 
between the system and the environment, even though the process of decoherence 
(caused by the coupling to the environment) is the ultimate source of 
entropy increase. Over the last few years, the argument we presented above
has been investigated and confirmed, using numerical simulations (see 
\cite{HuSh,Sarkar,Patt99,Pastawski99,Monteoliva}). 
Figure 6 presents clear evidence 
showing that in the chaotic regime the entropy production rate approaches
the value set by the Lyapunov exponent (data correspond to studies of a
quantum particle moving in a harmonically driven double well potential
\cite{Monteoliva}). 

\begin{figure}
\qquad\qquad
	\includegraphics[height=0.7\hsize]{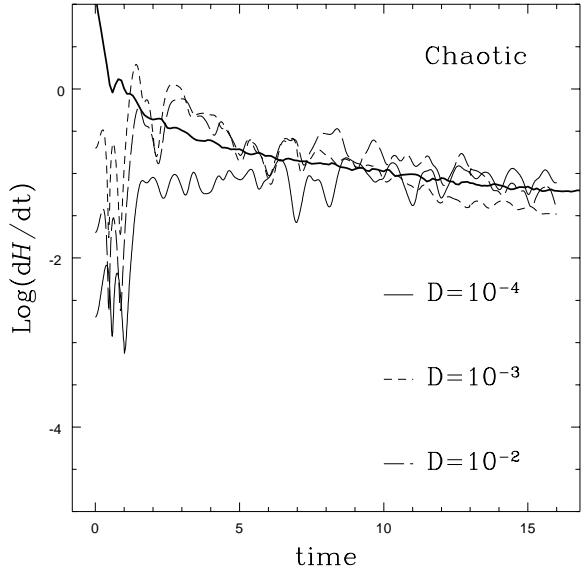}
\caption{Entropy production resulting from 
decoherence for a classically chaotic 
system 
becomes, after an initial transient, independent of the value of the 
diffusion constant and set by the Lyapunov exponent. See \cite{Monteoliva}}
\end{figure}

This independence is indeed remarkable, and leads one to suspect that the
cause of the arrow of time may be traced to the same phenomena that are
responsible for the emergence of classicality in chaotic dynamics, 
and elsewhere (i.e., in quantum measurements). In a sense, this is of 
course not a complete
surprise: Von Neumann knew that the measurements are irreversible 
\cite{vonNeumann32}. And Zeh \cite{Zeh90} emphasized the close kinship 
between the irreversibility of the ``collapse'' in quantum measurements 
and in the second law, cautioning against circularity of using one to 
solve the other. However, what is surprising is that both 
the classical-looking result ultimately has quantum roots, and that
these roots are so well hidden from view that the entropy production rate 
depends solely on the classical Lyapunov exponents.

Environment may not enter explicitly into the entropy production rate, 
Eq. (\ref{(6.4)}), but it will help determine when this asymptotic formula 
becomes
valid. The Lyapunov exponents will ``kick in'' as the dimensions of the patch
begin to exceed the critical sizes in the corresponding directions,
$\ell^{(i)}(t)/\ell_c^{(i)} \rangle 1  $.
The instant when that happens will be set by 
the strength of the interaction with the environment, which determines 
$\ell_c$. This ``border territory'' may be ultimately the 
best place to test the transition from quantum to classical. One may, for 
example, imagine a situation where the above inequality is comfortably 
satisfied
in some directions in the phase space, but not in the others. In that case,  
the rate of the entropy production will be lowered to include only these
Lyapunov exponents for which decoherence is effective.

\subsection{Decoherence, einselection, and the entropy production.}

The significance of the efficiency of decoherence goes beyond the 
example of the 
solar system or the task of reconciling quantum and classical predictions
for classically chaotic systems. Every degree of freedom coupled to the
environment will suffer loss of quantum coherence. Objects that are more 
macroscopic are generally more susceptible. In particular, the ``hardware''
responsible for our perceptions of the external universe and for keeping 
records of the information acquired in the 
course of our observations is obviously
very susceptible to decoherence. Neurons are strongly coupled to the 
environment and are definitely macroscopic enough to behave in 
an effectively classical fashion. That is, they have a
decoherence timescale many orders of magnitude smaller than the relatively 
sluggish timescale on which they can exchange and process information. 
As a result, in spite of the undeniably quantum nature of the fundamental 
physics involved, perception and memory have to rely on the information stored
in the decohered (and, therefore, effectively classical) degrees of freedom. 

An excellent illustration of the constraint imposed on information 
processing by decoherence comes from the recent discussions of the possibility
of implementing real quantum computers. Decoherence is viewed as perhaps 
the most serious threat to the ability of a quantum information processing 
system to carry out a superposition of computations \cite{Landauer,Bennett}.
Yet, precisely such an ability to ``compute'' in an arbitrary superposition 
would be necessary for an observer to be able to ``perceive'' an arbitrary 
quantum state. Moreover, in the external universe only those observables that  
are resistant to decoherence and which correspond to ``pointer states'' are
worth recording. Records are valuable because they allow for predictions, and
resistance to decoherence is a precondition to predictability 
\cite{Zurek91,Zurek93}.

It is too early to claim that all the issues arising in the context of the 
transition from quantum to classical have been settled with the help of 
decoherence. Decoherence and einselection are, however, rapidly becoming
a part of a standard lore \cite{GellmannHartle,Giulinietal}. 
Where expected, they deliver classical
states, and ---as we have seen above--- guard against violations of the
correspondence principle. The answers that emerge may not be to everyone's 
liking, and do not really discriminate between the Copenhagen Interpretation
and the Many Worlds approach. Rather, they fit within either mold, 
effectively providing the missing elements ---delineating the 
quantum-classical border postulated by Bohr (decoherence time fast or 
slow compared to the dynamical timescales on the two sides of the 
``border''), and supplying the scheme for defining distinct branches 
required by Everett (overlap of the branches is eliminated by decoherence).

\section{How to fight against decoherence: Quantum error correcting codes.}

It is clear that decoherence is a process that has a crucial role
in the quantum--to--classical transition. But in many cases,   
physicists are interested in understanding the specific causes 
of decoherence just because we want to get rid of it. 
Thus, decoherence is responsible for washing out the quantum interference 
effects we would very much like to see as a signal in some experiments. 
This is the type of situation one is clearly facing in 
quantum computation (and in the physics
of quantum information in general). A quantum computer is a gigantic
interferometer whose wave function explores an exponential number of 
classical computations simultaneously. Coherence between branches of 
the computer wave function should be maintained because the existence of 
quantum interference between these branches is the basic reason
why these computers can outperform their classical counterparts. 
Thus, decoherence in this context is a major problem. 

An obvious way of try to prevent decoherence from damaging quantum 
states is to reduce the strength of the coupling between the 
system and its environment. However, it is never possible to 
reduce this coupling to zero and eliminate decoherence in this way. 
Remarkably, in recent years new techniques that enable the active
protection of the information stored in quantum states from the degrading 
effect of the interaction with the environment have emerged. They
come under the name of ``Quantum Error Correcting Codes'' (QECC) and were
invented by people working on quantum computation \cite{Shor95, Steane}.  
They are based on remarkably simple and beautiful ideas and
could be found to be useful in other areas of 
physics. For this reason, we believe it could be interesting to 
include this final  main theme to give a 
simple--minded presentation of the methods that could enable us in 
principle to ``fight against decoherence'' preserving quantum states. 

\subsection{How to protect a classical bit} 

To introduce the basic idea of Quantum Error Correcting Codes  it
is better to start with a short discussion of the simplest ways in which
one can protect classical information. Suppose that we have 
a single qubit $b$ that lives in a noisy environment. Because of the effect
of the noise we will assume that the bit has a probability $p$ to 
flip after some time. Therefore, if we look at the bit after this 
time, the probability of the bit being unaltered by the noise is 
$1-p$ and therefore, the information is degraded. Can we protect this 
classical bit? The answer is ``yes'' and the way to do it is by using  
an error correcting code. The simplest such procedure is
based on the brute force use of redundancy as follows.
We can ``encode'' this one bit of information using more carriers, 
mapping the state of the bit into many identical copies (i.i. $b\rightarrow
(b,b,\ldots,b)$). If we do this, we can recover the initial information 
after the noise occurred by voting on 
and adopting as our result the one that gets
the majority of votes. In this way we also discover which carriers were 
altered by the noise (i.e., the minority) and recuperate the information. 
Of course, this works if the error probability is small enough. 
To be precise, let us assume that we encode the information in three 
carrier bits (this is the simplest 
repetition code). The probability that no flip occurs is $P(no\ flip)=
(1-p)^3$, and the others are simply $P(one\ flip)=3p(1-p)^2$, 
$P(two\ flips)=3p^2(1-p)$, and $P(three\ flips)=p^3$. Thus, 
the above error--correcting strategy (encoding
one into three bits and voting at the end) increases the probability
of keeping the information intact from $1-p$ to $1-3p^2 +2p^3=1-O(p^2)$,  
which is close to unity, provided $p$ is small enough. This example 
illustrates the simplest classical error correction code. Of course, 
much more sophisticated codes exist, and we are probably not doing 
justice to the beautiful theory of classical error--correcting codes 
(see \cite{McWilliams}) by using this naive code as an example. 
However, we think it is enough for the purpose of our discussion. 

\subsection{How to protect a quantum bit}

The basic question then becomes if it is possible to generalize this simple 
procedure to quantum mechanics. One may be tempted to guess that 
this task is impossible because a quantum version of the naive repetition 
code described above could never work as a consequence 
of the nonclonability of 
quantum states. Also, the fact that measurements drastically affect the 
state of quantum systems \cite{WoottersandZurek,Dieks}
is somehow suggestive of the difficulties of 
implementing an error--correcting quantum strategy naively translating  
the classical error--correcting ideas. However, these expectations were 
proven to be incorrect when in 1995 Peter Shor created the first 
quantum error--correcting code \cite{Shor95}. 
His work, once again triggered a lot of activity and over the last four 
years the theory of Quantum Error Correcting Codes was fully 
developed. So far, there have been some experimental demonstrations 
showing the workings of these codes (only in NMR experiments) 
but in our view, the interesting ideas of QECC still are
waiting for  phycisists to give a definite answer to whether or not 
they will be useful for other purposes than the ones that originally 
motivated them. For this reason, we find it interesting to bring these
issues to this review. 

Let us now describe how it is possible to create QECC. For this, we 
consider a quantum bit prepared in an arbitrary quantum state $\Psi=
\alpha|0\rangle +\beta |1\rangle $. To be precise, we will first 
describe how  noise affects the state of the qubit whose state
we want to protect. We will first consider the simplest case of a 
noise that just produces ``dephasing''.
We  assume that the noise introduces a random phase
with a probability $p$ or leaves the state intact with a probability $1-p$. 
Although this is not the most general kind of operation that a noisy 
environment can produce on a quantum system, we will later show that 
this is not a restrictive assumption and that the treatment we present
here can be generalized to include all of the effects that the noise can 
produce. So, for the moment we will just consider this ``dephasing''
noise. The dephasing can be simply described by the action 
of a $\sigma_z$ operator on the state of the system. 
In this chapter, we will adopt the following notation. The Pauli matrices
$\sigma_{x,y,z}$ are simply denoted as $X,Y,Z$. Thus, if the
initial state of the system is $\Psi_0=\alpha |0\rangle + \beta |1\rangle$
the final state (after the noise has occurred) is described by a 
density matrix as, 
\begin{equation}
\rho_{out}=(1-p)\rho_{in} + p Z\rho_{in}Z\ ,\nonumber
\end{equation}
where $\rho_{in}=|\Psi_0\rangle\langle\Psi_0|$. It is easy to see that
the interaction with the noise degrades the quantum state, causing the
loss of quantum coherence. As a measure of this degradation, we can 
compute the ``fidelity'' of the process that is simply given by the 
overlap between the ideal state and the actual state. Using the above 
form for the density matrix, we find out that  fidelity is reduced
to $F=Tr(\rho_{out}\rho_{in})=1-4p|\alpha\beta|^2$. Thus, fidelity 
is reduced by an amount that is linear in the error probability $p$. 
Another measure of the degradation is given by the loss of purity of
the final state that can be measured, for example, by $Tr(\rho_{out}^2)=
1-8p(1-p)|\alpha\beta|^2$. In what follows, we will present a method
that enables us to protect the quantum state in such a way that the 
fidelity (or the loss of purity) does not decay linearly with the error
probability but it does so quadratically. 

So, let us present a way to protect the state of our qubit from the 
effect of a dephasing environment. As in the classical 
case, we will use many carriers to protect one qubit of information (in 
our example, we  use three qubits to protect one). But  
the use of redundancy has to be more subtle in the the quantum case. The 
key idea is to encode the logical states into entangled states of the three 
qubits in such a way that when an error occurs, the logical states are
mapped into other orthogonal subspaces (one subspace for each error we 
want to correct). If this is the case, we can learn about the error by 
measuring an observable that just tells us in what two--dimensional subspace
the state is in. In this way, we learn what the error was without getting
any information about the state itself. Once we know the error, we
can correct it and start the process all over again. This idea is 
illustrated clearly (we hope!) by the three--qubit example. In this case, 
we can use the following encodinf for the logical states:
\begin{eqnarray} 
|0\rangle_L&=&{1\over 2}(|000\rangle+|110\rangle+|101\rangle+|011\rangle)
\nonumber\\
|1\rangle_L&=&{1\over 2}(|111\rangle+|001\rangle+|010\rangle+|100\rangle),
\label{0L1L}
\end{eqnarray} 
(the subscript $L$ is used to denote the logical states). The ``encoding''
process is simply the mapping of the physical states of the three independent 
carriers onto the above entangled logical states. This task is the first
one that one 
has to do to protect the information and is represented by a unitary 
operator (the encoding operator $E$). One takes the qubit whose quantum 
state is to be protected and applies an operation to it together with the 
other two carriers we use. This operation maps the initial state into 
the encoded state, i.e.,  
$E|\alpha|0\rangle+\beta |1\rangle)|00\rangle=\alpha|||0\rangle_L
+\beta|1\rangle_L$. Later in this section we will describe ways in 
which the encoding operation can be implemented. 

The reason why (\ref{0L1L}) is a good encoding can be seen as follows. 
It is a simple exercise to show that when we apply an error operator 
to any of the two logical states (i.e., when we act with a 
$Z$ operator on any one of the qubits) we obtain mutually orthogonal 
states. Thus, one can show that $|0\rangle_L\perp Z_i|0\rangle_L\perp
|1\rangle_L\perp Z_i|1\rangle_L$ for $i=1,2,3$, i.e., that the two logical 
states and their ``erroneous descendants'' are a set of eight mutually 
orthogonal states that constitute a basis of the complete Hilbert space 
of the three qubits. 
Therefore, the total Hilbert space can be decomposed in the direct sum of 
four two--dimensional subspaces. The ``logical subspace'' $H_L$,
which is generated by the two vectors $\{|0\rangle_L,|1\rangle_L\}$,
has three ``erroneous descendents'' which are simply $Z_iH_L$, and the 
total Hilbert space is the direct sum of $H_L$ and $Z_iH_L$ ($i=1,2,3$).
As a consequence, there is an observable that we could measure to 
determine in which one of the four subspaces the state is in. In so doing, 
we discover the error and can correct it trivially. 

To complete our
description, we just have to exhibit what this observable whose measurement
reveals the error is. To do this, it is interesting to look at the symmetries
of the logical states (\ref{0L1L}). It is clear that these states are
eigenstates of the operators $M_1=X_1X_2$ and $M_2=X_2X_3$ with eigenvalue $+1$
(thus, $|0\rangle_L$ is an homogeneous superposition of all states with
an even number of ones and $|1\rangle_L$ contains all states with 
an odd number of ones; therefore these states are invariant when we flip 
any two states, which is precisely what the $X_iX_j$ operators do). 
Moreover, it is easy to show that $M_1$ and $M_2$ are two commuting hermitian 
operators whose eigenvalues are $\pm 1$ (this follows from the fact that these
operators square to the identity, i.e., $M_i^2=1$). Moreover, it is  
simple to show that all the ``erroneous descendents'' of the logical 
subspace are also eigenspaces of $M_i$. For example, the subspace $Z_1H_L$ is 
formed by linear superpositions of the vectors $\{z_1|0\rangle_L,|1\rangle_L\}$
that are eigenstates of $M_1$ and $M_2$ with eigenvalues equal to $-1$. This
follows from the fact that as the error--operator $Z_1$ anticommutes with 
$M_1$ and $M_2$, it transforms eigenstates of these operators into 
eigenstates with a different eigenvalue (i.e., if 
$M_i|\phi\rangle=|\phi\rangle$, then $M_iZ_1|\phi\rangle=-Z_1|\phi\rangle$).
Therefore, if our goal is to find out in which of the four two--dimensional
subspaces the state is in, we just have to measure the two operators 
$M_1$ and $M_2$. The result of this measurement is always represented
by a set of two numbers that are $\pm 1$ (the two eigenvalues of $M_i$) 
and each of the four possible alternatives (that are known as the 
error syndromes) identify uniquely one of the four subspaces ($H_L$
corresponds to the syndrome $(+1,+1)$, $Z_1H_L$ to $(-1,-1)$, $Z_2H_L$ to
$(-1,+1)$ and $Z_3H_L$ to $(+1,-1)$.

It is also interesting to think about what kind of physical procedure 
we should follow to perform this kind of measurement. As discussed, 
we need to measure the operators $M_i$ that are constructed as tensor 
products of Pauli matrices. However, it is very important to realize that
we must do this {\bf without} measuring individually the factors appearing
in these products! Thus, in our case, we need to measure only $M_1=X_1X_2$ 
and $M_2=X_1X_3$, but we cannot do this by measuring the three operators
$X_i$ individually. If we were to do this, we would be measuring a complete
set of commuting observables and causing the system to collapse into a 
particular state. Instead, quantum error correction needs measuring, not
a complete set of observables but only enough observables to gain information
about the error without destroying the coherence in the state of the 
system (thus, we want our measurement to project the state into a 
two--dimensional subspace and not to collapse it into one ray). 

It is not hard to find a systematic way to devise a strategy that will  
enable us to measure any operator that is the tensor product of Pauli
matrices without measuring the individual factors. To do this, it is
clear that because the observables we measure are collective, we should  
induce an interaction between the qbits in such a way that after the 
interaction, the result of the measurement is ``written'' on only one 
particle. For example, suppose that we have two particles and we want
to measure the operator $M=X_1X_2$. Suppose also that we find a unitary 
operator $D$ satisfying the condition $Z_2D=DM$. This condition implies that
the operator $D$ will transform an eigenstate of $M$ with eigenvalue $m$
(that can only be $\pm 1$) into an eigenstate of $Z$ with eigenvalue $m$. 
Therefore, if we want to measure $M$, we can first apply the unitary operation
$D$ and then measure $Z_2$ (in other words, $D$ is the operator that changes
basis from $M$ to $Z_2$ eigenstates). Thus, now we just need to construct
this operator. This can be done by using a simple quantum circuit. 
In fact, the quantum circuit for the operator $D$ is shown in Figure 7.
We just have to apply a Hadammard rotation to each qubit and then do a 
{\tt c-not} using the first qbit as the control and the second one as the 
target. To show that this is the correct circuit for $D$, we just have
to show that the relation $Z_2D=DM$ is satisfied. For this purpose, we
apply $M=X_1X_2$ to the left of the circuit and start moving the $X_i$ 
operators to the right. As these operators satisfy that $RX=ZR$, 
they transform into $Z$ operators when they pass through the Hadammard 
rotations. Then, the $Z$ operator 
in the control goes through the end of the circuit
but the one acting on the target generates an extra $Z$ in the control
qubit that cancels the first one. Therefore, this implies that the 
circuit satisfies the required identity. Using this simple idea, is 
possible to design simple quantum circuits that can be used to measure 
any collective observables built as tensor products of Pauli matrices. 
Moreover, this can be generalized to any number of qubits. For example,
the circuit to measure $M_1=X_1X_2$ and $M_2=X_1X_3$ is given in Figure
6 and consists of three Hadammard rotations (one in each qubit) followed
by two {\tt c-not} gates with the first qubit acting as the control. It is 
easy to see that if  we measure the second
and third qubits after the circuit, we learn about 
the syndrome and therefore find out what the error was. 

To recover from the error, we just have to apply a simple operation to 
the remaining qubit that we do not measure (the first one in our example).
This qubit contains the quantum state up to some unitary transformation
that we can undo. To find out how to recover from the error, the 
idea is simply to see what the circuit does to the errors themselves. 
In fact, it is easy to show that the operator $D$ associated with the
decoding circuit appearing in Figure 7 satisfies that $Z_1D=DX_1X_2X_3$, 
and that $Z_2D=DX_2$, $Z_3D=DX_3$. Therefore, this means that if the 
encoded state is affected by a $Z_1$--type error, the resulting state
after decoding will have the last two qubits set to one (we already 
knew that this was the syndrome corresponding to this error), and the 
first qubit will be affected by an $X$ rotation that we should undo. 
On the contrary, the other two errors ($Z_2$ and $Z_3$) do not require
any corrective action. 

\begin{figure}
\qquad
	\includegraphics[height=0.3\hsize]{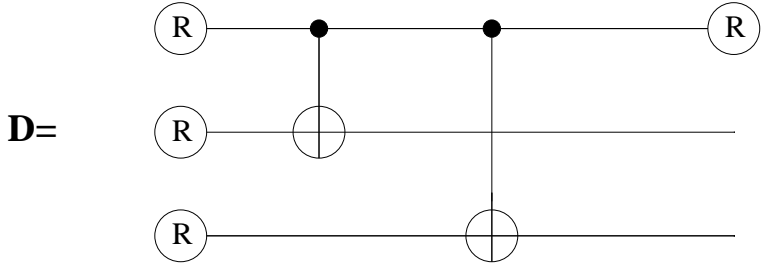}
\caption{Decoding circuit for the three qubit quantum error--correcting
code}
\end{figure}

So, to summarize, the error--correcting procedure is the following: (1) We
encode the qubit in three carriers applying the encoding circuit shown
in Figure 7. (2) After the errors act on the system, we decode the state,
detect the syndromes, and apply corrective operations. (3) We refresh 
the syndrome qubits (resetting them to the zero state) and encoding again. 
It is clear that measurement of the syndrome is not really necessary 
because 
it can always be replaced by a corrective operation performed by means
of a quantum circuit (in our case a {\tt c-c-not} that is controlled by the 
second and third qubits). The essential part of this method is the 
refreshing of the syndrome qubits that is the part responsible for taking
away the ``entropy'' generated by errors. 

Two final comments are worth making before giving a more formal presentation.
First, we should remark that our discussion so far assumed that errors were
applied by some agent that acted on a single (unknown) qubit. However, we
can extend this method to consider a situation in which there is a probability
$p$ for any one qubit to be affected by a $Z$--type error. In this case, 
the state of the three qubits before the decoding and corrective 
circuit is applied is given by the following density matrix: 
\begin{eqnarray}
\rho_{out}&=&(1-p)^3\rho_{in}+p(1-p)^2\sum_iZ_i\rho_{in}Z_i\nonumber\\
&+&p^2(1-p)\sum_{i\neq j}Z_iZ_j\rho_{in}Z_iZ_j+p^3Z_1Z_2Z_3\rho_{in}
Z_1Z_2Z_3\ ,
\end{eqnarray}
After we apply the decoding and corrective procedure to this density
matrix, it is clear that the first two terms will now be simply proportional
to $\rho_{in}$. Thus, in this way we have 
completely eliminated the term that 
is linear in the error probability $p$. The final state differs from the
ideal one only through terms that are quadratic in the error probability. 
Therefore, the fidelity of the whole process will be given by $F=1-O(p^2)$. 
On this linear--to--quadratic change in the dependence of $F$ on $p$ relies
the whole power of quantum error correction (which clearly only has a good
chance of working at this level, without concatenation, 
if $p$ is small enough). 

Finally, we could worry about not having considered more general classes 
of errors. However, it should be clear by now that the general idea 
described so far could be generalized to include more general operations.
It is important to realize that to take into account all possible effects 
the environment could cause on a qubit, we should protect not only against
phase errors (associated with $Z$ operators) but also against bit flips 
(associated with $X$ operators) and on a combination of both (associated
with $Y$ operators). It is clear that if we are able to protect against
three types of independent errors, we could also fight efficiently against
arbitrary unitary (or nonunitary) errors that can always be written
in terms of operators that are linear combinations of these three 
elements and of the identity. So, the question is how to invent codes that
protect against arbitrary errors affecting any one of the carrier qubits. 
A code like this was first presented by Peter Shor \cite{Shor95} and 
can be constructed using our previous three qubit QECC as a building block.
In fact, Shor encodes one qubit using nine carriers organized in three 
blocks of three qubits each. The logical states are a product of three 
factors like the ones shown in 
(\ref{0L1L}). This code has the following eight symmetry operations 
(the previous one had the two symmetries, $M_1$ and $M_2$): First, we 
easily find six 
symmetry operators that generalize the previous $M_1$ and $M_2$ 
in the three blocks of three qubits. Second, we find two other 
independent symmetries corresponding to the fact that 
the three blocks are repeated:  $M_7=Z_1Z_2Z_3Z_4Z_5Z_6$ and 
$M_8=Z_1Z_2Z_3Z_7Z_8Z_9$. It is easy to show that each of the $27$ 
different errors that can affect the nine carriers corresponds to a 
different syndrome (and therefore maps the logical states into 
orthogonal subspaces). The decoding should be done by measuring the 
above eight operators that reveal the syndrome and allow us to know 
the error that took place enabling us to correct it. A decoding circuit
for this code can be easily constructed following the same discussion 
presented above for the three qubit. 
It is interesting to note that the code presented by Shor is by no 
means the most efficient way to correct errors. In fact, we notice that
we are using an enormous Hilbert space of dimension $2^9=512$, but we 
would only need a space with enough room to accommodate for all the 
subspaces where we would map independent errors (in this case we 
require for this purpose only $2(1+3\times 9)=56$). Smaller codes have been 
developed, and the smallest one that corrects general one--qubit errors
requires five qubits \cite{perfect}, 
because $n=5$ saturates the identity $2^n=2(3n+1)$.
This is the so--called ``perfect'' QECC and has the following symmetry 
operators $M_1=Z_2Z_3Z_4Z_5$,  $M_2=Y_1Z_3X_4Y_5$,  $M_3=Z_1X_2Z_3X_5$,
and $M_4=Z_1Y_3Y_4Z_5$. To show that these symmetry operators constitute
a good QECC requires showing that all independent errors produce a 
different anticommutation pattern with the $M_i$ operator (this is 
left as an exercise). The construction of an encoding--decoding 
circuit for this code can also be done by generalizing the ideas we 
have described before.

\subsection{Stabilizer quantum error--correcting codes}

A more formal description of the principles underlying the theory 
of quantum error--correcting codes can be given (following the presentation 
of \cite{PazZurek98} we restrict ourselves to discuss a rather wide class of 
codes known as stabilizer codes (for more general codes and for a 
more thorough discussion of QECC we refer the reader to 
\cite{KnillLaflamme, Calderbanketal, GottesmanPhD}). 
We can consider codes that protect $k$ 
qubits by encoding them into $n$ carriers. Here, the code space ${\cal H}_k$ 
(or logical space) is a $2^k$ dimensional subspace of the total 
Hilbert space of the $n$ carriers. ${\cal H}_n$ is a tensor product 
of $n$ two--dimensional factors and has a natural basis whose elements
are product states of the individual carriers. 
This is the ``physical basis" that can be formed with the common eigenstates
of the operators $\{Z_1,\ldots,Z_n\}$ (for convenience, we label states of this 
basis not by the eigenvalues of the corresponding operators, which are 
$\pm 1$, but by the eigenvalues of the projectors onto the $-1$ subspace, 
which are $0$ or $1$: thus, the label $z_j=0$ ($z_j=1$) corresponds 
to a $+1$ ($-1$) eigenvalue of the operator $Z_j$). Furthermore, we order 
the $n$ carriers in such a way that the last $k$ qubits are the ones 
whose state we encode, and the first $n-k$ are the ancillary carriers. 
Therefore, states of the physical basis are of the form 
$|s,z\rangle_P=|s\rangle_P\otimes|z\rangle_P$ (where the strings 
$s=(s_1,\ldots,s_{n-k})$, $z=(z_1,\ldots,z_k)$ store the corresponding 
eigenvalues and the subscript $P$ is used to identify the 
states of the physical basis). 

An error--correcting code is a mapping from the physical product 
states $|0\rangle_P\otimes |\Psi\rangle_P$ onto the code space 
${\cal H}_k$, which is formed by entangled states of $n$ carriers. 
A rather general class of codes can be described in terms of their 
stabilizer group (see \cite{GottesmanPhD}). 
The stabilizer of the code is an Abelian group formed by all operators 
that are tensor products of Pauli matrices and have ${\cal H}_k$ as an 
eigenspace with an eigenvalue equal to 
$+1$. Every element of the stabilizer, which 
is a finite group with $2^{n-k}$ elements, can be obtained by 
appropriately multiplying $n-k$ generators, which will be denoted as 
$M_1,\ldots, M_{n-k}$. The elements of the stabilizer are completely
degenerate in the code space ${\cal H}_k$ (since all states in ${\cal H}_k$ 
are 
eigenstates with eigenvalue $+1$ of all $M_j$). To define a basis in the 
code space, we choose $k$ extra operators $L_1,\ldots,L_k$, which being tensor 
products of Pauli matrices commute with all elements of the stabilizer. 
These operators $L_{j'},\ j'=1,\ldots,k$ are the ``logical pointers" 
because  
they define the directions in ${\cal H}_k$ associated with the logical states  
$|0\rangle_L,\ldots, |2^k-1\rangle_L$ (logical pointers belong to the 
group of operators 
that commute with the stabilizer, known as the normalizer). 

The $n-k$ generators of the stabilizer together with the $k$ logical 
pointers are a Complete Set of Commuting Operators (CSCO) whose common 
eigenstates form a complete basis of the Hilbert space ${\cal H}_n$. 
Elements of this ``logical basis", labeled by $n$ quantum numbers, are 
denoted as $|m,l\rangle_L$, where the bit strings $m=(m_1,\ldots,m_{n-k})$, 
and $l=(l_1,\ldots,l_k)$ identify the corresponding eigenvalues, and the 
subscript $L$ refers to logical states. The CSCO formed by the generators 
of the stabilizer and the logical pointers defines a prescription for 
decomposing the original Hilbert space of the $n$--carriers into a 
tensor product of a $2^k$--dimensional logical space ${\cal L}$ and a 
$2^{n-k}$--dimensional syndrome space ${\cal Y}$. In fact, elements of 
the logical basis (which are entangled states of the $n$-carriers) are 
tensor products of states belonging to ${\cal L}$ and ${\cal Y}$: 
$|m,l\rangle_L=|m\rangle_L\otimes|l\rangle_L$. Encoded states, which belong 
to ${\cal H}_k$, are also product states of the form 
$|\Psi\rangle =|0\rangle_L \otimes\sum_l c_l |l\rangle_L$. 

The code protects quantum states against any error $E_a$ whose action on 
states of the logical basis is to change the logical syndrome and, 
eventually, rotate the logical state in ${\cal L}$ in a syndrome--dependent 
way,
\begin{equation}
E_a\ |m\rangle_L\otimes |l\rangle_L=
e^{i\phi_{ma}}\ |m+c_a\rangle_L\otimes U_a |l\rangle_L\ .
\end{equation}
Here, $U_a$ is a unitary operator acting on the collective logical space 
${\cal L}$, 
and $\phi_{ma}$ is a phase that may depend on the syndrome and the error. 
The error $E_a$ changes the syndrome from $m$ to $m+c_a$ where $c_a$ is 
the bit string storing the commutation pattern between the error and the 
generators of the stabilizer (the $j$th bit of this string is one if 
the error anticommutes with $M_j$ and is zero otherwise). The reason for 
this is that when acting on a logical state, 
the error $E_a$ changes the eigenvalue of the operator $M_j$ 
only if $\{M_j,E_a\}=0$. The label $a$ used to identify errors 
is arbitrary and, 
for the case of nondegenerate codes (which are the only ones we will 
consider here) 
it is always possible to label errors $E_a$ using simply the commutation 
pattern $c_a$ (i.e., we can choose $a=c_a$). 

To correct against the action of any of the errors $E_a$ (or against any 
linear superposition of them) one
can first detect the error by measuring the collective syndrome (i.e., 
measuring
the observables $M_j$, $j=1,\ldots,n-k$) and later recover from
the error by applying the corresponding operator $U^\dagger_a$.  
This detection--recovery process can be conveniently described as a 
quantum operation defined by the following mapping from the erroneous density 
matrix $\rho_{in}$ into the corrected one $\rho_{out}$,
\begin{equation}
\rho_{out}=\sum_{m=0}^N R_m\rho_{in} R^\dagger_m\ ,\label{recovermap}
\end{equation}
where the sum runs over all syndromes ($N=2^{n-k}-1$), and the recovery 
operator for each syndrome is 
\begin{equation}
R_m=|0\rangle_L\  _L\langle m|\otimes U^\dagger_m\ .\label{recoverop}
\end{equation}
By construction, these operators satisfy the identity 
$\sum_{m=0}^N R^\dagger_m R_m = I$.

Because our description of  error detection--recovery process  is entirely 
formulated on the logical basis, it does not involve so far any reference to 
encoding or decoding operations that can be simply defined as a change of 
basis. The encoding operator $C$ is a unitary operator mapping the physical 
basis, formed by product states of the $n$ carriers, onto the logical basis, 
formed by entangled states. Accordingly, $C$ transforms the operators 
$Z_i$ (whose eigenvalues define states on the physical basis) into the 
operators $M_j$, $L_{j'}$ (that label states on the logical basis). 
Thus, the encoding operator $C$ is such that $Z_j=C^\dagger M_j C$, 
$j=1,\ldots,n-k$, and $Z_{n-k+j'}=C^\dagger L_{j'} C$, $j'=1,\ldots,k$. 
Taking this into account, the action of the operator $R_m$ can be described, 
in the physical basis, as the following sequence of operations: i) decode 
the state, ii) measure the syndrome in the physical basis by measuring $Z_j$ 
in the first $(n-k)$ carriers, iii) if the result of the measurement is the 
string $s$, apply the syndrome--dependent recovery operator $U^\dagger_s$ 
resetting the syndrome back to zero, and iv) encode the resulting state. 

Finding a stabilizer code correcting a given set of errors is a rather hard 
task that involves designing generators having appropriate commutation 
patterns with the errors. Once the generators are found and the logical 
pointers are chosen, an encoding or decoding operator can be constructed 
(strategies for designing encoding or 
decoding circuits from the stabilizer are 
known; see \cite{CleveGottesman,Pringe}). The recovery operators depend on 
the encoding or decoding strategy and can be explicitly found from the 
encoding circuit by running errors through it. 

As we mentioned above, the simplest code protecting $k=1$ qubit 
using $n=3$ carriers correcting against phase errors in any of the 
carriers can be understood as a particular example of this general stabilizer
code class. In such a case, the basic errors to correct  
are $E_1=Z_1$, $E_2=Z_2$ and $E_3=Z_3$. The stabilizer of the code can 
be chosen to be generated by $M_1=X_1X_2$ and $M_2=X_1X_3$. 
The commutation pattern associated with each error is $c_1=11$ 
(because the error $Z_1$ anticommutes with both $M_1$ and $M_2$), 
$c_2=10$, $c_3=01$ (note that we could relabel the errors ordering
them according to their commutation pattern). The decoding 
circuits exhibited in Figure 7 has the properties 
\begin{equation}
C^\dagger Z_1 C=X_1X_2X_3,\ \ C^\dagger Z_2 C=X_2,\ 
{\rm and}\ \ C^\dagger Z_3 C= X_3.
\end{equation}
These properties entirely determine the action of the errors $Z_i$ in the 
logical basis. For example, the last identity implies that
$E_3|m\rangle_L|l\rangle_L=|m+c_3\rangle_L|l+1\rangle_L$. 
Thus, the error $E_3$ not only changes the syndrome but also 
modifies the logical state by flipping it. This means that the recovery 
operator for this error is $U_3=X$. Analogously, we can find how the other 
errors act on the logical basis, showing that $U_1=U_2=I$. 

\section{Discussion}

We have seen ``decoherence in action'' in a variety of settings. Our aim
was not to review all of the studies of decoherence done in recent 
years. Thus, we left aside from our review the discussion of some very 
interesting physical problems where the role of environment-induced 
decoherence is relevant. For example, in  cosmology, the way in which 
decoherence can account for the quantum to classical transition of density
fluctuations (and of spacetime itself) has been ---and still is--- a 
matter of debate (see \cite{Cosmology} for an incomplete list of relevant 
papers). Fortunately, there are also other areas where 
decoherence can be analyzed and tested in the laboratory. 
Among them, the use of systems of trapped and cold atoms 
(or ions) may offer the possibility of 
engeneering the environment (effectively choosing the pointer states) 
as proposed in \cite{Engeneering}. Trapped atoms inside cavities were
discussed \cite{Davidovich} and the relation between decoherence
and other cavity QED effects (such as Casimir effect) was 
analyzed \cite{Dalvit}. On the 
mesoscopic scale, the nature of decoherence may receive increasing
attention 
specially in the context of BEC both as a key ingredient in the
phenomenological description \cite{BECdeco} and as a threat to the
longevity of BEC Schr\"odinger cats \cite{DDZ}. Moreover, 
the nature of decoherence is being studied experimentally in the context
of condensed matter systems (see, for example, \cite{Webb}). 

The aim of this section is to describe briefly what is (and point 
out what is not) accomplished 
by decoherence, and to show how it facilitates understanding the transition
from quantum to classical. 
Environment-induced superselection is clearly the key interpretational
benefit arising from decoherence. The 
quantum principle of superposition does not 
apply to open quantum systems. States in the Hilbert space are no 
longer ``equal''. Under a broad variety of realistic physical assumptions, 
one is now forced to conclude that for macroscopic objects only a small subset 
of states can ever contribute to the ``familiar classical reality''. Only
the einselected pointer states will persist for long enough to retain useful
(stable) correlations with ---say--- the memories of the observers, or, more 
generally, with other stable states. By contrast, their superpositions will
degrade into mixtures that are diagonal in the pointer basis.

The precondition for ``perception'' (as in ``perception of classical 
reality'')
is the ability of the state to persist, or to evolve in a more or less
predictable manner during 
a time interval over which the observer is monitoring 
it. This time interval can occasionally be quite short, but it should not 
be as unreasonably short as the typical decoherence time for the macroscopic
systems. Thus, the only states that have a chance of being perceived as 
``real'' are the preferred (pointer) states. Indeed, given the limited accuracy
of the observer's efforts, it may be more precise to say that broad 
superpositions of pointer states are definitely ruled out.

It is important to emphasize that the environment-induced superselection leads
to a probability distribution that is 
diagonal in the preferred basis, and not to 
a single pointer state. Thus, the uniqueness of perceptions of the observer 
has its roots in the stability of the correlations between the states of the
macroscopic objects in the outside world and the records in the observer's 
memory (which, incidentally, must also use preferred states of, say, 
neurons to store records of the observations). 

The information possessed by the observer is not an abstract, esoteric
entity. Rather, ``information is physical'' \cite{Landauer} and ``there
is no information without representation'' \cite{Zurek93}. In practice, this
means that the state of the observer is in part determined by what he knows
about the rest of the universe. Thus, the physical existence of long-lasting 
records underlies the essence of the process of perception. Observers will
be aware of their own records, and of the external universe in a state 
consistent with these records. This viewpoint known as the {\it existential
interpretation} \cite{Zurek93,Zurek98b} accounts for the apparent 
collapse, but 
is consistent with either the Many Worlds or Copenhagen Interpretation.

The nature of the preferred states is dynamically negotiated in 
the course of 
the interaction between the system and the environment, but, as we have
already seen, the self-Hamiltonian of the system plays an important role.
Truly realistic models are difficult to treat, but lessons of the 
predictability 
sieve applied to simple models allow one to infer with some confidence that,
in general, pointer states will be localized in position. After all, most 
interactions depend on distance. Thus, localization is an inescapable 
consequence \cite{Zurek82,Zurek91}. Nevertheless, as we have already seen in 
perhaps the most relevant exactly solvable case of a decohering harmonic
oscillator, preferred states tend to be localized in both position and 
momentum and can be regarded as quantum counterparts of classical points.

Investigation of the coexistence of decoherence with chaos is an example of 
a bit more complicated case. There, we have seen that localization is 
effectively enforced (even if such systems cannot be treated analytically, and
extensive numerical studies are required).

An exciting ``corollary'' of decoherence in the setting of quantum chaos 
is the quantum derivation of the classically anticipated entropy production
rate, given by the sum of positive Lyapunov exponents. This suggests a quantum
origin of the second law of thermodynamics. Indeed, it seems that the 
resolution of the two outstanding puzzles of physics ---the arrow of time 
and the apparent classicality--- may originate from the same essentially 
quantum source, from decoherence and einselection. 

The study of decoherence and einselection over the past two decades has yielded
a new paradigm of emergent, effective classicality. It leads to a new 
understanding of the quantum origins of the classical. To be sure, not all
of the interpretational questions have been settled, and much further work
is required. Nevertheless, as a result of this paradigm shift, 
the quantum-to-classical transition has become a subject of experimental 
investigations, while previously it was mostly a domain of philosophy. 

\section{Acknowledgment}

This research was supported in part by NSA. We gratefully acknowledge 
Isaac Newton Institute in Cambridge, England, where much of this manuscript 
was prepared. JPP was also supported by grants from ANPCyT (PICT 01014) and 
Ubacyt (TW23).

%\end{thebibliography}


\thebibliography{}

\bibitem{Zurek81} Zurek, W. H., {\it Phys. Rev.} {\bf D 24}, 1516-1524 (1981).

\bibitem{Zurek82} Zurek, W. H., {\it Phys. Rev.} {\bf D 26}, 1862-1880 (1982).

\bibitem{Zurek93} Zurek, W. H., {\it Progr. Theor. Phys.} {\bf 89}, 281-302 
(1993).

\bibitem{ZHP93} Zurek, W. H., Habib, S., and Paz, J. P., {\it Phys. 
Rev. Lett} {\bf 70},
1187-1190 (1993); Anglin, J. R., and Zurek, W. H., {\it Phys Rev.} 
{\bf D53}, 7327-7335 (1996).

\bibitem{Gallis} Gallis, M. R., {\it Phys. Rev.} {\bf A53}, 655-660 
(1996); Tegmark, M., and
Shapiro, H. S., {\it Phys. Rev.} {\bf E50}, 2538-2547 (1994).

\bibitem{Bruneetal} Brune, M., Hagley, E., Dreyer, J., Ma\^itre, X., 
Maali, A., Wunderlich, C., Raimond, J-M., and Haroche, S., {\it Phys. 
Rev. Lett.} {\bf 77}, 4887-4890 (1996).

\bibitem{Raymer} Cheng, C. C., and Raymer, M. G., {\it Phys. Rev. Lett},
{\bf 82}, 4802 (1999)

\bibitem{Myattetal} Myatt, C. J., et al., {\it Nature}, {\bf 403}, 269 (2000).

\bibitem{XXX} Ammann, H., Gray, R., Shvarchuk, I., and Christensen, N., 
{\it Phys. Rev. Lett.} {\bf 80}, 4111 (1998).

\bibitem{Raizen} Klappauf, B. G., Oskay, W. H., Steck, D. A., and Raizen, M. G.,
{\it Phys. Rev. Lett.} {\bf 81}, 1203 (1998); Erratum in {\it Phys Rev. Lett.}
{\bf 82} 241 (1999).

\bibitem{Bennett} Bennett, C. H., {\it Physics Today} {\bf 48}, No. 10 (1995);
Bennett, C. H., and DiVincenzo, D. P., {\it Nature} {\it 404},
247 (2000).

\bibitem{WoottersandZurek} Wootters, W. K., and Zurek, W. H., {\it Nature}
{\bf 299}, 802 (1982). 

\bibitem{Dieks} Dieks, D., {\it Phys. Lett.} {\bf A 92}, 271 (1982). 

\bibitem{Zurek91} Zurek, W. H., {\it Physics Today} {\bf 44}, 36 (1991).

\bibitem{Tegmark99} Tegmark, M., {\it Phys. Rev.} {\bf E 61}, 4194 (2000). 

\bibitem{Zurek98a} Zurek, W. H., {\it Physica Scripta} {\bf T76}, 186 (1998), 
also available at quant-ph/9802054. 

\bibitem{Zurek98b} Zurek, W. H., {\it Phil. Trans. R. Soc. Lond.} {\bf A356},
1793 (1998), also available at quant-ph/9805065. 

\bibitem{vonNeumann32} von Neumann, J., ``Measurement and reversibility'' 
and ``The measuring process'', chapters V and VI if 
{\it Mathematische Grundlagen der Quantenmechanik}, (Springer, Berlin, 1932);
English translation by R. T. Beyer 
{\it Mathematical Foundations of Quantum Mechanics}, (Princeton Univ. Press, 
Princeton, 1955).

\bibitem{Monroeetal} Monroe, C., Meekhof, D. M., King, B. E., and 
Wineland, D. J., {\it Science}, {\bf 272}, 1131-1136 (1996).

\bibitem{Everett57} Everett III, H., {\it Rev. Mod. Phys.} {\bf 29}, 
454 (1957). 

\bibitem{Zurek94} Zurek, W. H., pp. 175-212 in {\it Physical Origins of 
Time Asymmetry}, Halliwell, J. J., P\'erez-Mercader, J., and Zurek, W. H., 
eds. (Cambridge University Press, Cambridge, 1994).  

\bibitem{Zurek83} Zurek, W. H., ``Information transfer in quantum 
measurements'', pp. 87-116
in {\it Quantum Optics, Experimental Gravity, and the Measurement Theory},
P. Meystre and M. O. Scully, eds. (Plenum, New York, 1983).

\bibitem{Bohm51} Bohm, D., {\it Quantum Theory}, 
(Prentice-Hall, Engelwood Cliffs, 1951).

\bibitem{RauchetalPhysScripta98} Rauch, H., {\it Physica Scripta} {\bf T76},
24 (1998).

\bibitem{Pfauetal96} Pfau, T., et al., {\it Phys. Rev. Lett.} {\bf 73},
1223 (1994).

\bibitem{Chapmanetal96} Chapman, M. S., et al., {\it Phys. Rev. Lett.}
{\bf 75}, 3783 (1995).

\bibitem{Zurek00} Zurek, W. H., manuscript in preparation (2000). 

\bibitem{Lloyd96} Lloyd, S, {\it Phys. Rev.} {\bf A 55}, 1613 (1996)

\bibitem{Schumacher96} Schumacher, B., {\it Phys. Rev.} {\bf A 54}, 
2614 (1996).

\bibitem{Landauer} Landauer, R., {\it Phil. Trans. R. 
Soc.} {\bf 353} 367 (1995); also, in  
{\it Proc. of the Drexel-4 Symposium on Quantum Nonintegrability: Quantum -- 
Classical Correspondence}, D. H. Feng and B.-L. Hu, eds. (World Scientific, 
Singapore, 1998); Unruh, W. G., {\it Phys. Rev} {\bf A51}, 992 (1995)
Chuang, I. L., Laflamme, R., Shor, P., and Zurek, W. H., {\it Science},
{\bf 270}, 1633-1635 (1995).

\bibitem{Zeh73} Zeh, H. D., {\it Found. Phys.} {\bf 3}, 109 (1973). 

\bibitem{Zeh90} Zeh, H. D., {\it The Physical Basis of the Direction of 
Time}, {Springer, Berlin, 1989).

\bibitem{Albrecht92} Albrecht, A., {\it Phys. Rev.} {\bf D 46}, 5504 (1992).

\bibitem{Albrecht93} Albrecht, A., {\it Phys. Rev.} {\bf D 48}, 3768 
(1993).

\bibitem{WallsMilburn}
Walls D.F. and Milburn G.J., {\it Quantum Optics}, (Springer Verlag, 
Berlin, 1994).

\bibitem{Convolutionless} Chaturvedy, S. and Shibata, F., 
{\it Z. Phys.} {\bf B35}, 297 (1979), see also Desposito, M. and 
Hernandez, S. H., {\it Physica} {\bf 227A}, 248 (1996). 

\bibitem{HPZ} Hu, B. L., Paz, J. P., and Zhang, Y., {\it Phys. Rev.}
{\bf D 45}, 2843 (1992).

\bibitem{Leggetal} Leggett, A. J., Chakravarty, S., Dorsey, A. T., 
Fisher, M. P. A., Garg, A., and Zwerger, W., Rev. Mod. Phys. 
{\bf 59}, 1 (1987).

\bibitem{PazMazag} Paz, J. P. pp. 213-220 in {\it Physical Origin of Time 
Asymmetry}, 
Halliwell, J. J., P\'erez-Mercader, J., and Zurek, W. H., eds. 
(Cambridge University Press,  1992).

\bibitem{Caldeira} Caldeira, A. O., and Leggett, A. J., {\it Physica} 
{\bf 121A}, 587-616 (1983); {\it Phys. Rev.} {\bf A 31}, 1059 (1985).

\bibitem{UZ} Unruh, W. G., and Zurek, W. H., {\it Phys. 
Rev.} {\bf D 40}, 1071-1094 (1989).

\bibitem{Haake} F.Haake and R.Reibold, Phys.Rev, {\bf 32},  2462, (1985).

\bibitem{HPZ2} Hu, B. L., Paz, J. P., and Zhang, Y., {\it Phys. Rev.}
{\bf D 47}, 1576 (1993).

\bibitem{Grabert}
Grabert, H., Shramm, P., and Ingold, G. L., {\it Phys. Rep.} {\bf 168}, 
115 (1988).

\bibitem{Romero} Davila Romero, L. and Paz, J. P., {\it Phys. Rev.} 
{\bf A 53}, 4070 (1997).

\bibitem{FeynmanVernon}
Feynman R. P.,  and Vernon F. L., {\it Ann. Phys.} {\bf 24},  118 (1963). 

\bibitem{PHZ} Paz, J. P.,  Habib, S., and Zurek, W. H., 
{\it Phys. Rev.} {\bf D 47}, 488 (1993).

\bibitem{Wigner}  Wigner, E. P., {\it Phys. Rev.} {\bf 40}, 749 (1932). For 
a review, see Hillery, M.,  O'Connell, R. F., Scully, M. O., and
Wigner, E. P., {\it Phys. Rep.} {\bf 106}, 121 (1984). 

\bibitem{Lindblad} Lindblad, G., {\it Comm. Math. Phys.} {\bf 40}, 
119-130 (1976).

\bibitem{Garraway} Garraway, B. M., {\it Phys. Rev.} {\bf A 55}, 4636 (1997), 
{\sl ibid} {\bf A 55}, 2290 (1997).

\bibitem{AnasHu} Anastopoulos, C. and Hu, B.L., e--print 
quant-ph/9901078. 

\bibitem{Zurek86} Zurek, 
W. H., pp. 145-149 in {\it Frontiers of Non-equilibrium Statistical 
Mechanics}, G. T. Moore and M. O. Scully, eds. (Plenum, New York, 1986).

\bibitem{ZHP}
Zurek, W. H., Habib, S., and Paz, J. P., {\it Phys. Rev. Lett.}, {\bf 70}, 
1187, (1993).

\bibitem{energy} Paz, J. P.,  and Zurek, W. H., {\it Phys. Rev. Lett.}
{\bf 82}, 5181 (1999).

\bibitem{Decodeco} Anglin, J. R., Paz, J. P., and Zurek, W. H.,  
{\it Phys. Rev.} {\bf A 53}, 4041 (1997).

\bibitem{FlemingGallis} Gallis, M. R., and Fleming, G. N.,  
{\it Phys. Rev.} {\bf A 42}, 38 (1990); {\bf A 43}, 5778 (1991); 
 Gallis, M. R.,  {\it Phys. Rev.} {\bf A 48}, 1023 (1993).

\bibitem{Wisdom} Wisdom, J., Peale, S. J., and Maignard, F. {\it Icarus} 
{\bf 58}, 137 (1984); see also Wisdom, J., {\it Icarus} {\bf 63}, 272 (1985).

\bibitem{Laskar} Laskar, J., {\it Nature} {\bf 338}, 237 (1989).

\bibitem{Sussman} Sussman, G. J., and Wisdom, J., {\it Science} 
{\bf 257}, 56-62 (1992).

\bibitem{ZP94} Zurek, W. H., and Paz, J. P., 
{\it Phys. Rev. Lett.} {\bf 72}, 2508-2511
(1994); {\it ibid.} {\bf 75}, 351 (1995).

\bibitem{ZP95} Zurek, W. H. and Paz, J. P., {\it Physica} {\bf D83}, 300 
(1995).

\bibitem{Chirikov} see selected papers in Casati, G., and Chrikov, B., 
{\it Quantum Chaos} (Cambridge University Press, Cambridge, 1995).

\bibitem{Berman} Berman, G. P., and Zaslavsky, G. M., {\it Physica} 
(Amsterdam) {\bf A91}, 450 (1978).

\bibitem{HSZ98} Habib, S., Shizume, K., and Zurek, W. H., {\it 
Phys. Rev. Lett}, {\bf 80}, 4361 (1998).

\bibitem{Ottetal} Ott, E., Antonsen, T. M., and Hanson, J, {\it Phys. Rev. 
Lett.} {\bf 35},
2187 (1984); Dittrich, T., and Graham, R., {\it Phys. Rev.} {\bf A 42}, 
4647 (1990), and references therein.

\bibitem{Caves} This point of view is related to the one expressed 
by C. Caves and co--workers who emphasize
on ``hipersensitivity to perturbations'' as the defining aspect of 
quantum chaos. See Caves, C., and Schack, R., {\it Hypersensitivity to 
perturbation: An information-theoretical characterization of classical 
and quantum chaos}, in {\it Quantum Communication, Computing, and 
Measurement}, edited by Hirota, O., Holevo, A. S., and Caves, C. M., 
(Plenum Press, New York, 1997), pp. 317-330. 
This criterion was introduced by A. Peres (see Peres, A. {\it Quantum 
Theory Concepts and Methods}, Kluger (1995)). 
 
\bibitem{HuSh}  Shiokawa, K., and Hu, B. L., {\it Phys. Rev} {\bf E 52}, 2497
(1995). 

\bibitem{Sarkar}  Miller, P. A., and Sarkar, S. {\it Phys. Rev.} {\bf E 58}, 
4217 (1998); {\bf E 60},  1542 (1999). 

\bibitem{Patt99} Pattanayak, A. K., {\it Phys. Rev. Lett.} {\bf 83},  4526
(2000).

\bibitem{Pastawski99} Pastawski, H.,  Usaj, G., and Levstein, P. 
{\it ``Quantum
chaos: an answer to the Boltzmann--Loschmidt controversy?''}, preprint 
Famaf (2000); for interesting related experimental 
work using NMR techniques see also Pastawski, H.,  Usaj, G., and  
Levstein, P., {\it Chem. Phys. Lett.} {\bf 261} 329 (1996).

\bibitem{Monteoliva} Monteoliva, D.,  and Paz, J. P., (2000) {\it to appear}. 

\bibitem{GellmannHartle} Gell-Mann, M., and Hartle, J. B., in {\it 
Complexity, Entropy, and the
Physics of Information}, Zurek, W. H., ed. (Addison-Wesley, Reading, 1990).

\bibitem{Giulinietal} Giulini, D., Joos, E., Kiefer, C., Kupsch, J., 
Stamatescu, I.
-O., and 
Zeh, H. D., {\it Decoherence and the Appearance of a Classical World in 
Quantum Theory}, (Springer, Berlin, 1996).

\bibitem{Steane}
Steane A., 1996, {\it Phys. Rev. Lett.} {\bf 77}, 793. 
Steane A., 1996, {\it Proc. Roy. Soc. Lond.} {\bf A452}, 2551. 

\bibitem{Shor95} Shor P., 1995, {\it Phys. Rev.} {\bf A 52}, 2493.

\bibitem{McWilliams} Mc Williams and Sloane, {\it 
``Theory of Error Correcting Codes''} (Elsevier, Amsterdam, 1977).

\bibitem{perfect} Laflamme R., Miquel C., Paz J.P. and Zurek W.H., 
{\it Phys. Rev. Lett.} {\bf 77}, 198 (1996).

\bibitem{PazZurek98} Paz, J. P., and Zurek, W. H., {\it Proc. Roy. Soc. 
London} {\bf A 454}, 355 (1998).

\bibitem{Calderbanketal}
Calderbank, A. R., Rains, E. M., Shor, P. W., and Sloane, N. J. A., 
{\it Phys. Rev. Lett.}, {\bf 78}, 405 (1997).

\bibitem{GottesmanPhD}
Gottesman, D. 1998, Caltech PhD Thesis, quant-ph, see also 
``Stabilizer codes and quantum error correction", Preprint quant-ph/9705052; 
{\it Phys. Rev.} {\bf A 54} 1862 (1996).

\bibitem{KnillLaflamme}Knill, E. and Laflamme, R., 
Preprint quant-ph/9608012; {\it Phys. Rev.} {\bf A 55}, 900 (1997). 

\bibitem{CleveGottesman} Cleve, R.,  and Gottesman, D.,  {\it Phys. Rev.}
{\bf 56} 76 (1997).

\bibitem{Pringe} H. Pringe, MsC Thesis (unpublished), Buenos Aires 
University (1997)

\bibitem{Cosmology} Halliwell, J. J., {\it Phys. Rev.} {\bf D 39} 2912 (1989);
Kiefer, C., {\it Class. Quantum Grav.} {\bf 4} 1369 (1987); 
Paz, J. P., and Sinha, S., {\it Phys. Rev.} {\bf D 45} 2823 (1992); 
{\it ibid} {\bf D 44} 1038 (1991); for more recent discussion see 
Lombardo, F., Mazzitelli, F. D., and Monteoliva, D. {\it Phys. Rev.} 
{\bf D} (2000) {\it to appear}.  

\bibitem{Engeneering} Poyatos, J. F.,  Cirac, J. I. and  Zoller, P., 
{\it Phys. Rev. Lett.} {\bf 77} 4728 (1997). 

\bibitem{Davidovich} Davidovich, L.,  Brune, M.,  Raimond, J. M., and 
Haroche, S., {\it Phys. Rev.} {\bf A 53} 1295 (1996). 

\bibitem{Dalvit} Dalvit, D., and Maia Neto, P., {\it Phys. Rev. Lett.} 
{\bf 87} 798 (2000); {\it see also} quant-ph/0004057. 

\bibitem{BECdeco} Anglin, J., {\it Phys. Rev. Lett.} {\bf 79}  6 (1997).

\bibitem{DDZ} Dalvit, D., Dziamarmaga, J., Zurek, W. H., {\it Phys. Rev.} 
{\bf A } (2000) {\it to appear}.  

\bibitem{Webb} Mohanty, P., Jariwada, E. M. Q., and Webb, R. A., 
{\it Phys. Rev. Lett.} {\bf 77} 3366 (1995); Mohanty, P., and 
Webb, R. A., {\it Phys. Rev.} {\bf B55}, R13 452 (1997).
\end{document}